# The Evaluation Gap in Astronomy
# – Explained through a Rational Choice Framework

Julia Heuritsch

julia.heuritsch@hu-berlin.de
Humboldt Universität zu Berlin, Research Group "Reflexive Metrics", Institut für Sozialwissenschaften, Unter den Linden 6, 10099 Berlin

# Abstract

The concept of evaluation gaps captures potential discrepancies between what researchers value about their research, in particular research quality, and what metrics measure. The existence of evaluation gaps can give rise to questions about the relationship between intrinsic and extrinsic motivations to perform research, i.e. how field-specific notions of quality compete with notions captured via evaluation metrics, and consequently how researchers manage the balancing act between intrinsic values and requirements of evaluation procedures. This study analyses the evaluation gap from a rational choice point of view for the case of observational astronomers, based on a literature review and 19 semi-structured interviews with international astronomers. By taking a close look at the role of institutional norms and different forms of capital – such as funding, publication rates and granted telescope time – at play in astronomy, light can be shed on the workings of the balance act and its consequences on research quality in astronomy. We find that astronomers experience an anomie; they want to follow their intrinsic motivation to pursue science in order to push knowledge forward, while at the same time following their extrinsic motivation to comply with institutional norms. The balance act is the art of serving performance indicators in order to stay in academia, while at the same time compromising research quality as little as possible. Gaming strategies shall give the appearance of *compliance*, while institutionalised means how to achieve a good bibliometric record are used in *innovative ways*, such as salami slicing or going for easy publications. This leads to an overall decrease in research quality.

## Declarations

**Funding:** This study was performed in the framework of the junior research group "Reflexive Metrics" funded by the BMBF (German Bundesministerium für Bildung und Forschung; project number: 01PQ17002) and affiliated with the Humboldt Universität zu Berlin.

**Conflicts of interest:** The author *declares that there is no conflict of interest.*

# Content

# Introduction

**Astronomy (or astrophysics)** is an interesting field to study the effects of evaluation procedures on its knowledge production processes. This is due to a variety of reasons. First, astronomy is relatively well separated from other scientific fields (Gläser et al., 2017; p.989). Its characteristic methodology and epistemology, that guide the knowledge production process, are unique. Because of its observational nature, astronomy was "confronted with the attribution of antirealism by Ian Hacking" (Anderl, 2015; p.1). Anderl (2015; p.2) however disregards this attribution, explaining that the methodology of astronomy "resembles the criminology of Sherlock Holmes: the astrophysicist must look for all possible traces and clues that may help to illuminate what has happened in a given region of the universe". Thereby, astronomy strives to understand the "causal history of singular events" in order to "find general causal relations with respect to classes of processes or objects" (ibid. p.7). In order to find those general causal relations, astronomy has to rely on the "'cosmic laboratory': the multitude of different phenomena and environments, naturally provided by the universe." Hence, epistemology of astrophysics is strongly based on the observation of distant objects "on the use of models and simulations and a complex treatment of large amounts of data" (ibid. p.1) in order to make causal inferences. Therefore, astronomy's methodology "can cope with the missing possibility of direct interaction with most objects of research" (ibid. p.1). Second, while astronomy is "traditionally viewed as the 'purest science,' driven by curiosity and with no practical application" (Baneke, 2019; p.24), Roy & Mountain (2006) point out that there are spinoffs and mutual technological developments from astronomy that also benefit society. In fact, astronomy, as one of the oldest sciences (Fara, 2010), has always had "an important place in the development of humanity" (Fernández 1998; p.61), because of its ability to solve practical problems such as navigation through the position of heavenly bodies. In order to compete funding, astronomy increasingly has to demonstrate its economic value (Baneke, 2019). Third, Basu & Lewison (2005), point out that astronomy is a suitable model for bibliometric evaluations, because almost all work is written in English and "conducted in the public domain and without commercial constraints on publication" (p.234). This allows for a good coverage of publications by journal databases (Gläser et al., 2017). Fourth, astronomy has an interesting practical organisation, which is very internationalised and democratised (Taubert, 2019), opening the discussion to historical directions and sociological analyses. For example, "the generation of empirical astrophysical knowledge relies on the distribution of observing time at usually internationally operated observatories" (Anderl, 2015, p.2; cf. Heidler, 2011). Another example is astronomy's interesting data sharing culture (Zuiderwijk & Spiers, 2019). It arises from the vast amount of data astronomers have to deal with and encouraged by their intrinsic motivation to efficiently build on other's knowledge, while at the same time discouraged by competition. All in all, astronomy as the "pure science" with an interesting methodology & epistemology, an internationalised community which includes collaboration and competition, generating outcomes that are easily bibliometrically evaluated, is an interesting field to study what effects such evaluations have on its knowledge production process.

When studying effects of evaluation procedures on knowledge production processes, one needs to consider that indicators, such as publication & citation rates, do not merely describe, but also prescribe behaviour (Desrosières, 1998). ***Reflexive Metrics***[1] (cf. Fochler & De Rijcke, 2017; Heuritsch, 2019a) acknowledges this. Because it is a relatively new field in the sociology of science, there have been a handful of studies on the efficacy of indicators in the field of astronomy, but not yet a comprehensive study on what effects indicator use has on knowledge production in astronomy.

---

[1] https://www.dzhw.eu/en/forschung/projekt?pr_id=605 & https://www.sowi.hu-berlin.de/en/lehrbereiche-en/wissenschaftsforschung-en/Research/young-investigator-group-reflexive-metrics?set_language=en

For example, Kurtz & Henneken (2017; p.703) note that "one of the principal goals of individual evaluations of individuals, whether bibliometric or not" is to predict "future performance" of those individuals. The authors acknowledge that "in all of science there is no such thing as the one 'true' indicator", because "all measures have error, and all measures are indirect" (ibid. p. 695). Having made that point, they proceed to studying the indicators' efficacy in predicting an astronomer's scholarly performance by examining different methods of evaluation, such as peer review, citation ranking, and download ranking. Their results show that all those methods are "approximately equally able to predict future performance, but have quite different properties, and yield different results (in terms of the exact individuals chosen)" (ibid. p. 708). They conclude that the tested measurements show large statistical scatter and may only in combination with each other and when keeping the context in mind serve as proxies.

An older study by Tatarewicz (1986) examined what effects federal funding has on scientific careers and publishing activity in the case of the development of planetary science during the space race. Davoust & Schmadel (1991) study the increased publication activity of astronomers over time and with increasing age. The authors point out that an obvious subject of further investigation is the correlation between publication activity and progress in astronomy, acknowledging that those factors are not equal as might be assumed by an indicator like publication rate.

McCray (2000), Atkinson-Grosjean & Fairley (2009) and Baneke (2019) performed studies on the ***moral economy*** of astronomy. The term draws back to Thompson (1993), for which a moral economy is related with a mentalité – "the expectations and traditions – that structured and mediated interactions between the consumers and producers of life's basic needs" (McCray, 2000, p.686). This includes what rights people have and how (non-)economic relations are regulated through social norms. Kohler (1999) adapted that concept to the activities of experimental scientists. Based on Thompson (1993) & Kohler (1999), Atkinson-Grosjean & Fairley (2009; p.148) write that a moral economy is a "system of shared values, traditions, and conventions about ways of doing, being, knowing, and exchange" held by a moral community. While the authors define "community" as something held together by "group values rather than overarching social structures or institutions" (ibid. p.148), Baneke (2019) states that the moral economy of a scientific community "includes scientific, institutional and […] moral values" (p.3). Atkinson-Grosjean & Fairley (2009) distance themselves from a material determinism, where "material conditions dictate the course of events, but rather that they demarcate the bounds of the possible" (ibid. p. 150). Moral economies are shaped by material, societal and cultural constraints and conditions – contrary to Merton's (1938) view, where the scientific institution is "distinctly separate from the wider social environment and operates within internally established universal norms" (Atkinson-Grosjean & Fairley, 2009; p.149). One of the functions of a moral economy is to "provide provisional maps for navigating the messy, contingent spaces where societal and scientific values are negotiated" (ibid. p.169). As the authors point out, a moral economy in science may be sensitive to external conditions, such as the "grants-based structure of 'normal' academic inquiry" (ibid. p.149). This raises questions about "research ethics, scientific authority, unintended consequences, power differentials, and cost-benefit ratios" (ibid. p.149). According to McCray (2000), in astronomy required resources to conduct scientific work include large sums of money to fund the construction of telescopes. In the USA these were, historically, built by private institutes or universities instead of the state, resulting in privileged access to those who were affiliated to the respective facility. Only after the National Science Foundation (NSF) started to provide federally-funded observatories in the mid-1950s, access to telescopes was made available to a larger population of astronomers. While this was a first step to democratise the telescope time allocation in astronomy, Atkinson-Grosjean & Fairley (2009) and McCray (2000) point out that the distinction between those who have privileged access to telescopes based on their institutional affiliations (the "Haves") and those who have to compete for telescope time (the "Have-Nots") resulted into two different moral economies in astronomy.

Taubert (2019) extensively describes the publication and communication infrastructure prevalent in astronomy. He elaborates on the relationship between that infrastructure and the ***two codes of science***: "truth" as primary code and "reputation" as secondary code, claiming reputation for a scholarly achievement happens through publication and its attribution through citation (Taubert, 2019). According to Taubert, an important function of reputation is to reduce the complexity of the vast amount of the scientific literature and to make the selection of information easier. Next to this orientation function, reputation is also used to measure and compare scholarly performance (Taubert, 2019) through publication and citation rates. On a side note he mentions that those kind of instruments of New Public Management can have "unintended consequences" (a term which draws back to Merton, 1936) and coupling measuring scholarly performance with incentives can have reactive effects on publishing behaviour of scientists.

As pointed out by Heuritsch (2019a) the performativity of numbers; i.e. the fact that they are reactive and do not merely describe, but also prescribe behaviour (e.g. Porter, 1995; Desrosières, 1998; Espeland & Stevens, 2008) is acknowledged by the term ***constitutive effects***, which was first used by Dahler-Larsen (2014). His conceptual move away from "unintended consequences" avoids assumptions of intention and instead puts the emphasis on the shaping role of indicators. By contrast, a concept, which sheds light on unintended effects of indicator use, is the "***evaluation gap***" (Wouters, 2017). The evaluation gap refers to the "discrepancy of what is being measured by indicators and the quality of the scientific content, as perceived by the researchers of the field" (Heuritsch, 2019a; p.148). As pointed out by the author, unintended effects of such a gap may be the deployment of a number of questionable research practices, such as "goal displacement" and "gaming" (e.g. Laudel & Gläser, 2014; Rushforth & De Rijcke, 2015). Dahler-Larsen (2014) treats both concepts as exclusive, because depending on paradigmatic foundations, "constructivists may be immediately comfortable with the idea [of constitutive effects], while rationalists and functionalists may still find value in the idea of unintended consequences" (ibid. p. 982). However, Heuritsch (2019a) questions whether those concepts necessarily need to be alternatives or whether they can be reconciled in order to better understand what effects indicator use has on knowledge production processes.

In order to study those effects in the field of astronomy, Heuritsch (2019a) performed a case study of international astronomers at Leiden Observatory and explored the intrinsic and extrinsic motivations to perform research. On the one hand, astronomers are driven by intrinsic values such as "curiosity, truth-finding and 'pushing knowledge forward'" (Heuritsch, 2019a; p.176). On the other hand, what indicators such as publication & citation rates value as good research(ers) gives an extrinsic motivation to perform research. Because "astronomer's values are based on the realist account that astronomers generally hold" (ibid. p.176), their intrinsic values are hardly co-constituted by research evaluation processes. This means that a discrepancy remains between the astronomer's intrinsic and extrinsic values, giving rise to an evaluation gap. This gap in turn "has a variety of constitutive effects on knowledge production, ranging from research agendas, researcher's behaviour and identities to research content" (ibid. p.177). Because their intrinsic values remain as their "ideals", but they must survive performance evaluations in order to stay in academia, astronomers "hold two opposing notions of science: the 'ideal' one which corresponds to their intrinsic values and the 'system' notion." (ibid. p.177). Consequently, a third notion arises – the balance act. Astronomers try to manage a balance act between fulfilling the requirements of the evaluation system, on the one hand, and not compromising their internal notions of what good quality research means to them, on the other hand.

This paper builds upon the study conducted by Heuritsch (2019a). It expands on the effects that indicator use has on research quality in astronomy by employing a rational choice framework. ***Rational choice theory*** (RCT) is a sociological theory of action, which explains how macro phenomena emerge out of micro phenomena and how, in turn, they can influence micro phenomena (Esser, 1999). This is reflected by the so-called *Coleman Boat* (based on Coleman, 1990) as depicted in *Figure 1*.

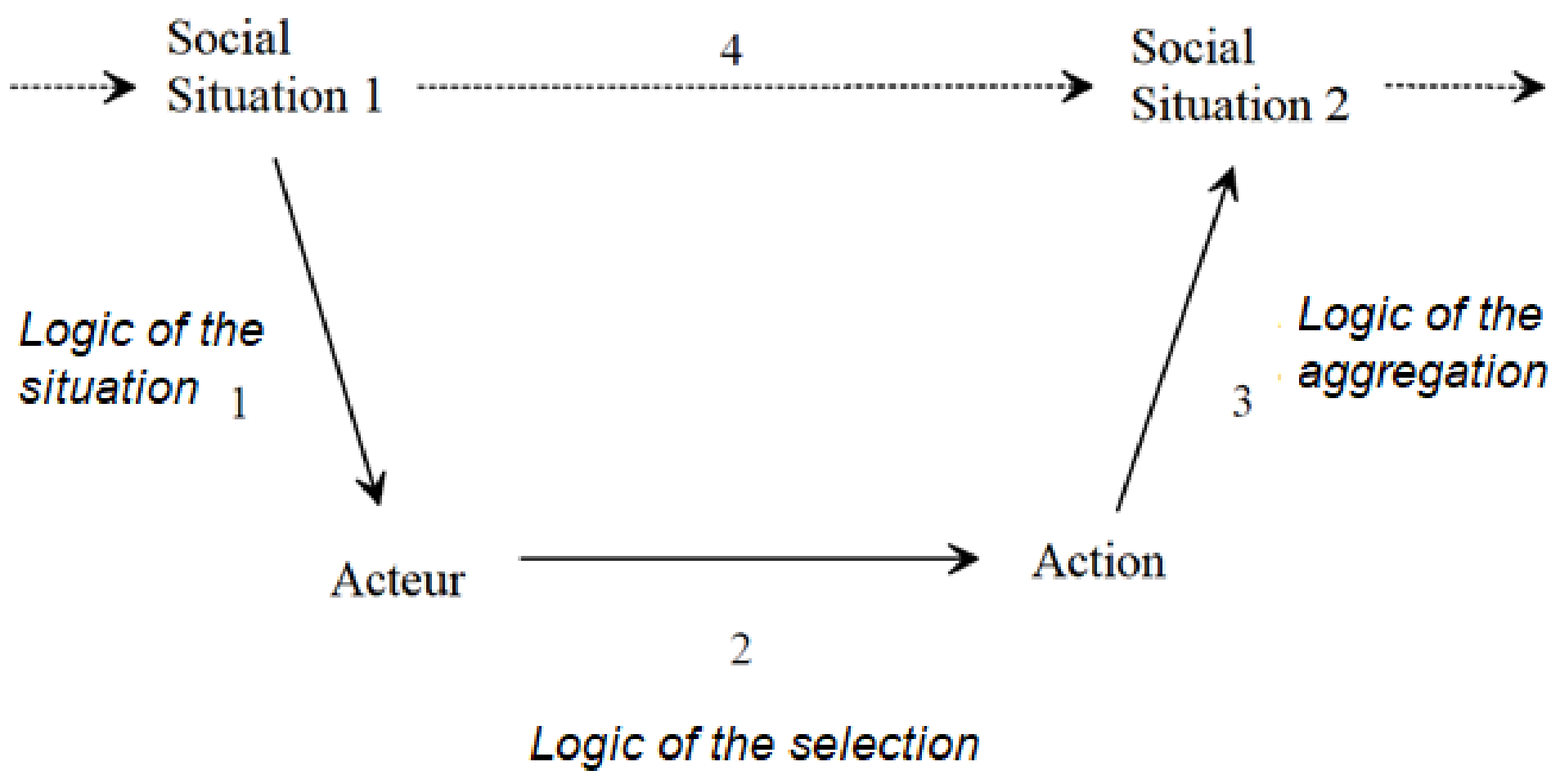

*Figure 1: The Coleman Boat depicting the foundational model of the sociological explanation of actions and resulting collective phenomena (modified by the author, based on Esser, 1999)*

According to RCT at any point of making a decision, an actor finds himself in a situation (*Figure 2*), which is comprised of four constituents: three external and one internal. The external constituents are 1.) Material opportunities at present, 2.) Institutional rules, such as norms and 3.) A cultural reference frame, such as symbols and shared values. The internal constituent is the identity of the actor, comprised of its values, wishes, skills and personality.

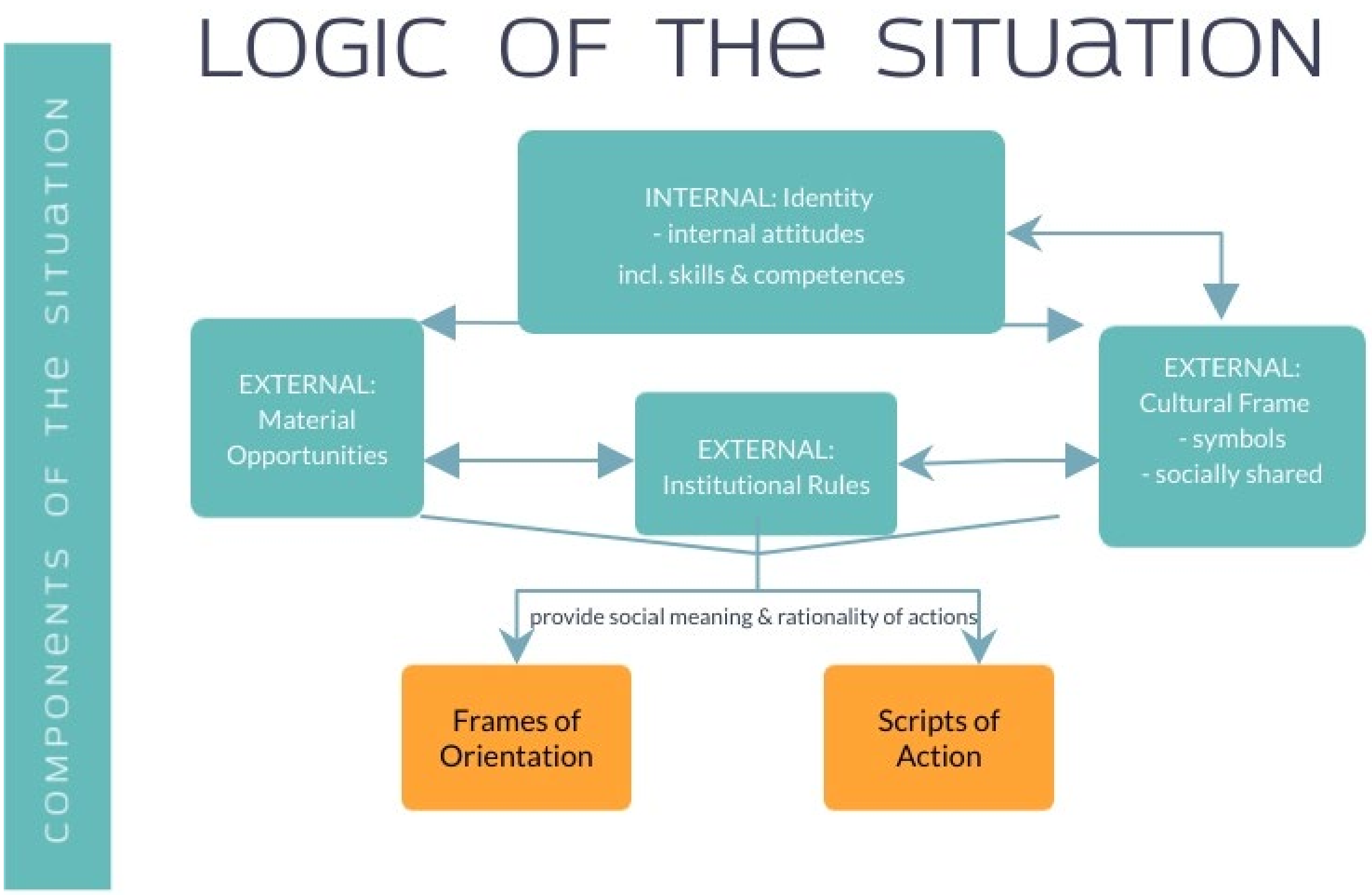

*Figure 2: Constituents of the Logic of the Situation (created by the author based on Esser, 1999)*

Reconstructing the constituents of an actor's situation results in the *logic of the situation*. This is the *first step* in finding a sociological explanandum for the resulting macro phenomenon (see *Figure 1*). The *second step* is finding *bridge hypotheses,* which translates the logic of the situation into the variables that the actor's behaviour can depend on. An *action theory* is now needed to explain the action on the basis of those variables. The arguably most general and detailed action theory is the so-called *expected utility theory* (EU theory; *Figure 3*), where the actor chooses the action amongst other alternatives, according to which EU value is the largest. The EU-theory is a causal action theory that is comprised of the following steps (Esser, 1999): 1.) An actor perceives being in a situation as defined by the 3 constituents (*see Figure 2*) which may either trigger an action or makes the actor consciously decide to take action. An action is always requires a choice between alternatives, whether one is consciously aware of them or not. 2.) Every action has consequences, which may be anticipated or to be avoided. 3.) The actor evaluates those consequences according to their attractiveness. The more factors the actor includes, the more accurate the evaluation. For example, the actor may not only consider the more obvious "goodness" or "badness" of the outcome, but also what kind if psychological costs attaining the outcome will have (e.g. sunk cost effects, framing effects, opportunity cost effects, ownership effects, certainty effects, reference point effects, availability effects, representation effects and satisficing; Esser, 1999). 4.) When assessing possible consequences of an action one essentially makes a prediction. Hence, an outcome can only be attained with a certain probability. An actor therefore needs to assess with what expectation an outcome may be attained. 5.) Both, the attractiveness and the probability of the occurrence of an outcome play important roles in choosing which action is the most attractive to take. In mathematical terms, the product of both factors determine the *utility* of an action. The higher the utility, the better the action. Note, that utility as defined here is something entirely subjective; the same outcome may have an entirely different utility for different actors, as it depends on their respective situations. 6.) The actor performs the action with the highest utility[2]. Often, critics of RCT voice their concerns that these 6 steps are way to cognitively demanding as to happen in practice at every action a human takes. Indeed, that is true. However, consciously exercising through all 6 steps is only one *mode of selection* of many others. This reflected mode requires the most cognitive effort an is usually chosen, when important decisions with long lasting consequences have to be taken (like choosing a job, what to study, which house to buy). Other modes, like the normative mode or the affect-driven (in everyday speech "irrational") mode do not require such cognitive efforts and allow for a shortcut through so-called "scripts". Scripts are pre-defined instruction manuals of how to act in certain situations on the basis of cultural and/ or institutional norms. As we will see in *section 1.3,* it is the very function of institutional norms to guide behaviour and reduce the complexity of situations and possible actions. What selection mode to choose depends on the *frame of orientation*. By framing the situation, an actor defines what is important or relevant. Hence, a frame of orientation is a model of the situation, often including the definition of certain goals, which gives an actor orientation. The attentive reader will have noticed that the choice of the frame of orientation and the related selection mode is an action like any other, following the EU-theory. Hence, it also follows from the logic of the situation what frames and scripts can be used.

The *final and third step* is to find *transformation rules* which explain how the individual actions of many actors aggregate to the sociological phenomenon in question. This step is also called the *logic of aggregation*. The result is a sociological explanation of how one social phenomena causes another, by reconstructing the causal mechanisms, which happen on a micro-level relative to the macro-level of the collective phenomena. Hence, exercising all 3 steps delivers a causal explanation for how *Social Situation 1* leads to *Social Situation 2,* which otherwise could not only be observed and described on a phenomenological level (Step 4 in *Figure 1)*.

RCT provides the appropriate tools to reconstruct the structural conditions for the evaluation gap observed by Heuritsch (2019a) and can explain how constitutive effects shape research behaviour. As Heuritsch (2019a) pointed out, Dahler-Larsen (2014) describes constitutive effects merely as something that occurs to passive actors. By also accounting for the internal constituents of an actor's situation, RCT can provide a more integrated theory on the consequences of indicator use in research evaluation on knowledge production in astronomy. In particular, with RCT we can study the balance act astronomers undergo in order to stay true to their intrinsic values and what that compromise means for research quality.

As Langfeldt et al. (2020; p.132) point out, "we know little about how formal peer review and research evaluation interact with more general notions of research quality, or how notions are impacted by the increasing availability of quantitative indicators of research performance". RCT provides such a "meaningful framework that can link the institutional conditions and diversity to the empirical manifestations of quality and its criteria" (ibid. p.131). The resulting integrated causal theory about effects that indicators and incentives can have on research quality in astronomy is the aim of this paper. By understanding the current situation and moral economy of astronomy, this and future studies may inform policy makers about the structural conditions that need to be taken into account and fostered – in practice – in order to support astronomers to do their research to their best of their capabilities and conscience.

---

[2] Due to inner conflicts (i.e. cognitive dissonances; see *section 2.2*), such as the approach-avoidance conflict, the avoidance-avoidance conflict or the approach-approach-conflict (Esser, 1991), an action with a seemingly lower utility may be chosen. However, that is just an indication for hidden costs not having been accounted for during the evaluation step, and not a proof for the EU theory not to work in practice, like many critics of RCT would argue.

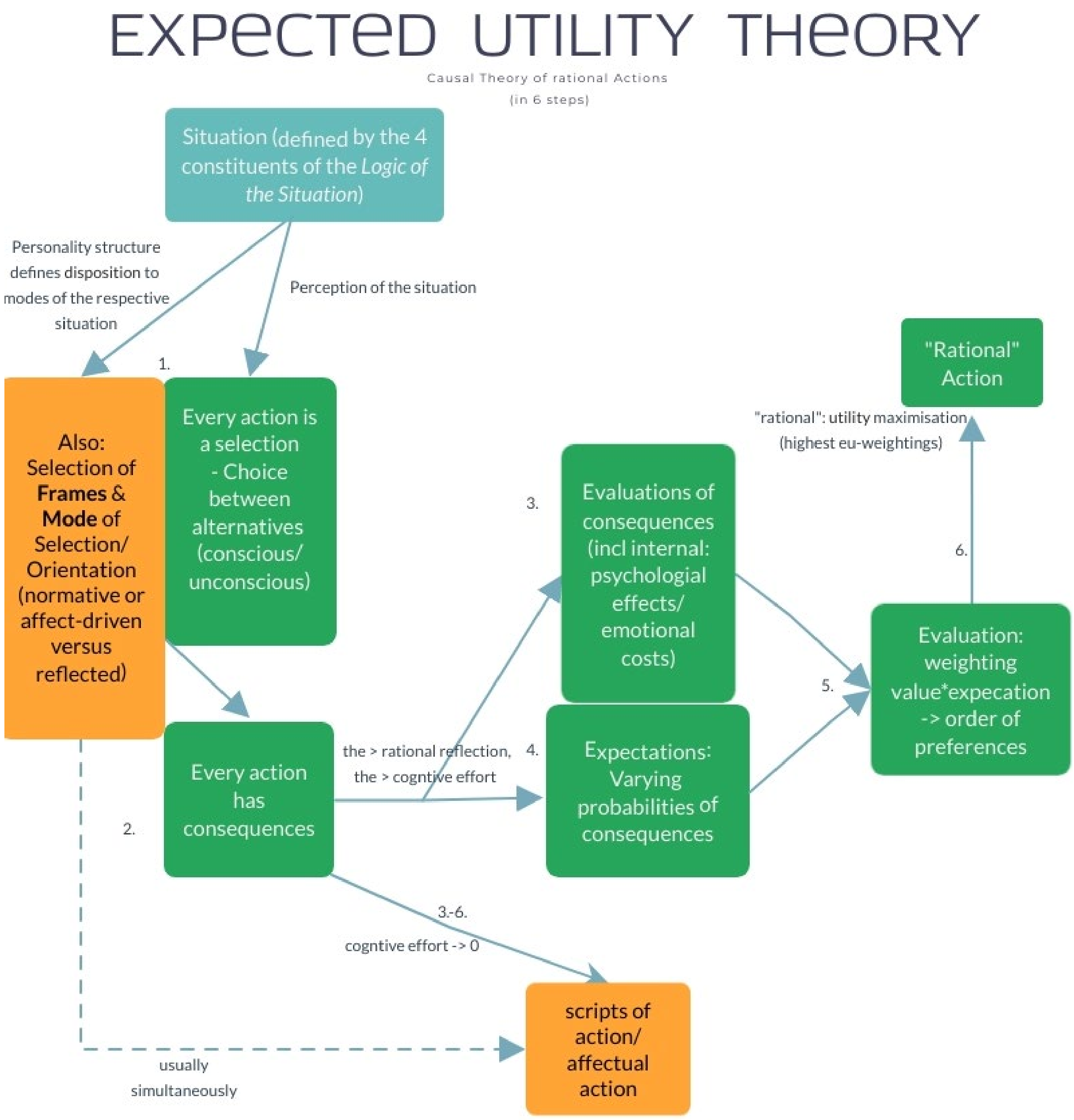

*Figure 3: Expected Utility Theory (created by the author based on Esser, 1999)*

# Methods

This research consists of 19 semi-structured interviews with international astronomers and a comprehensive review of astronomy-related science studies literature. A detailed demography of the interviewees can be found in *Table 1*. The interviews were held via Skype and in person at three different conferences, the European Week of astronomy and Space Science in June 2017, the General Assembly of the International Astronomical Union in August 2018 and the International Astronomical Congress in October 2018. At the conferences, a letter of invitation was provided to the organisers who distributed it via the conference-apps and/ or news updates. In addition to those astronomers who responded, the chairs of sessions which are related to observational astronomy were asked to pass on the invitation text, which resulted in interviews via Skype.

*Table 1: Interview Sample Structure and its Represantivity – Comparables[3] represent those astronomers who have an IAU membership. IAU, the International Astronomical Union, is the biggest professional astronomical society with a total of 13574 members. From the comparables, we can infer that our sample obtains a good representativity, with female astronomers slightly overrepresented. Only established (>60 years) astronomers are underrepresented. However, this may be due to the fact, that astronomers may remain (active) members of the IAU even after retirement. The list is sorted in chronological order with respect to the date of the interview.*

| Label | Gender | | Academic Position (Age) | | | Place of Work | |
|---|---|---|---|---|---|---|---|
| | female | male | Early Career (<35) | Postdoc/ Established (35-60) | Established (>60) | Western | Non-Western |
| Int-Faculty1 | 0 | 1 | 0 | 1 | 0 | 1 | 0 |
| Int-PhD1 | 0 | 1 | 1 | 0 | 0 | 1 | 0 |
| Int-PhD2 | 1 | 0 | 1 | 0 | 0 | 1 | 0 |
| Int-Faculty2 | 0 | 1 | 0 | 1 | 0 | 1 | 0 |
| Int-PhD3 | 0 | 1 | 1 | 0 | 0 | 1 | 0 |
| Int-Journal | 0 | 1 | 0 | 1 | 0 | 1 | 0 |
| Int-Faculty3 | 0 | 1 | 0 | 0 | 1 | 1 | 0 |
| Int-Faculty4 | 0 | 1 | 0 | 0 | 1 | 1 | 0 |
| Int-Faculty5 | 1 | 0 | 0 | 1 | 0 | 0 | 1 |
| Int-Faculty6 | 0 | 1 | 0 | 0 | 1 | 1 | 0 |
| Int-Faculty7 | 0 | 1 | 0 | 1 | 0 | 0 | 1 |
| Int-Faculty8 | 0 | 1 | 0 | 1 | 0 | 1 | 0 |

[3] https://www.iau.org/administration/membership/individual/distribution/

| | | | | | | | |
|---|---|---|---|---|---|---|---|
| Int-Faculty9 | 1 | 0 | 0 | 1 | 0 | 1 | 0 |
| Int-Postdoc1 | 1 | 0 | 0 | 1 | 0 | 1 | 0 |
| Int-Faculty10 | 0 | 1 | 0 | 0 | 1 | 1 | 0 |
| Int-Faculty11 | 0 | 1 | 0 | 1 | 0 | 1 | 0 |
| Int-Postdoc2 | 1 | 0 | 0 | 1 | 0 | 1 | 0 |
| Int-Faculty12 | 0 | 1 | 0 | 1 | 0 | 1 | 0 |
| Int-Postdoc3 | 0 | 1 | 0 | 1 | 0 | 0 | 1 |
| **Sample total** | 5 | 14 | 3 | 12 | 4 | 16 | 3 |
| **Sample in percent** | 26% | 74% | 16% | 63% | 21% | 84% | 16% |
| **Field[3] in percent** | 18% | 82% | 4% | 57% | 39% | 80% | 20% |

Building up on the research of Heuritsch (2019a), this study focusses on observational astronomers, which is why only observational astronomers were chosen for the interviews. This is because, to produce data, observational astronomers need to compete for limited observation time at telescopes and then be lucky to have the right weather conditions. Once granted, observation time does count like received funding in an astronomer's CV. However, non-detections are more common than detections and a substantial amount of non-detections are not publishable (Patat et al., 2017). Hence, observational astronomers face the risk to fall through the cracks of metrics in every step of their knowledge production process. Although this study focusses on observational astronomy, for the sake of simplicity we use the terms "astronomy" and "astronomers". Besides, many of the findings are applicable to astronomy in general.

Accounting for the fact that observational astronomers need observational data to do their work, interview questions address the challenges astronomers face when acquiring those data. Other questions concern themes such as evaluation procedures, which indicators are important therein, what obstacles astronomers face when doing their research, gaming strategies and what motivational factors drive them in doing their research.

Interviews were 50 to 100 minutes in length, transcribed by a company and coded in MaxQDA according to Mayring's qualitative content analysis (Mayring, 2000). Deductive category application was employed on basis of the codes used in the previous study (Heuritsch, 2019a) and further developed during analysis.

In addition to the interviews, this paper contains a comprehensive literature review. This is comprised of the publications of the 18 most important science studies & STS journals and proceeding series in the Scopus in-house database of the competence centre for bibliometrics covering publications until April 2018, which contain the keyword "astro" as the start of a word in their titles (to not include words like gastronomy, catastrophes etc). The resulting sets consists of 60 entries of 11 different sources. In addition, the data base was searched for publications which cite at least 2 of the set (14 results) or which were cited by at least 2 of the set (83 results). After cleaning doubles, manual checks of relevance and reiterating the process in the online version of Scopus (to include more recent publications) 43 relevant papers were downloaded to be reviewed for this study.

Furthermore, an instant survey was conducted with the online Pingo[4] tool at the department of physics at the University of Dallas (Texas, US). 34 astronomers participated in the survey which asked 5 questions related to the notion of ontology and epistemology that astronomers hold. An answer was usually comprised of one or a few words. In order to quantify them, such that we can determine what answers were given most, we coded similar answers into the same category (e.g. replicable, replicability and reproducibility were coded into "replicability"). Naturally, this survey was rather exemplary instead of representative. Nevertheless, we will use the results in *section 1.1.2* to further illustrate our findings from the interviews.

This paper will be structured as follows: The subsequent chapters will provide the results. Building on the empirical evidence from the interviews and findings from the literature review the results will be framed according to RCT. The 1st Chapter will reconstruct the *logic of the situation*. This includes the internal constituent such as the astronomers' values and definition of scientific quality. It further contains the external constituents as in material opportunities like funding, telescope time and grants; institutional norms like rules for publication and the cultural frame. The summary of this chapter will include the *bridge hypotheses* about how the logic of the situation guides and motivates astronomers' research behaviour. The 2nd Chapter outlines the *action theory* in the case of astronomy, including the balance act. The 3rd Chapter is dedicated to the *logic of aggregation*, explaining how indicator use in astronomy, through shaping astronomers' research behaviour leads to an evaluation gap, where research quality is sacrificed for quantity. The final chapter will contain discussion and conclusion.

[4] https://pingo.coactum.de/

# Chapter 1: Logic of the Situation

As we outlined in the introduction, determining the *logic of the situation* with its internal and three external constituents (material opportunities, institutional norms & cultural frame; see *Figure 2*) is the first step in finding the sociological explanandum (see *Figure 1*). The logic of the situation will in a second step be translated into the *bridge hypotheses* that are the variables which the actor[5] bases their behaviour on. In fact, those variables are the actor's motivational factors that go into an actor's decision of how to act.

Zuiderwijk & Spiers (2019) summarise many existing concepts of ***motivation***. Put simply, motivation is the reason for "why we do the things we do" (Keller, 2009; in Zuiderwijk & Spiers, 2019; p.2). There have been many studies on the role of motivations in decision-making in various contexts and from various perspectives. Different scientific fields, such as genetics, cognitive science, psychology and sociology have performed such studies. Behavioural psychology, for example, assumes that an "adequate theory of human behaviour can be developed by examining the reaction of people to environmental stimuli" (ibid. p.2). Motivations can be also categorised in many ways. For example, there may be hedonic and utilitarian motivations. There may be *intrinsic* and *extrinsic* motivations; "Intrinsic motivation refers to performing an activity simply because it is interesting, brings enjoyment and is satisfying, as opposed to extrinsic motivation, which refers to doing an activity because it leads to an external outcome (e.g. fulfilment of role, public support)" (Entradas et al., 2019; p.74). Usually, our situations and decisions are complex, such that we are not only driven by a single motivational factor. As Zuiderwijk & Spiers (2019; p.229) put it: "motivation is typically influenced by the complex interaction of multiple factors, both intrinsic and extrinsic, rather than by a single factor." This is congruent with the RCT framework, where various motivational factors can be derived from the complexity of one's situation. Therefore, we simply define **intrinsic motivational factors as those that stem from the *internal component*** of the actor's situation and **extrinsic motivational factors as those arising out of the three *external components***. Given that the four components of an actor's situation – and hence the resulting motivational factors – are interwoven and interact with each other (see *Figure 2*), it is difficult to do those interactions justice in a paper that is written in a linear way. In the attempt to structure the paper as logically as possible and to not include too much redundancy concepts may be introduced before we can explain them adequately in a later section. Important concepts and keywords of individual paragraphs are marked in ***bold and italics***.

## 1.1 Internal Component

### 1.1.1 Curiosity & Notion of Truth

The personality of the astronomer – their values, wishes and attitudes – comprise the internal component of the *logic of the situation*. The internal component represents the astronomer's intrinsic values regarding science and determines their intrinsic motivation to perform science. Heuritsch (2019a) shows that astronomers are driven by curiosity about the truths of the universe. This study expands on the astronomer's intrinsic values in order to make sense of what internal factors drive their actions.

[5] Note that in this paper an "actor" is an individual astronomer.

Almost all interviewees (9 implicitly & 3 explicitly: Int-PhD3, Int-Postdoc1 & Int-Faculty2) mentioned curiosity and a liking for problem solving as their main reason to be an astronomer. In particular, they are excited about the opportunity of "understanding the universe and the world we live in" [e.g. Int-Faculty10, Int-Faculty11] and the intellectual challenge this comes with. They enjoy the process of finding ***causal explanations*** for natural phenomena, regardless of whether or not an application results from it.

> "I wanted to understand phenomena that seemed unusual or that seemed surprising or intriguing, all of those. And then actually, uh, try to find physics explanations. So, so we have to understand the phenomena and the processes. And then often you build a scenario, how the astrophysical circumstances could develop in time to lead to the phenomenon, which we first found as intriguing or surprising." [Int-Faculty4]

> "You know, it is much more difficult to understand the human body or life, than the sun or the stars. So starting from the idea that with the physical laws you can understand how the universe works, was, uh, a incentive to study this." [Int-Faculty10]

> "Of course, it's astronomy yeah, it will not make your life better, it will not make you live longer, uh, whatever, yeah, uhm, cure the cancer. It's-, it's just astronomy, yeah. You do it because you are *curious*, you want to solve problems, understand something, how things work and-, and so on." [Int-Postdoc1]

Another interviewee compares research with "***detective work***":

> "Um, it's interesting because you are *exploring* [...] You're trying to answer and solve mysteries [...] Uh, it's *exciting* and it's like, uh, *detective work* [...] Uh, yeah, and like a lot of the times when you observe something it's something that no one else knows of like you're the only person at that moment that knows about it." [Int-Postdoc3]

Int-Faculty12 & Int-Faculty11 mention what they love about astronomy is that it is a fusion of maths and physics and as such represents an ***interface between disciplines involving collaborations*** with not only astronomers, but also other scientists. Astronomy draws heavily on other scientific fields. Examples are theoretical physics, particle physics, organic and anorganic chemistry, biology, computer science, mathematics and statistics, mechanical and electrical engineering, geosciences. Astronomers therefore interact with scientists from these fields and there is professional mobility between these fields.

The night sky with its untouchable and abstract phenomena evokes a particular ***fascination*** of astronomy:

> "I'm always attracted to, to looking at the night sky. [...] I was attracted to the, um, ridiculous immensity of it all. It's just, it's just the-, this is mind-blowing, um, expanses of, of time and space, which we can contemplate ourselves as kind of what makes a mockery of your existence on earth, but you still can attack that by being able to understand what you can of it. And so, this was the purest of pure endeavour in my view." [Int-Faculty8]

> "And the beauty of the starry night, the beauty of the starry skies captured my imagination very early on. [...] this power of seeing the things." [Int-Faculty1]

> "It's just nice, that we can, uh, well analyse an object, uh, so far away and kind of abstract for many even for us." [Int-Postdoc1]

Astronomers refer to the fact that not without reason is astronomy one of the oldest sciences (Fara, 2010), because humans have always wondered what is out there and what place we take in the universe. The beauty and mysticism of the night sky has inspired humans since the beginning of time: "But I was convinced since ever that astronomy was the basic motivation for a human being to evolve" [Int-Faculty10]. At first, humans had mythical explanations for astronomical phenomena, later they developed scientific hypotheses. Int-Faculty8 concludes that the history of science is closely tied to the history of astronomy. Int-Postdoc1 & Int-Faculty9 refer to the romantic image of astronomy. Int-Postdoc1 enjoys "doing science like in the old ages" and Int-Faculty9 explains how "the romantic side of it" grew, because in school she could only discuss astronomy with "nerdy friends". Int-PhD3 emphasises that even people who are not interested in science are fascinated by astronomy, because it's about the most fundamental questions of the universe. For him it was a "logical choice" to study astronomy as he wanted to "go deep and understand everything". Others also got "bitten by the bug of astronomy" [Int-Faculty7] because of teachers who introduced them to astronomy or astronomical events that sparked their interest. They started with astronomy as teenagers in their private time, by e.g. joining astronomical amateur associations or having their own telescopes [e.g. Int-PhD2, Int-Faculty1, Int-Faculty8 & Int-Faculty9]. However, not everybody knew from when they were teenagers that they would like to do astronomy, but their curiosity about the most fundamental laws of the universe drove them into it [e.g. Int-PhD1, Int-Postdoc2].

Astronomers choose their field despite the fact that it ***doesn't promise a great salary*** (cf. Taubert, 2019), especially compared to jobs in engineering, for example. Int-PhD2 explained that she was hesitating between astronomy and engineering, because she was advised that engineers earn more money, but like Int-Postdoc1 she found it "too technical and too applied" and she's rather into basic research. Int-Faculty7 made the same choice and Int-Postdoc2 emphasises she didn't care much about money or security as compared to the "purpose" that research in astronomy gives you. Tackling the fundamental questions of the universe gives her a "great motivation". Int-Faculty10 calls himself lucky that he was paid doing his hobby:

> "I am always, uh, saying that the feeling that I have been doing what I liked, understanding the- the universe and I was paid for that I think is unbelievable, somebody giving you a salary to do what I like to do. […] So I've spent all my life working- I mean getting money to play, to play, play with physics, to play with instruments, to build instruments."

The interviews underscore the fact that astronomers are ***realists*** who believe that there is an objective truth and strive to approximate it with theoretical models. This may not come as a surprise since astronomy is a paradigmatic, hard science (Heidler 2011). However, just like in my previous research (Heuritsch, 2019a) it is important to make this explicit, since the astronomer's attitude towards truth and reality ***has essential implications for their motivation to conduct science and what quality*** research means to them. Before diving into those, we present the exemplary results of the Pingo survey regarding the astronomer's notion of truth at this point.

As a second question after what quality research means to them (which is treated below in *section 1.1.2*), the audience was asked "What does 'reality' mean to you?". 62 answers were received in total (34 people participated, but they could give more than one answer to this question). As apparent in *Figure 4* “Truth” was the term that was answered the most. When coding the answers into categories (*Table 2*), we see that research needs to be solid and testable in order to represent reality. It needs to be based on empirical data or verifiable by means of empirical data and understandable. Answers like “life” and “presence in the here and now” are likely to be explained by the fact that the University of Dallas is a catholic university.

*Table 2: Most given answers to the question “What does ‘reality’ mean to you?”*

| Q2: What does "reality" mean to you? | |
|---|---|
| **Number of mentions** | **Code** |
| 6 | Measurable/ Testable/ Quantifiable/ Can be proven |
| 5 | Truth |
| 5 | Solid/ Firm/ Grounded |
| 4 | Life |
| 4 | Empirical Data |
| 4 | Presence in the here and now |
| 3 | Tangible |
| 3 | Personal Experience/ perceivable |
| 2 | Understandable |
| 2 | Outside one's mind |

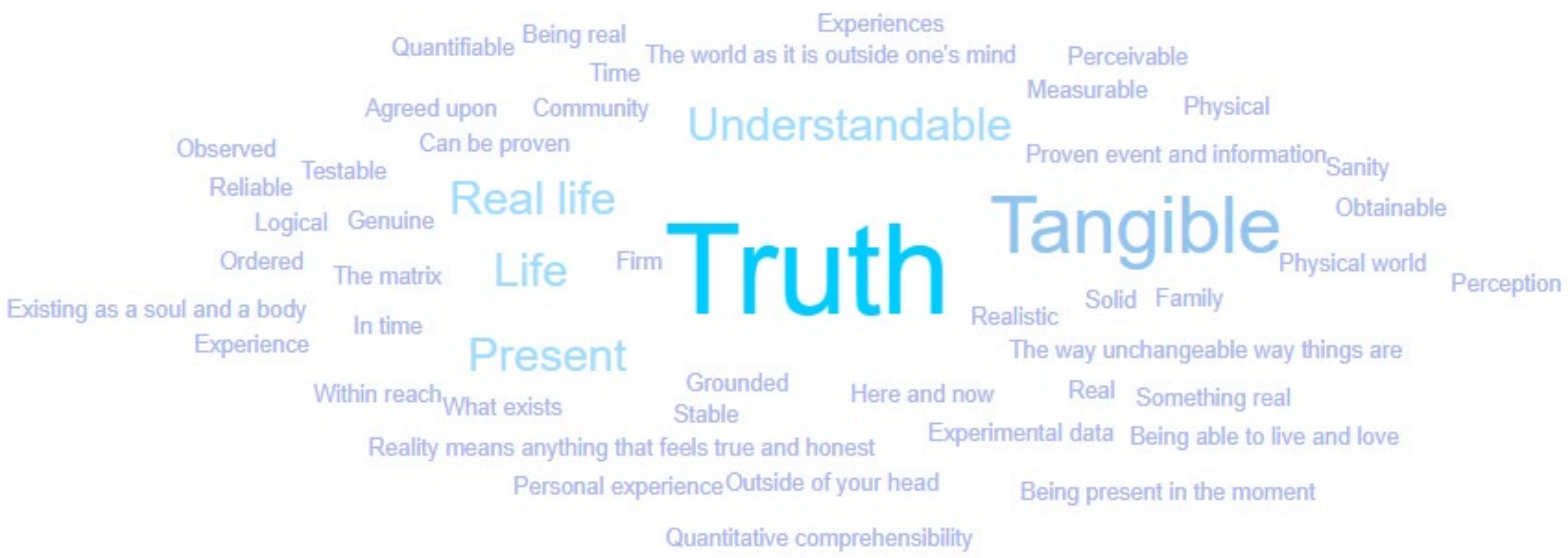

*Figure 4: Semantic cloud resulting from the question “What does ‘reality’ mean to you?”*

The next question was “What does ‘truth’ mean to you?”. The audience replied with 68 terms. The reason for why “real” was mentioned so often (*Figure 5*) might be due to the priming of the former question. When coding into categories (*Table 3*) we interprete that astronomers relate “truth” with undeniable, reliable, verifiable facts that are based on honest and unbiased research.

*Table 3: Most given answers to the question "What does 'truth' mean to you?"*

| Q3: What does "truth" mean to you? | |
|---|---|
| **Number of mentions** | **Code** |
| 9 | Real, Reality |
| 7 | Undeniable, reliable |
| 6 | Honest |
| 6 | Proven, verifiable |
| 4 | Facts, objective, correct |
| 3 | Unbiased |
| 2 | good data, discoverable |
| 1 | Understandable |

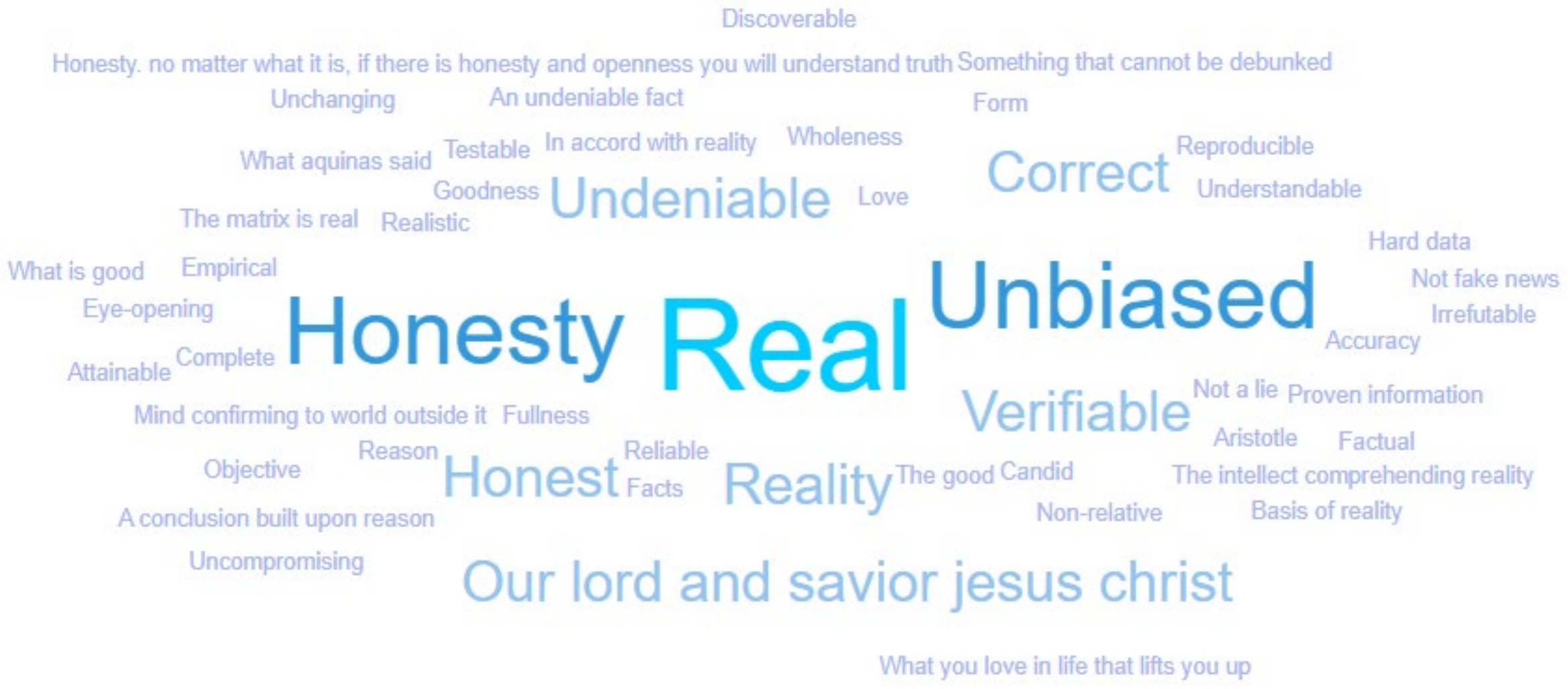

*Figure 5: Semantic cloud resulting from the question "What does 'truth' mean to you?"*

Next, the audience was asked "What do you believe?" (Q4) and were given two options: "Absolute truths exist in the world" and "Any truth is constructed by humans.". The former was chosen by 91% (30 people) and the latter by 9% (3 people). When then asked "What do you rather believe in?" (Q5) 85% (29 people) chose the answer "The universe exists out there and with our measurements we try to approximate the laws that govern it". 15% (5 people) chose "We are constructing an image of our universe (which is held up by our theories)."

While the results of the Pingo survey are only illustrative, they underscore the astronomers' realist attitude. Please note that realism does not mean that one believes we can, in practice, observe everything as it is (cf. Kant's thing-in-itself; Anderl, 2015), but rather that we can approximate the thing itself. Astronomers are aware of that:

> "But there are fundamental questions for humankind that perhaps we cannot fully answer. Nevertheless, I know, we are doing great progresses and it's all worth doing it." [Int-Postdoc2].

### 1.1.2 Research quality (3 criteria)

As mentioned above, astronomers' intrinsic values regarding studying the universe and their realist attitude shape what they define as ***good research quality***. The first Pingo question addressed that topic – the audience was asked "What are criteria for good quality research in your opinion?". 84 answers were given, which are represented by (*Figure 6*). The results (*Table 4*) match with the three quality criteria that Heuritsch (2019a) found:

1. "Asking an important question for the sake of understanding the universe better and to push knowledge forward": The codes "new", "finding causal variables" and "progress" match with this criterion.

2. "Using clear, verifiable and sound methodology": "replicability", "good data", "accuracy, being thorough", "honest", "unbiased" and "ethical" are categories that fit this criterion.

3. "Clear communication of the results in order for the community to make use of them": Codes like "comprehensive" and "understandable" match with this criterion. The latter was mentioned also when asked what reality and truth means to the audience (see above).

*Table 4: Most given answers to the question "What are criteria for good quality research in your opinion?"*

| **Q1: What are criteria for good quality research in your opinion?** | |
|---|---|
| **Number of mentions** | **Code** |
| 7 | Replicability/ Reproducibility |
| 6 | Good Data |
| 6 | Societal impact (usefulness, applicability, accessibility) |
| 5 | Support (by mentor), "Uplifting Atmosphere" - Community, good team |
| 4 | Quantitative models |
| 4 | New |
| 4 | Accuracy, being thorough |
| 3 | Honest |
| 3 | Finding causal variables, testing hypothesis |
| 3 | Unbiased |
| 3 | Being organised |
| 3 | Focus on Goal |
| 2 | Understandable |
| 2 | Ethical |
| 2 | Comprehensive |
| 1 | Ability to fail |
| 1 | Progress |

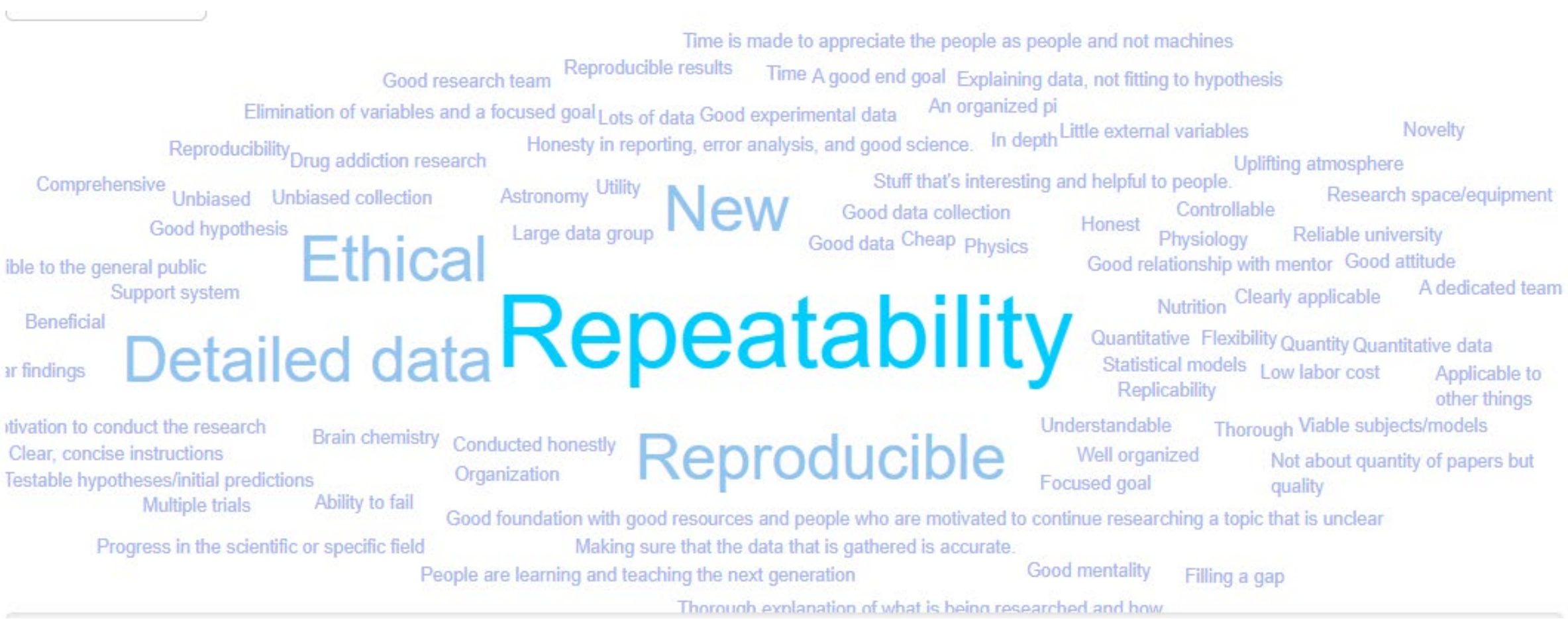

*Figure 6: Semantic cloud resulting from the question "What are criteria for good quality research in your opinion?"*

The following sub-sections present the results from interviews with respect to what research quality means to an astronomer in a top down coded manner – according to the 3 quality criteria.

<u>1.1.2.1 Originality</u>

The astronomer's main drive to conduct science is their curiosity about how the universe works. It is not surprising then that one of their criteria for quality research is to "ask an important question" (Heuritsch, 2019a; p.159), which pushes knowledge forward. In other words, a research question needs to have some aspects of ***originality*** in order to bring humanity forward.

> "[…] having the knowledge is the reason why I'm not still in a cave you know with a hammer trying to catch animals … (I: laughter). Knowledge is the basic, uh, trigger for the human evolution." [Int-Faculty10]

On the one hand, by definition, groundbreaking research, is the best candidate for pushing the frontiers of science. Several interviewees mentioned research that has "***intellectual impact***" [Int-Faculty11], so research which changes the way we think about a problem as highly valuable.

> "Probably those [papers are of high quality] which have the 'wow' factor. When you read the paper and then, 'Mmh, this is interesting. Wow. They have finally achieved something.' […] Again, the 'wow' factor when you are using-, I mean, as an observer, I would say, everything what is the, you know, most accurate or the deepest, or you are pushing the extremes. When you look into the phase space of unknown into a corner, which was never visited before. That's again something, which I would consider as high quality." [Int-Faculty1].

Int-Faculty2 also emphasises that good research gives answers one cannot find anywhere else and that often the most interesting discoveries were unexpected. Int-Faculty9 argues that game-changing science is more compelling and people in the community should encourage each other to "stretch themselves". On the other hand, interviewees admit that impactful discoveries are rare and research which makes "steps towards how we look at things" [Int-Faculty11] are also relevant. Those ***incremental steps*** are also mentioned to be important for science [e.g. Int-Faculty9, Int-Faculty12] – whether it's a positive or negative result, what matters is the addition to knowledge.

> "But you may have positive or negative results but have to be an advancement if compared with the yesterday." [Int-Faculty10]

> "I mean, you attempt to solve a mystery. Of course, not always the ultimate solution of grand mystery, but something like that. And then you have various strategies to solve the scientific question. […] I had this metaphor that doing scientific research is most of the time *very boring thing*. Like turning up little stones on the seashore. And underneath most of the stones, you don't find anything interesting. […] So, most of the time the typical researcher is *doing routine things* and out of those routine things, *sometimes exciting things come out*, but it's *not plannable*." [Int-Faculty1]

Int-Postdoc1 also admits that papers which are long and revealing an incremental addition to knowledge often may be quite "boring", however, that doesn't mean it's not good science. In fact, she doesn't like to discriminate importance on the basis of how big the addition to knowledge is, because preferences are biased.

> "Uhm, so, uh, I-, I don't like this kind of criteria like-, uh, that say-, that saying that something, that's really cool or this science is more important than this one. Uh, everyone likes what he likes, what she likes and, uh, everyone is biased.", "You can always say something but, uh, I'm never honest while answering this question. Because I just know what people may expect so, okay, I'm saying this, just feel, let's say politically correct. But I hate questions like this [what is important], it's very-, it's all biased what-, what you like." [Int-Postdoc1]

Int-Faculty10 stresses the importance of ***scepticism*** and ***perseverance***: as a researcher, one never ought to stop studying and questioning:

> "To never think what you know is enough, never think what you know is true. Because I'm ready to accept anything weird.", "But what I want to say is that you have not to be sure that you are right. You have to question your discovery. You have to question what you think is right, uh. And if you can, you can build up knowledge and give direction."

1.1.2.2 Clear, verifiable and sound methodology

Given the astronomer's realist attitude with a strong sense of what "truth" means, it is also no surprise that one of the 3 main quality criteria of good research is that research may not contain any errors. This starts with good quality data [e.g. Int-Postdoc2 & Int-Faculty3], since good science can hardly be based on bad data. When one collects the data themselves, this involves working "seriously" with the data [Int-Faculty5], i.e. checking adjustments and apertures and equipment. Biases [e.g. Int-PhD1, Int-Postdoc1] need to be removed as much as possible and assumptions laid out explicitly [e.g. Int-Postdoc2 & Int-Faculty8]. Sound methodology [e.g. Int-Faculty2] needs to be applied for the analysis. The methods need to be well-outlined and interpretations well-justified, so as to become aware of biases [Int-Postdoc2]. If there is an error in the analysis the value of the publication virtually equals zero [Int-Faculty2] and one wouldn't use it for one's own research [Int-Postdoc2]. That is why a good piece of research is tested for errors and "when you have the slightest doubt you need to investigate" [Int-PhD3].

To fulfil those criteria, skills in using tools and data analysis are important [Int-Postdoc2; see *section 1.2.2*], although ***integrity*** and ***transparency*** are often mentioned as being more vital.

> "I happen to believe that a person's moral compass, their integrity is their most important quality even more than their ability to research or to teach." [Int-Faculty3]

> "You can tell a good paper I think from the-, well, you can see the transparent process by which data gets, uh, gets put into it or data gets used, um, turned into analysis through the methodology which you, you set up. And if there's any inconsistencies and that shouldn't be there at all. […] Um, there's a transparency, uh, in, uh-, and ***honesty***, so you don't hide it behind the numbers." [Int-Faculty8]

1.1.2.3 Clear Communication & Transparent Sharing of Data and Reduction Code

The degree to which a piece of research is replicable is not only dependent on how transparently the analysis was described and whether one can access the data it is based on, but also on how well the paper is written. The presentation of one's research should be done in an accessible way [Int-Faculty2], neither hastily written nor containing too many details [Int-Postdoc2 & Int-Faculty10]. A good paper "reads like a novel" and "leaves no questions at the end" [Int-PhD1 & Int-Postdoc3]. This includes demonstrating why the research is important, in order to guide the reader through the story. According to Int-Faculty9 most astronomers don't do that as they think it's obvious why the reader should care, which makes the paper dry.

> "Um, a good scientific paper to me is one which, uh, one which clearly tells a story that is, uh, you can follow. […] Uh, and so being able to tell this coherent story and guide a reader through this, is a part of good science practice because […] I would like to see that as part of good practice in communicating the science that you're doing. […] Uh, in terms of quality, um, of, of, you know, scientific production that's, that's actually important because you are, you're giving the motivation as the reader for why the, the author is bothering, you know, taking some of your life, in trying to understand this particular thing." [Int-Faculty8]

To ensure readability, Int-Postdoc3 & Int-Faculty2 emphasise that the language plays a large role, because "I have something to say, I want my articles to be read. Not being cited, I want them to be read. My articles are readable and the language makes a difference." [Int-Faculty2]. Int-Postdoc1 points out that language literally plays a role, since she knows some Russian astronomers who don't publish in English, which is "bad, because then nobody can read it" and hence, the research cannot effectively push knowledge forward.

Zuiderwijk & Spiers (2019) found that astronomers have a quite open attitude towards ***data sharing***. The authors relate this attitude to the intrinsic motivation of astronomers to push knowledge forward. As a consequence, most astronomers share their data voluntarily (Zuiderwijk & Spiers, 2019; Heuritsch, 2019b). If they don't share their data, in most cases this is due to a bad infrastructure (Zuiderwijk & Spiers, 2019; Heuritsch, 2019b) and lacking incentives to spend the extra time (see *section 1.3.4*) and less so because of a fear of others being first, although this also happens (see *section 1.5.2*).

Finally, ***replicabilty*** was mentioned as a key quality criterion, which depends on both, Quality Criterion 2 & 3; The degree to which a piece of research is replicable is a function of how transparently the analysis was described, including sharing the reduction code used for it, of whether one can access the data it is based on, how understandably the paper is written and of whether or not the results contain errors.

> "I've also always had a fairly strong sense of quality. […] In the sense of, uh, I believe that any work that we publish should be, uh, described in sufficient detail […] That other people can repeat the steps. […] So-so on almost all the data. So I use a lot of public data and therefore my practices if I publish my own data, uh, then after my paper is published that, uh, data is also public. […] Uh, so that way we ensure that, uh, that it's complete. So for me, one-one of the best indicators of quality is ***provenance***, which means can I reproduce the other person's work?" [Int-Faculty7]

> "A good paper, I think it has to have enough, uh, information that someone can reproduce it. I mean they don't have to reproduce it but it should be a self-contained paper where someone does not have to ask you of what you did here in so order to reproduce it. They should just be able to read the paper and reproduce it." [Int-Postdoc3]

### 1.1.3 Intrinsic Motivation

The ***3 quality criteria*** are all directed to the goal of pushing knowledge forward. This goal is based on the astronomers' intrinsic values, which we can summarise in 3 dimensions: a cognitive dimension (intellectual interest, puzzle solving, curiosity), a dimension of subjective experience (having fun conducting research) and an aesthetic dimension (the beauty of astronomical phenomena). Astronomers' intrinsic motivation to perform research then is guided by their intrinsic values and quality criteria. Groundbreaking research is favoured, because it promises a large knowledge push, but incremental steps are perceived as valuable too. Sharing data enables faster progress and replicability. Quality research has a lot to do with consistency and describes a transparent process from the data to methods, analysis and interpretation. This involves being honest about the things one doesn't know and being reflective about biases and assumptions [Int-Faculty8]. Astronomers seem to agree on those different, but related aspects about research quality, however they do admit that it's not always easy to judge, a problem we will deal with in *section 1.3.5*.

> "I believe that professional evaluation should be based upon the quality of teaching and research. So how do you judge that?" [Int-Faculty3]

In summary, the astronomers' definition of quality are oriented around what Taubert (2019) calls the ***primary code of science*** – the striving for "truth". The ***secondary code*** of science is "reputation". We find evidence that astronomers wish for reputation. For example, Int-Faculty4 explained proudly that a theoretical model he developed carries his name now and young astronomers may look up to established astronomers who have made important discoveries. Praise is a social reward (Gagné et al. 2015) and hence triggers extrinsic motivation, rather than intrinsic, however as the authors point out, praise has been found to be positively correlated with intrinsic motivation; generally spoken, when we are recognized for our work, our work appears more fun. As we unfold the external components of the *logic of the situation* we will learn however, that there is more to reputation than praise.

Having said that, other researchers being interested in one's research is a motivational factor [Int-Postdoc2 & Int-Faculty4], which can be drawn back to the quality criterion of having "intellectual impact". Int-Faculty2 summarises what motivated him to choose his topic: curiosity, the fact, that not many people worked on it and that it promised impact on the community. It is not only the impact in the field, however, that motivates astronomers, but also ***societal impact*** in the form of value for humankind (Int-Postdoc2, Int-Faculty2 and Int-Faculty9; see *Table 4*; cf. Heuritsch, 2019a). Because "of their love for astronomy", many astronomers are heavily involved in outreach activities, even though most of them "do not receive (or look for) any outreach training or funding"[6]. There is a well-developed citizen science, a relatively large number of amateur astronomers driven by incentives such as the right to name asteroids and comets discovered. Further, there are highly developed educational programs, such the Universe Awareness (UNAWE) program aimed at world-wide primary schools, science fairs, well-developed science journalism and internet activities, and so on.

## 1.2 External Component: Material Opportunities

Material opportunities comprise one of the three external constituents of the *logic of the situation*. They represent the summary of all kinds of resources that an actor has control over in that particular situation (Esser, 2000b). Bourdieu (in Kreckel, 1983) introduced four different categories of capital: *economic*, *cultural*, *social* and *symbolic* capital. The results of the interviews and the literature review are top-down coded into these four categories and presented in the following sections.

### 1.2.1 Economic capital

Economic capital is closest to what we usually understand as capital in our daily lives. It consists of financial resources, time and material for one's work, such as computers, but also facilities. This section analyses what economic capital is needed for an astronomer to perform good quality research.

Not surprisingly, one of the most essential resources is ***funding***[7].Universities usually receive an annual performance-based budget from the government on the basis of certain parameters, such as number of supervised PhDs, number of acquired external grants and teaching hours [e.g. Int-Faculty8; cf. Heuritsch, 2019a]. According to Int-Faculty1 & Int-Faculty12 this is usually "barely enough" to pay salaries, "so, doing research requires independent funding. Without external grants, there would be no research" [Int-Faculty1]. PhDs are funded either through the university, the external grants from their supervisors or their own grants. Int-PhD2 & Int-PhD3 report that they already had to apply for their PhD grant in their 2$^{nd}$ year of their Master and it was dependent on their BSc & MSc ***grades***.

> "Of course, it is extremely competitive, so only the students with the highest grades will get it." [Int-PhD2]

At many universities in the US, the university funds mainly the PhD students and staff has to live entirely off external grants [e.g. Int-Faculty2]. In order to receive external grants one must write funding proposals. This must not only demonstrate that the proposed science is solid, but also compelling [Int-Faculty9]. For this it's hardly enough to promise an *incremental step*, but a *larger advancement* in the field [Int-Postdoc3 & Int-Faculty9]. Compelling research should be innovative, but not too risky:

---

[6] https://www.iau.org/news/pressreleases/detail/iau1813/

[7] As an amusing side note; Int-Faculty11 mentioned "Donald Trump is an obstacle" for funding in astronomy, since his administration has cut 5% of NASA's astronomy budget.

> “Usually they want something that is going to uhm you have to answer the question why is this important? Why we need to do this now and will the methods work? You know. Uh often the best proposals are innovative but not too innovative in that if you uh propose some things that the reviewers will not think of too risky.” [Int-Faculty11]

Funding agencies do not only look at the proposed research however, but also at the researcher’s past ***publications and citations*** those received.

> “Typically, funding agencies want publications. High quality publications defined as published in the top journals, with the highest impact factors, receiving large number of citations, and things like that.” [Int-Faculty1]

As outlined in Heuritsch (2019a), astronomers report that ***luck*** also plays a role in receiving funding. [Int-Postdoc1] emphasises that the funding system sometimes feels like a “lottery”, because many factors – that don’t have anything to do with the quality of the proposed research – go into a funders’ decision. It may depend on what topics are important at the time of application (see *section 1.3.3*) or if a position has opened, which needs to be filled in order to not lose the allocated funding. Sometimes the funding committee doesn’t include an expert of the field, the proposal is related with, which may also put the proposal at a disadvantage [Int-Faculty8]. Thurner et al. (2020) explain that this often happens for innovative ideas, since their evaluation need backgrounds from several fields to be understood. Int-Postdoc3 talks about the potential of “dominating personalities” on a funding panel leading to undemocratic decisions.

There are also other sources than external grants. Given that astronomy is a basic science, not promising applications or any other commercial benefit, there are not many opportunities to partner with companies in order to receive funding. Int-Faculty8 concludes that in New Zealand, there is only one funding stream astronomers can apply for – the “Blue Skyfunding Stream” for “pure research”. Philanthropy is also “becoming more important” [Int-Faculty12]. Some universities also offer special funding schemes to bring their citizens back to their country [Int-Postdoc1, Int-Faculty1 & Int-Faculty8]. Often those fellowships offer attractive salaries and packages for establishing research groups.

When observational astronomers, who work with big datasets and telescope facilities to obtain data, need to acquire funding in order to pay for a research group analysing the data, they can apply for mission-based funding [Int-Faculty11 & Int-Faculty12]. Often, the institute who owns or helped building the ***telescope*** can claim a certain percentage of the available observing nights [Int-Faculty8] or is at an advantage in competing for telescope time, because of inside knowledge [Int-Faculty7]. One may also collaborate with an astronomer who received funding for an observing mission or who has such a privileged access to a telescope [Int-PhD2].

Another important resource is ***data***. For observational astronomers, data are their “bread for every day” [Int-Faculty5]. To obtain data, astronomers have mainly three ways [e.g. Int-Postdoc1 & Int-Faculty12]: 1.) Applying for their own ***observation time*** at a suitable telescope (see *section 1.3.4*). 2.) Collaborating with somebody who has access to a telescope (see *section 1.2.3*). 3.) Using data from archives. Astronomers may reuse data from archives to obtain new insights (Zuiderwijk & Spiers, 2019) and complementing their own data [Int-Faculty7 & Int-Faculty12]. Next to their intrinsic motivation to push knowledge forward (see *section 1.1.2.1*), the authors also found that “benefits [of sharing data] mentioned included increased ***citations***, bigger profile and more ***visibility*** of the researcher, increased credit and ***Recognition***, […] and making it easier to ***publications*** in good journals" (ibid. p.235). As we will see in *section 1.5*, sharing data to obtain these forms of capital stems from an extrinsic motivation to get attention and advance one's career.

On the one hand, astronomers say that nowadays the archives contain a "vast amount of data" [Int-Faculty5, Int-Faculty7 & Int-Faculty8], which has "changed the sociology of the field" [Int-Faculty7]. While in the past an astronomer was forced to collect their own data or collaborate with people who have access to a telescope [Int-Faculty5 & Int-Faculty7], nowadays, archival data is easily accessible and usable. On the other hand, Int-Journal, an astronomer working for the journals of the American Astronomical Society (the AAS Journals), emphasises that linking data and analysing software to papers is an increasingly relevant challenge, given that there are various places for data to be stored and that neither storage, nor the format of the data is standardised well enough. This often makes it difficult to find and use data (Int-Postdoc1; cf. Zuiderwijk & Spiers, 2019; Heuritsch, 2019b). Furthermore, for data to be usable, it also needs to be good quality. Bad weather conditions during observations or data having been milked for producing papers may decrease the quality of archival data. On top of that, there are hardly any external incentives to share data, which often makes sharing data not worthwhile an astronomer's time:

> "For some interviewees, effort appears to have a strong effect on the motivation to openly share such research data, mainly because of the time it takes to openly share data and because there are too many other things to do […]. If certain rewards would be in place the efforts would play a smaller role" (Zuiderwijk & Spiers, 2019; p.235).

***Time*** is an essential resource in itself [e.g. Int-Faculty2, Int-Faculty8 & Int-Faculty11]. Time allows for freeing up the relevant mental capacity to absorb and process a lot of information. For young researchers time is limited, because they are "hampered by other stresses":

> "So, you're worried about how to get the next postdoc, how to get the next job, um, relationships, and things like that. So, always those things, you know, compete for your attention, wait, going out tonight, I have a family, that sort of thing." [Int-Faculty8]

For more established researchers the available time they can spend on research is limited by their various managerial tasks [e.g. Int-Faculty11]. Writing funding & observing proposals takes a lot of time [e.g. Int-Postdoc3 & Int-Faculty2], often promising little return rate [e.g. Int-Faculty8]. Establishing, supervising and maintaining research groups and collaborations also cost time [Int-Faculty1, Int-Faculty2 & Int-Faculty12].

In addition to funding, access to telescopes and/ or data and time, astronomers generally need ***technical equipment*** [e.g. Int-Postdoc1, Int-Postdoc2 & Int-Postdoc3]. This includes: good computing resources, given the large amounts of data that they need to process, a fast internet connection and software licences such as for programming languages in order to analyse the data. Last, but not least, astronomers need ***access to information***, such as journals and articles [Int-Faculty8], which is hardly restricted in astronomy, given that most pre-prints get uploaded onto the freely accessible ***ArXiv***.

### 1.2.2 Cultural capital

Cultural capital includes the ***education, the knowledge and skills*** that one has attained in order to navigate through a specific environment. In their studies, astronomers gain fundamental knowledge in maths and physics, including data analysis and programming skills. In their Bachelor and Master training, astronomy students follow courses in most of the fields that astronomy draws on (see *section 1.1*) physics; math and statistics; chemistry; computer science, including big data techniques and algorithms; and signal processing. A substantial part of these courses is often provided by the physics and mathematics departments, which is why astronomy students and students from these fields are well acquainted with each other. This contributes to the ease of professional mobility between fields. Interestingly, however, several interviewees mention [Int-Faculty2, Int-Faculty3 & Int-Faculty6] they find a matching personality and integrity ultimately more important than one's astronomical skills (*see section 1.1.2.2*).

Nevertheless, astronomers [Int-Faculty7, Int-Faculty11; cf. Heuritsch, 2019a] find it hard to move to another subfield. That is perceived as very risky. Funders are not willing to take the risk to support such a move, if you ***haven't published*** in that field yet, regardless their actual skills.

> "Okay so, uh even if you have a good track record in fields they don't like to fund you when you move fields. It's a tendency. to want to fund you then you're the you know you worked on X, you want to make an incremental progress and whatever it is you would do the next step and build on previous work you've done that's something that reviewers are more comfortable with because their confident you've done this in the past you can do it in the future." [Int-Faculty11]

Next to expertise in astronomy, successful navigation through an environment also entails ***knowledge about the norms and rules*** prevailing and prescribing the behaviour. They may be embodied in tacit or explicit form and are outlined in *section 1.3*. Embodying those norms happens through a process of socialisation that accompanies the studies and a career in academia.

It turns out, that at the start of a career in academia, astronomers are quite ***naïve*** with regards to the institutional norms. Int-Faculty9 emphasises how important it is to teach students what "real science" is like, because courses don't really cover that:

> "Uh, one thing I've tried to pass along when I have had students work with me is I do try to show them what *real science* is like because I think that is a barrier when-- I've seen interns, you know, go through programs where they get all these presentations given to them, and they get to go see neat things but they never actually get to see the *day-to-day workings* of science. […] I thought she might as well see that this is part of research, you know, try proposing for funding and justifying your work and filling in the forms. And things like that just so that people don't get, uh, you know, an overly idealized view, uh, of-of astronomy."

Int-Faculty1 reports how lucky he was that his supervisor explained to him "how the system works early on", because that helps "to optimise your chances". That is why he introduced a course for PhDs and Master students: "The art of scientific publishing". Int-Faculty1 recalls a "level of disappointment" when he learned that science worked different to what he thought:

> "I mean, there was a certain *level of disappointment* because I was very naïve and *idealistic*. Not naïve. Idealistic is the best word. I mean, we are all very idealistic when we start our career, okay, this is all about searching the truth, and searching the scientific truth, and you know that going into the unexplored terrains, and things like that. And then when reality hits back and okay, you have to *publish those bloody*

> *results* and then you have to write up your paper however you don't like it, that's when you realize that's this is a profession. […] Doing scientific research is just another profession, with its own internal rules and procedures."

Early career researchers [Int-PhD1, Int-PhD2 & Int-PhD3] confirm that they were surprised when they realised over a process of their studies, and partly their PhD, that "the system" doesn't work like the "ideal" of how they had pictured science (cf. Heuritsch, 2019a). We may wonder why the fact that astronomers' performance is measured in terms of indicators (sections *1.2.4 & 1.3.5*), is so surprising, given that most people in our Western society have grown up in a meritocratic system, where grades are important from early childhood onwards. However, we may explain this naiveté through the "romantic image" [Int-Faculty9] that people have in general about astronomy and the astronomers' strong intrinsic motivation to ***follow their curiosity and push knowledge forward***.

### 1.2.3 Social capital

Social capital are all human resources that an actor has access to, including their co-workers, their family, friends and network (Esser, 2000a). Int-Faculty10 points out that in order to do good astronomy, you need a good team and a supporting community. The scientific community was also mentioned as an inspirational factor (e.g. Int-Faculty8, Int-Faculty11 & Int-Faculty12). Most astronomers are part of a research group, which often consists of one ***supervisor, usually a faculty member, and several postdocs and PhDs***. As outlined in Heuritsch (2019a), PhDs and postdocs are the working horses of academia, since supervisors are often occupied by managerial tasks (see *section 1.2.1*). At the beginning of one's career a good supervisor who supports and guides the early career researcher is vital [Int-Postdoc1; cf. Fohlmeister & Helling, 2012]. The supervisor should also help the researcher to connect with other people, as to establish a network of ***collaborators***, which is important social capital in astronomy (cf. Fohlmeister & Helling, 2012). Collaborators enable access to and trading of other kinds of capital, such as funding, human resources, publications & authorship, data and access to telescopes, which is why Heuritsch (2019a) described the network of collaborators as a market place.

An illustrative example of such a trade of human resources against funding was given by Int-PhD3, whose supervisor advertised his skills to a collaborator in the US at a conference – proposing a collaboration in exchange of funding. Because Int-PhD3 didn't want to move to the US, they found other collaborational partners with EU funding, enabling him to become a PhD and hire a postdoc in addition. Int-Faculty2 describes his collaborations as a "safety net" that enabled him being a co-author on ***papers*** even when he was sick for a few years, resulting in a good-looking bibliometric record. Int-PhD3, now a 2$^{nd}$-year PhD, has already been an author on 4 papers due to collaborations.

Telescope time is easier to attain as well through collaborators, through those, "who have (privileged) access and/or hold guaranteed time" [Int-Faculty2]. Int-PhD2 explains what collaborators benefit from providing telescope time – being the PI on the successful telescope time proposal and having access to the resulting data:

> "That's the whole game you know. You 'use' them as PI on your proposal because they have privileged access (for example working at a Chilean University) […]. This may sound unfair in both directions, but you can see it as trade, as a win-win. That person has a successful proposal as PI, gets the data and will be co-author on your paper, you have to do all the work, but you get to observe (if the proposal gets accepted of course). Actually, the more I think about it, the more I believe that there are no disadvantages for those collaborators." [Int-PhD2]

An astronomer may also attain ***data*** directly from the collaborators, instead of the collaborators' telescopes. Int-Faculty7 gives an example where he was unlucky with the weather, didn't get useful data as a result and "his only option was to set up a collaboration with someone." [Int-Faculty10] works within a large collaboration, that enables him to "use all the data [he] can crab", which is necessary for his project.
One of the main functions of collaborators, next to the access to economic capital, is to share and develop new ideas. Many interviewees [Int-PhD2, Int-PhD3, Int-Postdoc1, Int-Postdoc2, Int-Postdoc3, Int-Faculty4 & Int-Faculty8] report that they need collaborators as ***sparring partners*** to exchange opinions, broaden perspectives, discuss questions and giving constructive feedback to improve one's research.

> "So, other people's interests and other people's willingness to listen to you, and to be your sparring partner, and to give you critique, I think is absolutely crucial for a successful beginning of a research career." [Int-Faculty4]

Int-PhD2 points out that in our digital and globalised world it has become really easy to collaborate with people. Collaborators may not only be mediated by supervisors or other collaborators, but also met at ***conferences*** (Int-PhD3 Int-Postdoc2 & Int-Faculty7; cf. Fohlmeister & Helling, 2012):

> "Again, if I look back at my career, personal contacts were always the defining moments and conference attendances. And especially giving talks, giving good talks at conferences. That was, at least, in my case, which was quite helpful over the years. Because, I mean, giving a boring lecture doesn't help you, but if you give a good talk, that helps because then people notice you. And during the coffee break, they go to you, they ask questions. And if you are good, then you can answer those questions. And that's how naturally collaborations arise. […] So, being a social, that's very much important." [Int-Faculty1]

Additionally, according to Isaksson & Vesterinen (2018; p.9), "for an international effort like astronomy", research evaluations lead to better results for astronomers who participate in "high profile international collaborations", because "research evaluations are usually built on comparisons of units."

### 1.2.4 Symbolic capital

Symbolic capital is acquired through recognition of achievements. It's a symbol for having obtained other forms of capital. For example, a transcript is symbolic capital, acknowledging one's incorporation of specific knowledge. Therefore, symbolic capital, although closely linked to it, "is not identical to incorporated cultural capital" (Bourdieu 2004, p.56). Bourdieu writes further that "symbolic capital is a set of distinctive properties which exist in and through the perception of agents endowed with the adequate categories of perception" (ibid. p.55).

***Recognition*** is a Mertonian concept: "When the institution of science works efficiently … recognition and esteem accrue to those who have best fulfilled their roles, to those who have made genuinely original contributions to the stock of knowledge" (Merton, 1973 [1957],: p. 639 in Bourdieu, 2004; p.11). Further, "the institution of science has developed an elaborate system for allocating rewards to those who variously live up to its norms" (Merton, 1973 [1957], p. 642 in Bourdieu, 2004; p.11). What those norms that need to be lived up to are will be treated in detail in *section 1.3*.

As we have seen in *section 1.1,* a big salary is not one of the astronomers' main driving factors. In fact, "it's part of the professional honour of a scientist that ***reputation*** is more important than money" (Franck, 1998; p. 38 in Taubert, 2019; translated from German). Only striving to find the truths of the universe is not enough in order to advance one's career – one also has to publish. Hence, the function of ***publications*** is not only to share the knowledge with the community, which would be part of astronomers' intrinsic motivation, but also to attract attention and generate reputation. That is why "***authorship*** (and the recognition that flows therefrom) is the undisputed coin of the realm in academia: it embodies the enterprise of scholarship" (Cronin, 2001; p.559). Because authorship is such an important currency and because the numbers of authors per paper have increased, Fernández (1998; p.71) refers to the concept of a "fractional author, i.e. the scientist that produces the $n^{th}$ part of a scientific paper." The position of the author determines not only, how much recognition the author gains; the first author almost gains full recognition for the published article, but at the same time the judgement of the scientific work is also "contaminated by the position of the authors" (Bourdieu, 2004; p.57). In other words, if the first author has a good reputation, the paper is judged as more valuable than if the author is not known in the community. We will dicuss the ***role of authors*** and how this role is linked with generating attention in more detail in *section 1.3.3*.

While reading an article is one way to indicate (at least silent) respect of an author, ***citing*** a paper adds to the value of the paper (Kurtz & Henneken, 2017). Next to authorship, citations are an important currency to demonstrate scholar achievement. The logic is easy to follow: the more important the contribution, the more often the paper gets cited. However, as discussed in various pieces of literature (for a list refer to Table 3.5. in Moed, 2005), citation rates do not accurately map the relevance of a scholarly article, or by extension, the performance of a researcher. In addition to the mapping issue and ***gaming*** that may result in misusing citations, there are some field-specific reasons for why citations are not necessarily an accurate reflection of scholarly value. First, in astronomy "citation rates are highest for papers with European and American authors and much smaller for papers from less-developed countries" (Trimble & Ceja, 2012; p.45). Second, citation rates depend on the subfield, and hence cannot be compared field-wide [Int-Postdoc2 & Int-Faculty9]. Third, if one works on a ***hot topic***, one has a higher chance to receive citations than on topics that are less fashionable at that time:

> "[…] citation rates are much larger for currently hot topics (exoplanets, cosmology), than for less hot ones (binary stars, for instance) […]" (Trimble & Ceja, 2012; p.45).
>
> "If you really are in a hot topic, you get so many citations. […] And then, what defines a hot topic also depends on the year. […] But it's also cultural." [Int-Postdoc2]

Fourth, since astronomy mainly conducts basic research, the value of findings may only become apparent years later (cf. Heuritsch, 2019a):

> "Uhm [impact is] something you judge when you read a paper uhm but also in some ways it's something you judge uhm sometimes best in retrospect. Sometimes uhm you know, you don't, sometimes read a paper you realize it is important. Sometimes the impact of a paper isn't clear for several years. So you know someone writes a paper

> and it opens up a new direction that other people build on so sometimes the most important papers uhm are not immediately impact." [Int-Faculty11]

Special to astronomy is that granted ***telescope time*** is capital in itself (Heuritsch, 2019a; Roy & Mountain, 2006). Just like ***funding grants, and rewards***, telescope time is also part of an astronomer's reputation currency.
The journal impact factor (***JIF***) is not really relevant in astronomy. Although frequently mentioned by interviewees (e.g. Int-PhD2, Int-Faculty1, Int-Faculty2 & Int-Journal), they point out – and we will see in *section 1.3.3* – that there are mainly three journals where astronomers publish and they count the same in terms of JIF. Only a publication in *Nature* or *Science* would generate more attention.
Those different aspects of currency that add up to an astronomer's reputation, which is the most important symbolic capital in academia, are measured by so-called ***performance indicators.*** Indicators, such as publication & citation rates, received telescope time, grants and rewards supposedly map the performance of researchers and the value of scholarly papers. This mapping, such as equating citations with impact, has the function to reduce complexity, not only to direct attention towards certain publications, instead of being lost in ***information overload*** (Taubert, 2019), but also to guide ***hiring and promotion decisions*** as we will see in *section 1.3.5*.

### 1.2.5 Capital – Summary

Astronomers acknowledge that there is a variable that influences all forms of capital – ***luck***. First of all, luck is needed when it comes to observations. The quality of attained data depends on the proper functioning of the instruments (Int-PhD2) and on the weather:

> "And, sometimes when you have a weather problem, you have three nights and you lose two because it's raining or it's cloudy." [Int-Faculty5]

> "[…] many hours (or nights) could be lost because conditions were not matched to the observer's expectations." (Roy & Mountain, 2006; p.26)

Especially successful, established astronomers admit that they have had a lot of luck in their career. Int-Faculty4, for examples reports that he was lucky that he was supervised by famous astronomers and having met the right people at the right time: "I think that's characteristic of careers, they are often determined more by people than by subjects." Int-Postdoc1 also points out how important it is to have a well-known supervisor who enables connections with other astronomers (*see section 1.2.3*). Int-Faculty6 reports that he was lucky that he got multiple job offers without even applying and that it was easier when he started his career, because "publication pressure was not as big as it is now". Besides, after WW2 "there was an enormous shortage of competent people", so Int-Faculty4 "was offered six jobs when [he] got [his] PhD in Cambridge without ever writing an application". The shortage was so large that Int-Faculty4 was promoted to an associate professor "without ever doing a day of work for that university." Luck is an especially important component, when it comes to competition, since often competition is so high that even good proposals are rejected [e.g. Int-Postdoc2, Int-Postdoc3, Int-Faculty11 & Int-Faculty12]. Int-Faculty9 explains that luck usually doesn't change anything about the lowest and highest ranked proposals. However, if only a few proposals can be funded and the cut off needs to be decided, then the decision can feel like "rolling the dice", because usually there is nothing wrong with 3 out of 4 proposals.

> "And sometimes it's kind of lottery, uhm, uhm so you really have to have lots of luck uh but like everywhere, uh. But, uh, yeah, well uh, you should just try and try. If it doesn't work this time, you have to try next time." [Int-Postdoc1]

In fact, Pluchino et al. (2018; p.1) found that "almost never the most talented people reach the highest peaks of success, being overtaken by mediocre but sensibly luckier individuals." They demonstrate that the *naïve meritocratic belief* (*see section 1.3.6*), where one's success is merely based on one's cultural capital, doesn't work out in reality, because luck plays a much larger role than hitherto assumed.

The naïve meritocratic belief is also the basis for the assumption that future achievements can be predicted by past successes. There is various literature, which study how talent and past successes are related to future achievements. For example, Kurtz & Henneken (2017; p.706) find that results "on the relation of peer review of funding proposals with future citation outcomes suggests that both peer review and bibliometric methods are similar, in that in the mean higher scores predict mean higher scores in the future". The question that rises when looking at correlations between past and future successes is, however, whether those successes are indeed due to great talents and achievements, as the naïve meritocratic belief would suggest, or whether we must rather attribute future successes to the **Matthew effect** (Merton, 1968). In the latter case, the naïve meritocratic belief creates a self-fulfilling prophecy, where future successes are not only based on talent and cultural capital, such as skills, but also other forms of capital and luck.

Astronomers call the Matthew effect also the "chicken-or-egg problem" (Heuritsch, 2019a; p.163) – in order to receive funding one needs to show a track record, but one needs funding to establish a track record. For Int-Faculty1, for example, this results in ***publication pressure***:

> "[…] typical funding agencies have various weights associated to your track record. But typically, at least 40%, 50% of the evaluation is based on your track record. And your track record consists of the number of your papers, the number of citations, all these things. So, if I want to get money then, I have to be able to claim that I'm still actively publishing researcher."

Trimble (1991; p.76) found that "receiving an astronomy PhD from a prestigious rather than a non-prestigious university is clearly predictive of a high probability of achieving a long-term career in research […]". Furthermore, we have seen in the sections above, that the possession of capital leads to acquiring more capital. For example, papers are judged as more valuable, when the 1$^{st}$ author has a good reputation (cf. Taubert, 2019). Collaborators don't only bring more collaborators, but also capital in various other forms. Int-Faculty8 pointed out, that while his salary doesn't increase when he receives an external grant, he climbs up the "promotion ladder" and can buy his time out of teaching and managerial tasks, resulting in more time for research, which means more papers and hence a better track record.

In summary, we have seen that, in order to conduct research, astronomers need:
1. *Economic capital*: Funding, time, access to telescopes, data, technical equipment such as computers, strong internet connection and software packages.
2. *Cultural capital*: Expert knowledge, programming and data analysis skills, (tacit) knowledge about institutional norms
3. *Social capital*: A good research team (including supervisor, PhDs and postdocs) and collaborators which bring more access to more capital

4. *Symbolic capital*: Recognition & reputation; reflected by various performance indicators, such as publications & authorship, citation rates & impact, allocated telescope time which count the same as grants & rewards

> "It's a combination of being very good [in terms of output] and knowing people." [Int-PhD2]

Furthermore, astronomers need luck, because luck increases the chances for success and the more successful, the more capital an astronomer receives due to the Matthew effect.

> "Anyway, I had a lot of luck. And that's the third feature [of what is needed to perform research], yes. So, interactiveness [collaborations] uh, freedom [in terms of time and taking risks], and also luck." [Int-Faculty4]
> "I mean, you cannot plan. So, that's why I always or I frequently say that to be successful in research, you have to be a very lucky, you have to be very willing to do or able to do a routine work, you may have some sort of smartness. But that's the least important, luck and the willingness to work hard. […] And a bit of being genius helps, but not always. At least, that's my experience." [Int-Faculty1]

So far, we have elaborated on the capital needed to perform good research, but not yet on its abundance. Resources are limited. Many telescopes are over-subscribed – often they have an *oversubscription rate*[8] of 30-50% (Int-Faculty7). Depending on the source of funding, success rates can also be quite low.

> "Most grants in the US are three years. […] three years and the success rates are below 20%. So you're constantly on the treadmill writing these grants trying to bring in the small, this, this sort of […] amount of money." [Int-Faculty12]

The fact that capital is limited leads to competition, which we will elaborate on in *Chapter 2*.

> "You know, the way science is done these days is creating competition. Being a successful researcher means being successful in a very competitive field because the resources are limited. All the resources are limited. Not just the financial resources, but even human resources. You don't have an infinite pool of bright PhD students. No. You are competing for the brightest students. So, because of that, there is a competition. Because of the competition, I myself, feel like being always in a kind of hurry. And that doesn't let you think properly, I believe. And that's a limitation. So, there is a limitation coming out of the lack of resources, limited resources. There is also limitation because of the competition because you have to be there, you have to publish. You have to be, you know, bright and excellent, and you have to be able to give proof of your excellence at the various levels. And then you don't have time to think. These are all different kinds of limitation, but I certainly feel like being limited." [Int-Faculty1]

---

8 http://www.eso.org/sci/observing/phase1/p96/pressure.html?fbclid=IwAR2mP2iit7tjo7G81xU6MxxpeImhAbn9tl5H2NldVQmPucq-_3QkfTO_S5A & https://www.astrobetter.com/wiki/Oversubscription+Rates?fbclid=IwAR3_u9QwJ0FSFWTH_8zGk3r4_7eu5AC3IlFvoGWUJoVlNoZyzV62d29rvGE

## 1.3 External Component: Institutional Norms

### 1.3.1 Introduction to the concept of institutional norms

According to Esser (2000c) institutions are rules with a claim on validity, which generate expectations (*Figure 7*). There are three forms of institutions: norms, roles and social scripts. Roles and scripts contain norms, but also open slots for improvisation. Institutions have three functions: 1.) the *orientation function*, 2.) the *organisational function* and 3.) the *endowment of meaning function*. First, by setting the rules for what is appropriate behaviour and defining what are desirable goals, institutional norms give an actor a sense of orientation. Second, because actors can usually rely on others following the same rules, institutional norms bring order in the otherwise not comprehendible and navigable social situations. This is also how institutional norms *parametrise* a situation. As opposed to non-social situations (so-called *parametric situations*), where an actor is "only" exposed to nature, social situations entail a *double contingency*. This means that an actor, in order to attain a goal, needs to anticipate the actions of others. By prescribing and restricting possible patterns of behaviour, institutional norms virtually reduce the double contingency to a simple one. Third, by defining what is good, what is bad and what is admirable, institutional norms endow actions with meaning. Meaning is entirely socially constructed and institutional norms and cultural frames are ways giving meaning that a larger population of people can identify themselves with in order to work or live together. This third function represents the internal anchor of the respective institutional norm. The more actors identify themselves with the meaning the norm provides, the higher its *legitimacy*. However, for those actors who cannot identify themselves with that meaning, the norm needs an external anchor, comprised of external incentives and enforcement mechanisms, such as sanctions. When an actor disagrees with the norm and the sanctions are lower as compared to the gain of not following the norm[9], the actor may shift from *compliant* to *deviant behaviour* (see *Chapter 2*).

### 1.3.2 Mertonian Norms

This section analyses the institutional norms that prevail in academic astronomy, whether astronomers incorporated them as tacit or explicit knowledge. Since the famous sociologist Merton was the first one to elaborate on norms that guide science, we will first introduce his concept.

Merton (1973 [1942]; p.268) argues that, because of historical reasons science "came to regard itself as independent of society […] as a self-validating enterprise which is in society but not of it". According to this structural-functional vision, its institutional goal is the "extension of certified knowledge" (Merton, 1973 [1942]; p.270), which implies certain moral imperatives and technical prescriptions. He describes four institutional imperatives that comprise the ethos of modern science, which are known today as *CUDOS – the Mertonian norms of science*:

1) **C**ommunism:
Science represents a community of enquirers who work together and share their results openly. Property rights in science are limited to recognition.

2) **U**niversalism:

[9] Which naturally is again an evaluation according to the EU-theory (*see Introduction, Figure 3*)

Science is impersonal and certification of knowledge is independent of ethnicity, gender, nationality, age, religion and even personal qualities.

3) **D**isinterestedness:
Scientists are not emotionally involved in their theories. This is to avoid bias and to ensure integrity. Merton does allow for "passion for knowledge and idle curiosity" however.

4) **O**rganised **S**cepticism:
Science asks questions about facts regarding any aspect of nature, irrespective of whether other parts of society approve of investigating those aspects of nature or whether those facts play in anybody's favour.

When we look back at *section 1.1*, the astronomers' intrinsic values and definition of research quality, we can infer that the Mertonian norms of science are implicitly present in the culture of astronomy. However, even in astronomy, "traditionally viewed as the 'purest science,' driven by curiosity and with no practical application" (Baneke, 2019; p.24), there is a shift to having to demonstrate economic value as well: "In order to apply, a project had to prove both its scientific and its economic value, stressing 'innovation' rather than pure science." (Baneke, 2019; p.22). However, whether a scientific field produces something useful for society, i.e. something with societal impact, is difficult to measure. The same goes for scholarly performance and "accountability", which is why those concepts are ever more linked to metrics, such as ***performance indicators*** (Espeland & Vannebo, 2008). Performance indicators introduce new norms to academia, partly contradicting the Mertonian ones. As we will see in *section 1.5*, having to fulfil targets set by indicators gives an extrinsic motivation to perform research, which may decreases research quality. Before we can show how, we must analyse the norms and rules prevailing in astronomy today.

### 1.3.3 System of Action – Communication & Publication Infrastructure

According to the structural-functional vision that science is part of society, science has the specific function to generate knowledge that counts as true (Taubert, 2019). Hence, the ***primary code*** characterising science is truth. According to the author we can classify a second functional differentiation within the disciplines: the division between the research system and the communication system, which both deliver complimentary contributions to the knowledge production process. Within the research system new truth claims are developed and (empirically) tested under consideration of the paradigmatic standards prevailing in the specific scientific field (Taubert, 2019). As elaborated in *section 1.1*, astronomy is a paradigmatic and hard science with a strong belief that we can at least approximate the facts of the universe. The more accurate theories and models can describe and predict natural phenomena, the more "true" they are. The latest theories and models count as true. The function of the formal communication system is to enable a circulation of proven truth claims within the scientific community (Taubert, 2019). The two systems are complementary with respect to knowledge production, because research can only produce new knowledge, when it can build on known facts, which is available thanks to the communication system (Taubert, 2019). We can see how the functions of the research and communication system are related to the astronomers' ***quality criteria 1 & 3***. Within the ***research system***, the goal is to push knowledge forward and within the communication system the goal is to present the truth claims in a way that it can be built upon. Following this classifications, we will now elaborate on the communication system, since it's the action system, that lies in the focus of scientific work (Taubert, 2019).

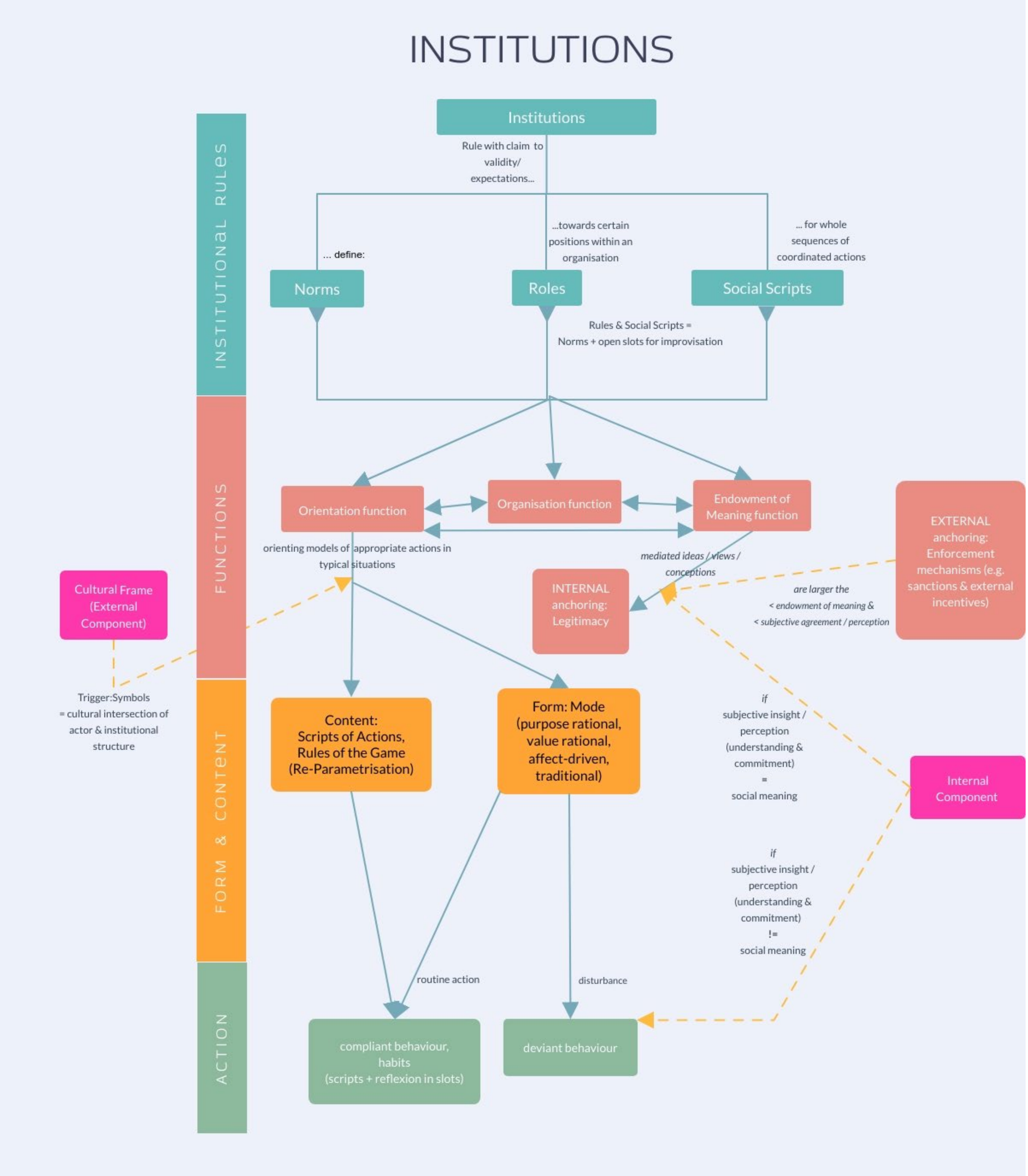

*Figure 7: Institutional Norms – This figure depicts the 3 functions institutional norms, such as norms, roles and social scripts have. They shape the actor's space of possible actions. Depending on the agreement of the actor with these norms (Internal Component; see Figure 2) and the effectivity of the internal and external anchoring the actor will show compliant or deviant behaviour. This figure was created by the author based on Esser (2000c)*

According to Taubert (2019), the ***communication infrastructure*** in astronomy is characterised by 5 main constituents: (1) a relatively small number of core journals, (2) an e-print server (ArXiv), (3) a database with abstracts and retro-digitalised metrics (ADS), (4) data repositories with star catalogues and observation tables (e.g. CDS) and (5) notifications which attest information regarding transient phenomena and "***hot topics***".

There are seven core journals in astronomy: the Astrophysical Journal (ApJ), die Astrophysical Journal Letters (ApJL), the Astrophysical Journal Supplement Series (ApJS), the Monthly Notice of the Royal Astronomical Society (MNRAS), astronomy & Astrophysics (A&A), the Astronomical Journal (AJ) und die Publications of the Astronomical Society of the Pacific (PASP) (Kurtz et al. 2005; in Taubert, 2019). Thereof, ApJ, MNRAS and A&A are the “3 biggest journals” (Taubert, 2019; p.228, translated from German). MNRAS is a British journal founded in 1827, ApJ is an American journal founded in 1895, and A&A is the youngest journal founded in 1969. According to Taubert (2019), the establishment of A&A is a good example for the good organisational structure within astronomy; through a collectively coordinated action in order to increase visibility within the highly internationalised astronomical community, journals of European countries ceased to establish one internationally European journal – the A&A. All three peer-reviewed journals are almost equivalent in value with regards to the communication system. However, some interviewees perceive ApJ as having greater impact in terms of readership, because the American journal generates more visibility [e.g. Int-Faculty1].

> “You have to publish your paper in one of the 3 major journals if you want your work to be visible, because the other journals don't count.” [Int-Postdoc1]

> “This is just part of the culture of astronomy right now. There’s not a hierarchy of journals – in biology and medicine and so forth, usually there is hierarchy of journals.” [Int-Faculty12]

The widely used green open-access ArXiv was developed in 1991 to solve the problem of most scientific journals; the ever-increasing publication rate (Ginsparg 1996; in Taubert, 2019). ArXiv is mainly used to publish pre-print articles or articles already published in one of the three main journals. While those articles are not necessarily peer-reviewed (yet), ArXiv functions as a freely available source of information, which underscores the astronomers’ intrinsic motivation to distribute knowledge efficiently (***quality criterion 3***).

The NASA Astrophysics Data System (ADS) is a central portal for astronomers to find various information and resources (Taubert, 2019): “The astronomy community usually turns to the Astrophysics Data System for bibliometrics” (Isaksson & Vesterinen, 2018; p.1). Next to containing the abstracts of publications of the three main journals, it often contains the full text as well as bibliometric information and the link to the ArXiv version.

The Centre de Données Astronomiques de Strasbourg (CDS) curates several astronomical databases, such as SIMBAD, VizieR and ALADIN, which are freely accessible. Its main function is to connect observational data with publications containing the results of the respective data analysis (Taubert, 2019).

After classifying the main components of the communication infrastructure, we turn to the main ***functions of the communication system***: registration, recognition, distribution and archiving (Taubert, 2019). Through the submission of an article an author claims priority (Merton, 1973 [1957]; in Taubert, 2019). Recognition happens through ***citations*** of research results, written in the form of a journal publication. Citations are not only dependent on the relevance of the results, but also on the reputation of first author or the research group that has written the paper (which leads to the ***Matthew Effect***, as described in *section 1.2.5*). Reputation is, as mentioned in *section 1.1.3,* the secondary code of science. “The appropriation of credit and allocation of responsibility thus go hand-in-hand, which makes for fairly straightforward social accounting” (Cronin, 2001; p.559). Distribution refers to the act of creating visibility of the published truth claim. In that respect does distribution serve the primary code, “truth”, the secondary code “reputation” and the astronomers’ ***quality criteria 1&3*** at the same time. Archiving guarantees the continuous preservation of knowledge over longer periods of time (Taubert, 2019). Archiving enables researchers to build on knowledge (***quality criterion 1***) and to evaluate the performance of others.

The communication system rests on the shoulders of the ***publication infrastructure***. Its functions are to regulate the publication process, including peer review, and to provide structures to access and provide orientation among those resources. The former function restricts publication behaviour and the latter enables research behaviour of scientists (Taubert, 2019). “Publication of scientific research is a social activity, with its rules and fashions” (Davoust & Schmadel, 1991; p.11). The publication infrastructure therefore sets many standards that influence ***evaluation procedures*** and reward structures. Hence, the publication infrastructure, but also other “changes in the conditions of scientific work and the associated reward structures have had significant impacts on our understanding of authorship” (Cronin, 2001; p.567).

The ***role of authors*** is to fulfil the functions of the communication system; “in relation to the registration function of the communication system, the author can be characterised as the first communicator about a new truth claim or research result” (Taubert, 2019; p.117, translated from German). That means that the publication infrastructure provides both; the registration, recognition and distribution function. Through the publication infrastructure, achievements can be attributed to specific researchers; the authors.

The ***role of the recipient*** is part of an author’s role. This is because citations do not only attribute value to the cited publication, but also to the citing one, since truth claims can only be legitimated with reference to other publications (Schimank 2012; in Taubert, 2019):

> “Okay, what I did was important. I've become one of those people who is obsolete but you have to reference me or else you look like you don't know what you're talking about. And it's exactly what I did when I wrote my dissertation is like, ‘Okay, I've got to reference this paper or else I'll look like I don’t know what I'm talking about.’” [Int-Faculty9]

Being an active author is an essential part of a research career. Publications are an essential form of capital (see *section 1.2.4*), without which a researcher can hardly build their résumé or receive funding.

> “And at the time I still had a Postdoc and I was doing uhm still doing some research. But my research career was sort of not continuing at that, you know in, like, writing papers, everything.” [Int-Journal]

After describing the communication and publication infrastructure, including their functions and the role of an author, we proceed with the ***criteria*** a piece of research needs to fulfil to be publishable. Research has to be novel, relevant to the field, scientifically excellent and a clearly communicated story [e.g. Int-Postdoc3, Int-Faculty3, Int-Faculty1, Int-Faculty8, Int-Faculty10 & Int-Faculty11] to pass peer review. As we saw in *section 1.1.2.1*, incremental steps are important.

> „And, uh, sometimes you don't have so good data, so good results. And anyway there should be something original or something that, uhm, will make uh, a progress in

> science. It is a small step but at least it has to be a small step forward in science." [Int-Faculty10]

However, ground-breaking research can be published "faster" [Int-Faculty8], which may bring confusion to what degree of novelty is publishable. For example, [Int-Faculty8] reports that when the field of exoplanets was young, one could publish a paper on every single discovery of a new exoplanet, whereas now there is a move towards publishing only one paper on several discoveries.

> "[…] some people have a feeling that everything that they publish must be-, must have some interesting twists in the, in the story. […] Me not so much, I think [a single exoplanet] is new and interesting by itself. It's, it's, it's another, um, data point, um, a population analysis. […] And publishing the results of 'I found another planet', and it has these characteristics, that's another data point in, in those statistical analyses, which is important to get those out." [Int-Faculty8]

Davoust & Schmadel (1991; p.19) analyse that a reason for why observations as such are not publishable anymore is that "papers are easier to write now than twenty years ago: instrumentation has improved and data of a given quality are thus easier to obtain, word processors and computer graphics are readily available and general efficiency of research has increased dramatically." This increased efficiency may have led to journals no longer accepting mere observations. As described in Heuritsch (2019a), it often depends on the subfield whether a paper ***describing the observation can be published*** without the analysis, but there may be confusion about that. Int-PhD2's supervisors also disagree on whether an observation can be published as a discovery in itself or whether the analysis should make up the bigger part of the paper with the data being complimentary.

> "Yeah you need to have new science, so as I said, with the observations I can apparently not submit a paper just about the observations which would be nice but no.", "There's no science, not enough science in it and so it would never get accepted. So, we discussed on a project I can do with the data." [Int-PhD2]

Obtained data by itself do not count as a publication. Hence, there is very little value for astronomers to spend the extra time into archiving them. A lot of telescopes do make observational data freely available after a proprietary period of 1 year, in which the group that obtained the data has priority access to analyse the data to get some science out of it (Heuritsch, 2019a). However, that is often not the reduced data that was used for the analyses, but the raw data. This makes ***replication*** more difficult and sometimes information gets lost.

> "But in many facilities, it turns out that large fractions of the data do not get published, for a number of reasons. One potential reason is the lack of resources. For example, people leaving the field, like for many post-docs. There's a bottle neck of post-docs going to faculty positions." [Int-Postdoc2]

According to Int-Journal, there are two barriers when it comes to publishing data in an accessible way. First, archives, like CDS, require the publication of a paper before uploading the data. That poses a "chicken-or-egg problem", which "creates some dissonance and some barriers". The second barrier is posed by insufficient services and infrastructure. Int-Journal refers to institutional repositories and libraries as "roach motels" (based on Salo, 2008); one can upload the data, but there is no interface, helping to get them out or making them findable.

"So, most repositories have no way of characterizing the data that they contain, they are just buckets of stuff. And so, the first barrier is when do I publish it with the data. And the second barrier is putting it into this roach motels, putting these buckets that doesn't do much for their discovery. And so, uhm, what we really need are places that provide interoperability services."

When being involved in large collaborations, the group may need to find an organised way of publishing the data and to give everybody the chance to work on it [Int-Faculty11 & Int-Faculty12]. However, this is very time intense, since there is little help of journals to curate this process:

"We've done things where we make, we make little websites to go along with the papers or whatnot, (I: Right.) but I would find these pretty unsatisfying. I would prefer a journal to curate this stuff electronically (I: Okay.) with, um, I really- I don't know. I think the field was building up a very large amount of, of, uh, user supplied information that unfortunately doesn't have very much archival longevity." [Int-Faculty12]

A lack of resources or incentives might hinder astronomers to publish their (reduced) data, however due to their intrinsic motivation to push knowledge forward, they often do see the value in sharing data with others (cf. Zuiderwijk & Spiers, 2019, [Int-Faculty8, Int-Faculty11]).

"And [archived data] will stay forever. So even if somebody has no use for the data now, uh, they may have use for the data in the future. (I: Yes. Yes.) So, it will be available. (I: Yes.) And that is very important. (I: Yeah.) That's why astronomical archives are so important. (I: Mm-hm.) You don't know when something will be useful, so suppose a supernova goes off in a galaxy (I: Yeah.) That galaxy certainly becomes very interesting. (I: Yes.) And people want to know what that galaxy looked like 40 years ago." [Int-Faculty7]

Int-Journal who has been working primarily on developing data reduction ***software***, before he started working for the AAS journals, reports that when he was a postdoc he was told that he should find a way to publish papers, because publication and citation rates are the most essential capital in ***evaluations***.

"You know thousands, tens of thousands of person hours, developer hours spent building this. And there's no, there's no citation. I mean the only citations that people use for it, if you want to talk about citation is, is a single author conference procedure. So one guy, is essentially accruing all the credit for all the developers who build this (one piece of) software […]." [Int-Journal]

Int-PhD3 also spent the first 2 years on developing a software package for data reduction and complained about the fact that he needed to find a way to make a paper out if it. He remarked that if the model doesn't fit the observations it wouldn't be publishable at all, which is a pity, since there is hardly any literature on what doesn't work. Patat et al. (2017; p.51) report that "30–50 % of the programmes allocated time at ESO do not produce a refereed publication" and that "approximately 80 % of the proposals show a publication rate of less than 60 %, irrespective of the instrument used to produce the data" (ibid. p.52). This is despite the fact, that all interviewees agree that ***negative results*** are important and should be publishable, because one simply doesn't know what the outcome of research is [e.g. Int-PhD1, Int-Postdoc2 & Int-Faculty1] and otherwise other astronomers spend "tremendous amounts of research time" to try out "completely the same stuff again" [Int-PhD3]. Interviewees explicitly distinguish negative results from failed research (cf. Heuritsch, 2019a).

> "Yes [negative results should be publishable]. Because it's about work you've done, and it might be useful for other people. Of course you don't have to claim that everything is perfect and fine if it's not. But it should be publishable. So it's only failed I would say if you give up." [Int-PhD2]

> "I mean, then you shouldn't spend too much time on publishing negative results, but shortly, briefly because that could be helpful for those who want to repeat your failed attempt. Sometimes of course, trying to check the failure can lead to a positive result. So, that's why I think failed research is that out of which no one learned anything." [Int-Faculty1]

Next to the difficulty of publishing software and negative results, there are also topics that are more "fashionable" than others (cf. Heuritsch, 2019a); the so-called ***hot or sexy topics***. The American National Academics of Science (NAS) decadal survey has a lot of influence on what topics are popular to be researched and funded in a coming decade: "the decadal survey committees present a prioritized and influential list of instruments and facilities for federal funding" (McCray, 2000; p.692). Funding agencies then constrain research by predominantly funding those topics that are deemed high priority (Baneke, 2019). That is how the decadal survey dictates, "how resources should be allocated, what facilities will best serve the community, and what the community's priorities are to be" (McCray, 2000; p.702). Furthermore, topics are sexy, when they can be easily sold to policy makers (like finding extra-terrestrial life; cf. Heuritsch, 2019a) and "research priorities are also influenced by forthcoming instruments" (Baneke, 2019; p.26).

After outlining the norms of what is publishable, we are now turning to the ***goals of publishing***. The most obvious goal is to communicate results to the community (***Quality Criterion 3***; Davoust & Schmadel, 1991), which is impeded when only certain results count as publishable. According to Taubert (2019) and Davoust & Schmadel (1991), another goal is attaining the certification for the completion of a research process. This is in line with the finding of Heuritsch (2019a) that astronomers do feel like their project has failed, if they can't publish on it – despite the fact that they would not equal failure with not being able to publish (*see above*). Astronomers are aware that they need to publish "as fast as possible" [Int-PhD1] to get recognised for their achievement, but do not necessarily view a paper as being the end result of a scientific research process:

> "[One should publish] as soon as possible. (I: Okay.) Okay. This is the opportunistic reply. The opportunistic answer to your question. But writing up a paper, I have a very strong opinion that writing up the paper is not the closing phase of your research. Just the opposite, on the contrary, it is the most intellectual part of the research because when you write-up your paper, that's when you sum up your thoughts. You describe your results and that's when you finally truly understand what you have achieved. […] And that's because during writing up the paper, you finally reach the meaning of your work. So, in other words, as soon as possible, that's important because we are in competition. But then again, when you understand your work, that's when you have to publish your results. But writing up the paper must be started earlier." [Int-Faculty1]

Nevertheless, do the journals "constitute the archives of scientific research: they record discoveries, data and theories and the authors who are responsible for them. They also serve as a forum to express new ideas about scientific facts" (Davoust & Schmadel, 1991; p.11). Despite the existence of ArXiv, astronomers depend on publishing in peer-reviewed journals, since that is how they generate visibility and recognition, which in turn has positive effects on their career prospects (Taubert, 2019). The peer review process leads to recognition of the piece of research, which is why publications are also ***symbolic capital***. Sharing data online, by contrast, may generate visibility, but not recognition (Taubert, 2019).

Interviewees generally report that they value ***referee's comments***, because they give the chance to improve a paper [e.g. Int-Postdoc1, Int-Faculty12; cf. Taubert, 2019] and that the comments are fair most of the time. However, receiving negative referee reports may be perceived as painful and so awaiting the feedback can be stressful [e.g. Int-Faculty8]. There are some referee reports which may sound personally attacking and that can be especially dispiriting for young scientists who haven't learned not to take referee comments personal [Int-Postdoc1, Int-Faculty8]. When coming from a different subfield or ***perceiving publication pressure***, the reviewer may not be able to make a proper assessment [Int-Faculty2 & Int-Postdoc2]. When referees do not agree with the interpretation of the results, sometimes there is no other choice than taking the paper to a different journal. Int-Faculty2 & Int-Faculty9 point out that it is the role of the referee to check for scientific correctness and to ensure integrity, but that hindering publication because of a disagreement with the interpretation may go too far. In her specific case this led to another group publishing the same research first.

### 1.3.4 System of Action – Obtaining & Handling Data

Observational astronomers need access to telescopes to obtain data and access to data archives to compliment their observations (*see section 1.2.1*). Deciding for and building telescopes often require large collaborations. That is why Baneke (2019; p.1) acknowledges that astronomical instruments and "big science projects are an integral part of the ***moral and institutional economy*** of modern astronomy". The author studies how the Dutch astronomical community chose its next big telescope project in the 1990s, focussing on "strategic arguments", rather than scientific ones. Baneke (2019) found three main strategic concerns; First, Dutch astronomers wanted to ensure future influence. Being involved in building instruments ensures ***privileged access***. Second, they wanted to gain visibility. With a flagship project political support can be gained and "the bargaining position within the international astronomical community" can be maintained (ibid. p.7). Third, the most important strategic concern addressed continuity. Large, long-term projects help to maintain resources such as staff and thereby expertise. According to Baneke (2019; p.6), the strategic arguments were "clearly for internal use" and differed from "the later 'public' arguments", which were more oriented towards scientific arguments. Given the astronomers' intrinsic values (see *section 1.1*), one would have assumed the opposite – that scientific arguments are used within the community and strategic ones applied when it comes to talking to politicians or funding agencies. However, according to Baneke (2019; p.6) those intrinsic values "guide the internal functioning of the community, not as strict (moral) directives nor as pragmatic guidelines, but rather as implicit, internalized standards of how the community should function in an ideal world." Astronomers are aware that in the real world, by contrast, capital such as access to telescopes, is important to be able to perform research (see *section 1.2.1*). International collaborations are important in two ways (Baneke, 2019): Large collaborations are needed to cope with the costs and scale of big science projects and also to negotiate access to telescopes. The latter is an essential resource, not only to obtain data, but also "also for attracting or keeping prestigious staff, another scarce resource, and for training students." (Baneke, 2019; p. 26). Since instruments are an important part of the epistemology in astronomy, using them shows students "how science is done" (Baneke, 2019; p.26). In summary, astronomical instruments have constitutive effects on how science is done; "they shape research opportunities, and they affect careers and institutions" (Baneke, 2019; p.26).

Now that we have established how essential telescopes are for the ***moral economy*** in astronomy and that data is important capital for an observational astronomer, we can turn to the process of ***data acquisition***. When an astronomer wants to apply for ***telescope*** time, one usually has to send a proposal to the telescope's selection committee (***if one doesn't have privileged access***). Int-Postdoc1 & Int-Faculty5 explain what is needed for the proposal; a compelling research question and an idea of what objects/ part of the sky to observe. Then you need to know what instrument/ telescope is suited for the project, the technical specifications and what dates are needed to observe the target object. In Int-PhD2's case there were three suitable telescopes and so she wrote a proposal for all of them, together with her Chilean and German collaborators, to have a better chance to get one of them accepted. To her surprise, all of them got accepted. However, then she was unlucky with the weather and a support astronomer at the telescope, who was new and didn't know enough about how to operate the telescope. Usually the competition for telescope time is much higher though.

> "Because in a semester there are 180 nights, but usually around 10% is for technical work. So, you have maybe 170 nights per semester. And it could happen that in a semester there are people asking for 300 nights." [Int-Faculty5]

What Int-Faculty5 describes here would be an oversubscription of roughly 2 and naturally, the more popular the instrument the higher the oversubscription rate. At the Very Large Telescope (VLT) the rate is 3 (1/3 of the proposals get chosen; Int-Faculty6) and ALMA, an ESO cutting edge facility, has a rate of 9. At Hubble telescope, 1 out of 6 or 7 proposals get accepted [Int-Faculty3]. According to Int-Faculty6, a high oversubscription rate isn't healthy, since a lot of really good proposals need to be turned down and that's not motivating for astronomers with good ideas. If the oversubscription rate is too high quality, scientifically relevant proposals need to be turned down and as a result, people feel time allocation only depends on luck:

> "So if it's about one in three or one in four [proposals accepted] then people say, 'Okay, uh, I need to do better but at least I know I have a chance.' (I: Right.) If it's one in ten, then most people will think, 'Well, it will now depend on luck and a few things that however good I write it, it may still not be good enough.'" [Int-Faculty6]

> "For a mature observatory where the astronomical community knows very well the capabilities and idiosyncrasies of the instruments onboard the observatory, typically one third of the proposals are, in my opinion and in that of many committee members with whom I have spoken, of exceptionally high quality. Unfortunately, the high oversubscription rate requires that a large percentage of these high quality proposals must be turned down for lack of available observing time." (Linksy, 2006; p.112)

Linksy (2006; p.112) observed that "peer review committees are generally very good at identifying with minimal bias the top one-third of the submitted proposals." He distinguishes four biases for discriminating amongst the lower two-thirds: (1) the *lack of expertise bias*, (2) the *subject-area bias*, (3) the *inadequate perspective bias*, and the (4) *lemming bias*. First, often astronomers with critical expertise cannot serve at selection committees due to their teaching responsibilities. Second, if most panel members aren't familiar with a proposed topic, the selection committee may not choose it, despite of the scientific value of the proposal. Third, a selection committee must also preserve the interests of an observatory and hence not choose projects which may be considered routine or else unexciting, but might be critical for a major project. We can also call the fourth type of bias, the "***sexy topic***" bias – "At any one time, certain topics are ripe for investigation because of new observational capabilities, theoretical advances, or new perspectives on previously dormant topics" and selection committees have the tendency to "go with the flow" (Linsky, 2006; p.114).

When astronomers are granted the telescope time, they either travel there for the time of the observation or obtain the data from remote (e.g. Roy & Mountain, 2006). In either case, often the help of technicians or support astronomers is needed, since telescopes nowadays are so large and complex that their handling often requires more staff and expertise. Afterwards, data need to be processed and reduced, before it can be analysed, which Int-Faculty5 calls "doing your science".

> "And after that, when you have the clean image, you start measuring. (I: Mm-hmm.) And that is not as automatic as you can think. Sometimes you can do it very quickly and sometimes you can't. Depends on-, depends on the quality of the weather or how was the telescope working." [Int-Faculty5]

Reducing, analysing and modelling data needs self-written software packages and code or developed by others. That may be time intensive either way, because when already existent software is used, that often needs to be adapted and searched for bugs or systematic errors [Int-Faculty12]. The data reduction and analysis process after obtaining the data from a telescope may not differ from the case where data is taken from an ***archive*** instead. Retrieving data from archives may pose several ***challenges***. First, contrary to obtaining data oneself, when using data from others, one needs to base one's scientific question on what one knows that is available [Int-Postdoc2 & Int-Faculty3]. Second, a lot of data collected up until a few decades ago hasn't been digitalised yet. Int-Faculty8, for example, works with that kind of data and remarks, that often data gets corrupted when transferring it from old to new devices and one needs to find other copies elsewhere to fill the blanks. In addition, the data may be too big to be sent via an internet connection and so it has to be mailed physically. Third, despite CDS, the central database, an infrastructure for different data sources, archiving data is under-standardised (see *section 1.3.3*) and not incentivised by the ***evaluation system***. Data becomes public usually a year after obtaining them with a telescope. However, in many facilities large fractions of data are also not published, due to a lack of resources or relevant people leaving [Int-Postdoc2]. The archives may contain the raw data, but not often the reduced data. Sometimes people are reluctant to provide their reduced data, either because of lack of time or because there is still the hope to get some science out of it [Int-Postdoc2, Int-Faculty10].

> "Um, I know that, uh, like some people, uh, who work with light curves they just publish the data on their database immediately. (I: Mm-hm.) Some people hold on to it. Uh, like, I don't think there is a general – a way, uh, a way of generalizing – […] Some people maintain databases, some people don't." [Int-Postdoc3]

Even if the data are public, one needs the tools/ software to analyse them [Int-Faculty10]. Astronomers provide more or less good documentation of the data [Int-Faculty11]. Hence, the "mechanical challenge" [Int-Faculty12] is to find, download and using suitable data. Fourth, when the (reduced) data isn't accessible, one needs to contact the responsible PI [Int-Postdoc1, Int-Postdoc3, Int-Faculty3 & Int-Faculty11], which may lead to delays and possible restraints in using the data [Int-Faculty8]. Int-Postdoc1 points out that this is the ethical way to go in any case, as just because the data is online, it doesn't mean that the research has been finished. The PI can be found via the associated paper; accessible data are usually referred to in a publication, since publication of data is not rewarding in itself (see *section 1.2.4*). That is also the reason why astronomers don't ***cite the data itself***, but the associated paper [Int-Postdoc1, Int-Postdoc2, Int-Postdoc3, Int-Faculty11 & Int-Faculty12], which brings most credit to the PI [Int-Faculty10]. Sometimes astronomers, especially large collaborations, specify how they want to be cited or mentioned in the acknowledgements [Int-Postdoc1, Int-Faculty11].

Int-Faculty10 explains that the citation culture is one of the reasons why, historically there has been a "fight" between astronomers involved in building instruments and those who are not. Building an instrument requires lots of funding, but doesn't enable writing papers (immediately). Additionally, those papers aren't cited as much, because people aren't interested as much in the technical details of an instrument. This puts instrument builders at a disadvantage as compared to observers and theoreticians when being evaluated for their performance. This brings us to the next section.

### 1.3.5 System of Action – Evaluation System

> "I was Director of Local, National, and International Institutes. So, I have suffered from evaluations all my life." [Int-Faculty4]

Roy & Mountain (2006; p.24) confirm that "the new paradigm of strict accountability" caused the "ground-based sociological paradigm" to shift "out of the domain of telescope and instrument building as a scientific endeavour, and pushed into the realm of a tightly managed project […]". That poses the question, how concepts like accountability, performance and quality in science are measured.

Let us first distinguish between evaluation of PhD students, astronomers aiming at tenure and tenured staff. The ***evaluation of PhD students*** differs by university, but generally we find that they have to write yearly progress reports and present their research in their research group on a regular basis [e.g. Int-Faculty1 & Int-Faculty11]. In Int-PhD3's case the report is only a formality that nobody really reads and Int-PhD1 doesn't know on basis of what criteria the report gets judged. Int-Faculty1 did badly in the first 3 reviews of his PhD, because he didn't have enough ***publications***. A PhD usually should take 3-4 years, depending on the country and the PhD student needs to publish a paper per year. If it's less, like in the cases of Int-PhD3 & Int-PhD1, where the requirement is two and zero papers, respectively, one doesn't have good chances to get a postdoc afterwards [Int-PhD2, Int-Faculty1]. In addition, PhD student often have to absolve course-work, including exams [Int-Faculty1].

In order to attain tenure, one must have obtained a good track record, which is made up of the so-called ***performance indicators*** that make up the astronomer's symbolic capital (*see section 1.2.4*). ***Publication and citation rates*** are one of the most important indicators in astronomy; they are the basis for the top currency – ***reputation***. As Int-Postdoc2 puts it: "Uh, without papers, uh, you don't exist as a scientist." A tenured astronomer usually also needs to demonstrate the ability to acquire funding, like external grants, telescope time and needs to be involved in teaching and other services for the department or the community, like serving on selection committees [Int-Faculty12]. They also need to supervise PhD students.

> "Publications, lectures, prizes. Uh the normal things in academia, letters of reference. … Yeah those are the indicators. Standard indicators." [Int-Faculty7]

Those achievements may not only determine whether one ***receives tenure***, but also a salary increase. Often, annual reviews about the achievements in the past year and the goals for the coming year decide over such a "promotion" [e.g. Int-Postdoc3, Int-Faculty1, Int-Faculty3, Int-Faculty11,]. For example, Int-Faculty1, the director of an astronomy institute, developed an Excel spreadsheet which contains the number of 1st author & co-author publications, the corresponding JIF, citation rates, research grant income, (PhD) student supervision, external professional work and services to the community, like refereeing journals, outreach (also mentioned by Int-Faculty2 & Int-Faculty5) and committee work. Int-Faculty1 admits that it's "bad" to evaluate the performance of his staff that way, but doesn't know how to determine a salary increase in a fairer way. By means of this calculation, he can also determine underachievers and help them become more productive in a system, where publish-or-perish is "a real thing". Further, Int-Faculty8 reports:

> "We put forward a portfolio of stuff that we would be doing. So, things like here are the general articles I've written. Here are the conferences I've attended, here are the conferences I've hosted. Um, here are the students I have, uh, supervised and graduated as PhD. Here are the big grants I've won. Here are the collaborations I'm in. Here are all the media appearances I've done. You know, everything you could think of, um, in-, which has some kind of value to the university is in, in this report."

The review may take the university a long time and results into grades for every individual, which range from A, the "high flyers" to C, the "under-performers" and "inactive". The university receives funding from the government according to the grades. The individual researcher can apply to use some of that money for salary and travelling, but not for resources. In addition to that portfolio, the university maintains an automated system, called "research outputs", which aligns the researcher's publications automatically and "keeps an eye on the bibliometrics". Researchers don't get to see that number, but they should make sure the report is up-to-date. Int-Faculty8 doesn't know what weight is given to the different bibliometric indicators when it comes to a promotion. Int-Faculty5 & Int-Faculty7 also report that the rules for climbing up to the next level aren't always exactly clear. In Int-Faculty7's experience with job applications in India, it is not transparent which metrics are valued. He was only asked to provide a list of his publications, a CV and a summary of his research projects, but he also included indicators, such as number of publications & citations and the H-index. Int-Postdoc1 mentions that she feels that appraisal situations are "quite fake and pointless" in the sense that no meaningful content is discussed. According to Int-Faculty7, when it comes to job applications, ***recommendation letters*** are more important than indicators, unless "they are completely way above the average or way below the average". At his faculty, the value of an individual recommendation letter depends on the career level of the applicant; recommendation letters of PhD applicants are given less weight, given their little experience. In the case of faculty, people who are qualified to talk about the candidates, are contacted, even if they are not on the list the applicant handed in. Int-Faculty3, Int-Faculty9 & Int-Faculty12 also stress the importance of recommendation letters in one's career. Several other interviewees report that ***qualitative assessments*** are performed in various forms (e.g. Int-Faculty 2& Int-Faculty3; cf. Taubert, 2019). For example, in the form of a "social assessment" [Int-Faculty10 & Int-Faculty12]. Shy people may fall through the cracks here, because they are not as active on conferences & social media. Int-Faculty12 points out that excellence can manifest on many different axes. That is why one needs to get the whole picture when evaluating a person for a position.

> "I shouldn't say but I think most of the parametric evaluations that are on the market are not enough." [Int-Faculty12]

Int-Faculty11 explains that indicators are imperfect, because one doesn't want to know how people did in the past, but how they are going to do in the future. Nevertheless, "at early career stages personnel decisions are essentially predictions about an individual's future achievements" (Kurtz & Henneken, 2017; p.695) and the hope is that the past performance, as reflected by the indicators, predicts their future one [Int-Faculty11 & Int-Faculty12]. Additionally, "scholarly ability, itself, is a quality (Pirsig, 1974) that can be perceived, but which is very difficult to specify exactly" (Kurtz & Henneken, 2017; p.696). That is why indicators define what counts, even though "they're very noisy and imperfect" [Int-Faculty11]. "Citations and publications are among the most important metrics in the evaluation of an astrophysics researcher's performance" (Zuiderwijk & Spiers, 2019; p.236) and this is reflected in the hiring processes, despite efforts in the direction of qualitative assessments. Int-Postdoc1 & Int-Faculty9 report that one's CV and publication rate are important for an application, because as a researcher it's important to show results in the form of a publication record.

> "One has to show her/his scientific productivity and the number of papers and citations is always taken into serious consideration for this." [Int-Postdoc2]

We conclude that, despite the "fuzziness" [Int-Faculty2] of indicators, and the awareness that qualitative assessments are as important as quantitative ones, astronomers admit that quality is difficult to measure and so they widely resort to bibliometric measures.

In order to understand what norms rule research in astronomy and how good research performance and quality is defined, we are also interested in the developments of institutional norms over time. We will first look at ***changes in the evaluation system*** and then at changes in the definition of quality research.

Int-Faculty11 & Int-Faculty12 don't think that the evaluation system has changed that much over time, at least in terms of how funding is attributed and on what criteria astronomers are hired. However, both state at the same time that there is more pressure on early career astronomers due to increased competition (cf. Heuritsch, 2019a). Davoust & Schmadel (1991) state that the job market in astronomy is saturated and Int-Faculty10 explains that there are more "middle-good" people in comparison to great astronomers than a few decades ago. He finished his degree with Cum Laude and yet it took him 2-3 years to find a permanent position, which made him desperate. Now people need to do this for 10-15 years: "So I think you have a much- much hard life than we had, uh, 40 years ago" [Int-Faculty10]. Int-Faculty4 also describes that getting research positions was easier just after WW2.

> "But you know, Julia, in 1963 the world was just beginning to wake up from the nightmare of the 2nd World War. And all the reconstruction that was going on. But many things like physical science, were growing. And there was an enormous shortage of competent people. So, I was offered six jobs when I got my PhD in Cambridge without ever writing an application. Yeah. So, which one did I choose? I chose one at a Canadian university, which started with a year leave of absence at half salary as an assistant professor. With no duties except to study philosophy in Amsterdam. And then to come at the end of that academic year to that university and then I was promoted to associate professor. Without ever doing a day of work for the university."

Possibly that is why, as opposed to Int-Faculty11 & Int-Faculty12, Int-Faculty6 thinks that today, the "numerical stuff" counts more than in the past, where recommendation letters were more important. According to Int-Faculty3, in the past there were also no formal student evaluations of teaching. We conclude that our empirical evidence is not sufficient to say much about the development of the evaluation system, but we do observe that established astronomers generally consider themselves as having been ***lucky*** and that today they would have had a harder time climbing up the career ladder. They feel like they could afford riskier projects (see *section 2.1*).

Let us turn to changes in the astronomers' definition of what ***good research quality*** is. Heuritsch (2019a) found evidence that it hasn't changed much over time and is very much rooted to their realist attitude and related to the ***Mertonian norms***. This is nicely illustrated by the differentiation between astronomy and Astrology, an "admittedly *a posteriori* science" (Almási, 2013; p.26) goes back to Tycho de Brahe, a popular astronomer from the 16th century. He distanced himself from Astrology and stressed the importance of "observational accuracy and correct theoretical conclusions" (Almási, 2013; p.25). His aim was to make "irrefutable knowledge claims" and "raising astronomy to the height of quasi-mathematical exactness" (Almási, 2013). While his aim was to "reinforce his astronomical credibility" (Almási, 2013; p.25), as we know from *section 1.1.2*, improved knowledge and empirical & methodological correctness, are still two of the astronomer's quality criteria.

In the context of good quality research, we now turn towards institutional norms about replicability and failure, since it is interesting to observe that the astronomers' individual opinions about those topics ***diverge*** from what happens in practice in academia, as we will see in *Chapter 2*.

***Reproducibility*** is not only a ***quality criterion*** that stems from the astronomers' drive for correct inferences and truth search, but also one that is institutionalised in science in general. Replicability is one of the reasons why astronomers voluntarily share data (Zuiderwijk & Spiers, 2019), allowing other researchers to check and build upon their results, which in turn improves research quality. [Int-PhD3, Int-Faculty1 & Int-Faculty2] stress how important it is to be able to perform "sanity checks" of how the results were derived and Int-Faculty1 states that without replicability a piece of research is worth nothing.

> "There's an important difference between a scientific publication and a news article. A news article is to summarise: here are the great results. A scientific publication has to lay out how the results were derived, in a such a way that you can follow through and reproduce what has been done. If you sweep things under the carpet, you're writing a news article; you're not writing a scientific publication." [Int-Faculty2]

As pointed out by Heuritsch (2019a), evaluation procedures do not only shape the landscape of success, but also the opposite; the landscape of failure, which is everything outside of the former. Following that logic, non-publishable research is perceived as failed in academical astronomy. This is despite the fact that, to an individual astronomer, research is only failed when it can't live up to their ***3 quality-criteria***. For an astronomer it is not failed research, if someone learns something out of it [e.g. Int-PhD1 & Int-PhD2]. That is also why it would be important if the publication of negative results was institutionalised (see *section 1.3.3*).

> "I would consider something as a failed research out of which no one learned anything. All the rest is useful research. So, if no one learned anything out of the research then that's a failure." [Int-Faculty1]

> "Failed research is research that wasn't conducted properly, not rigorously, data analysis that doesn't look properly at the evidence. I wouldn't call anything failed based on the outcome. Something will come out of it, and you will always have lessons learned […]. If everything goes as planned, I think that's the largest failure because it means that nothing new came out of it." [Int-Faculty2]

> "You count failures on a person basis or on result basis, because it can differ a lot. Because maybe the person worked very hard, and there's not really a result coming from that, which is just a pity, just the way it is. Otherwise, it could be a person who's maybe worthless in science, or not. Maybe he's worthless in science, but he's still produced a paper which is based on shitty data. Then it seems that it's more a failure to me." [Int-PhD3]

### 1.3.6 Conclusion – (Naïve) Meritocracy

The assumption that future performance can be predicted by past performance, such as making funding dependent on publication- & citation rates (see *sections 1.2.4 & 1.3.5*), is based on the *meritocratic paradigm*, which dominates "highly competitive Western cultures" (Pluchino et al., 2018; p.1). According to this paradigm, the distribution of success among people "is either a consequence of their natural differences in talent, skill, competence, intelligence, ability or a measure of their wilfulness, hard work or determination" (ibid. p.2). In the case of academia, success is measured by the bibliometric indicators, serving as a proxy. However, as the authors point out, the "importance of external forces" usually is underestimated in the meritocratic paradigm (ibid. p.1). That explains why "intelligence (or, more in general, talent and personal qualities) exhibits a Gaussian distribution among the population, whereas the distribution of wealth – often considered a proxy of success – follows typically a power law (Pareto law), with a large majority of poor people and a very small number of billionaires" (ibid. p.1). The authors demonstrate that the "hidden ingredient" that explains that discrepancy is randomness or, in order words, ***luck*** (ibid. p.1). They find that "the most successful individuals are not the most talented ones and, on the other hand, the most talented individuals are not the most successful ones" (ibid. p.7). In other words, the most talented people are "overtaken by mediocre but sensibly luckier individuals", so a "great talent is not sufficient to guarantee a successful career" and "the most successful individuals are also the luckiest ones" (ibid. p.1,14,8). One may refer to Pluchino et al. (2018; p.2) for a series of examples how success of scientists is determined by luck. The authors point out that evaluations of a researcher's talent happen a posteriori, which switches cause and effect, and results in "rating as the most talented people those who are, simply, the luckiest ones" (ibid. p.3). In the meritocratic paradigm resources are usually attributed to those who are evaluated as the most successful ones. Since they are not in fact the most talented ones, the risk of this paradigm is that the success of the luckiest individuals accumulates through a "positive feedback mechanism, which resembles the famous 'rich get richer' process" (ibid. p.15). This is known as the ***Matthew Effect***. The authors conclude, that the meritocratic paradigm must be challenged and its mechanisms of resource distribution reconsidered, such that luckier individuals are not given an advantage over more competent ones.

## 1.4 External Component: Cultural Frame

The *cultural frame* is the third external component of the logic of the situation. It consists of the symbols present in one's current situation that indicates which *frame* to apply (Esser, 1999). Frames help defining the situation, since they inform the actor what mode of action (*see Figure* 2&3) and what norms to apply. Symbols help to identify what (material) opportunities and restrictions apply and what alternatives are possible. Parsons & Shils (1954; in Esser, 2001) call the ensemble of norms, values and symbols the *cultural system*. This system is element, influencing the actor's values, and object, influencing the actor's wishes, of orientation at the same time. Therefore, the cultural system steers and restricts actions.

The *cultural frame* shapes the *cultural bond*, which is one of three aspects of the *structural bond* that connects actors involved in a *social situation* (Esser, 2000a). While strategic acting comes into effect when the *material bond* dominates (see *section 2.2*), and *normative bonds* structure various kinds of social relationships (such as those between researchers; *section 1.3*), the *cultural bond* leads to interactions, communication and co-orientation. This section will first review the symbols that define astronomer's interactions, before explicating the cultural bond that astronomers share.

As we have discussed in *sections 1.2.4 & 1.3.5*, ***bibliometric indicators*** are the prevailing symbols of performance and success in research. They serve as proxy for scientific achievements and -quality. Hence, they are used in resource allocation decisions, whether that is funding, job positions, telescope time or rewards. Especially publication- and citation rates are equated with an astronomers' impact and determine their reputation. Grades are important for getting a good PhD position, which is the starting point of an astronomer's career if they want to stay in academia. Similarly to other systems within the Western meritocratic paradigm, the academic system is very ***hierarchical***, with the positions symbolised by the titles such as PhD, Postdoc, associate professor and professor. These titles serve as symbols for having acquired more or less high cultural capital and hence are endowed with more or less power.

Let us turn towards the cultural bonds astronomers share all over the world. First of all, as we saw in *section 1.1*, astronomer's share a passion for space and enjoy the intellectual activity of solving the mysteries of the universe. Their ***common goal*** of pushing knowledge forward is curiosity driven and they have a mutual definition of what "truth" is in accordance of their realist attitude. Astronomy is a single-paradigmatic science and the concept of "truth" is deeply ingrained in the culture of astronomy. Discussions at the frontier of the field revolve around what theory or model approximates reality better. Those are the ones that then are regarded as "true" and the paradigm on which theories can be built upon.

Second, astronomy is a relatively small field with a ***tightly knit, highly internationalised global community*** (Heuritsch, 2019a). Training of young astronomers and work habits and culture in institutes and faculty departments are similar internationally. One major cause of this is the ***need for cooperation*** with regards to the construction and use of instruments: major telescopes are built, managed and used by multi-country collaborations. Roy & Mountain (2006; p.25) elaborate on the shift from the "gentleman astronomer to experimental teams" that happened in the past century. Traditionally, observing was the privilege of wealthy men, spending "lonely nights" at the telescope, from which astronomy derives its "non-insignificant element of romanticism" (cf. *section 1.1.1*). According to the authors, both, internal (culture) and external forces (such as funding agencies), changed the "whole sociology of what constitutes 'an observation'" (ibid. p.25). Internal forces stem from the “strong common culture” astronomers share through their publication infrastructure, “professional associations and societies, with consulting and evaluation panels that use a pool of international colleagues” (ibid. p.34). The globalisation of astronomy has “emerged from almost two centuries of exchanges between scientists from leading European, Asian and American institutions, the migration of scholars triggered by geopolitical changes, and the quest for pristine observing sites in distant parts of the world” (ibid. p.34).

> “The growing professionalization of science accompanied by a massive influx of graduate students into University research institutes, the revolution in communication, the pressure to publish in order to progress in a scientific career, and the growing complexity of knowledge are invoked as causes for the abandonment of the traditional individualism in science to a collective regime.” (Fernández, 1998; p.61)

According to Roy & Mountain (2006; p.34), “funding agencies are looking for strong national and international partnerships and prefer collaboration to competition”, although at the same time we observe that limited funding opportunities also lead to ***competition***. This paradox most likely results from the need for collaboration in big science projects, while competing for limited resources at the same time (cf. Heuritsch, 2018; see *section 1.5*).

Third, next to the need to collaborate, another cause of astronomy’s international socio-cultural homogeneity is a high degree of ***international mobility***. The Master graduates who want to do a PhD and the PhD graduates who want to try for a career as professional astronomers face a job market where supply exceeds the number of available posts substantially. The chance to get a PhD student position and, even more so, a postdoc can only be increased by the willingness to relocate to other countries (Fohlmeister & Helling, 2012). At a more senior level, having stayed abroad for a few years is mandatory for many tenure track appointments.

> “But if you’re not flexible to move to the other side of the planet, there you go leave the field, head chopped off.” [Int-Postdoc2]

Hence, working abroad for some years is a common feature in the early career stages of professional astronomers. However, “motherhood complicates things” (Fohlmeister & Helling, 2012; p.286) and indeed several interviewees – mothers and fathers – explain that family considerations may impede the astronomer’s career, because of restricted mobility [e.g. Int-Postdoc2, Int-Faculty4, Int-Faculty7, Int-Faculty8 & Int-Faculty9].

> “Well it doesn’t mean that I don't love science actually. I still love astronomy but, uh, you know for different reasons. And at some point in your life you are not sure, do you want to-, to do this or just-. You know at some point you think that family is more important. So I, uh, had a baby-, had a baby, so-.” [Int-Postdoc1]

> “The job market sometimes favours people that don’t have a family, or that can easily take their family along with them. And this creates biases of course. Because women have normally more difficulties in moving their families than men do.” [Int-Postdoc2]

Familiarization with international colleagues starts even before graduation, during Bachelor and Master training. Many students from countries without a strong tradition in astronomy do their training abroad. This does not only stimulate an international outlook of the students that study abroad, but also of those who study in their home country, because they meet and interact with the foreign students in courses. This strong international orientation of the training and career of astronomers strengthens the international socio-cultural homogeneity.

Despite the ***Mertonian norm of universality***, ***bias and discrimination*** is not only a problem when it comes to mobility: "Systemic discrimination on the basis of gender and race, among other ascribed identities, harms minoritized people. This is a structural problem in society, and astronomy is not immune to it" (Prescod-Weinstein, 2017; p.1). Most of the studies on (unconscious) bias we found concentrate on underrepresentation of women in astronomy (e.g. Caplar et al., 2017; Knezek, 2017; Lucatello & Diamond-Stanic, 2017; Rathbun, 2017; Schmidt & Davenport, 2017; Tuttle, 2017). Fohlmeister & Helling (2012; p.286) find that "prejudices [against women] are abundant, and are perceived as discriminating". Women "find networking in professional environments challenging as they usually juggle more than just their jobs and are still confronted with gender prejudice" and "are more likely than men to quit astronomy after they obtain their PhD degree" (ibid. p.1). Ethnicity bias also comes into play in various forms. As mentioned in *section 1.2.4*, Trimble & Ceja (2012; p.45) find that "citation rates are highest for papers with European and American authors and much smaller for papers from less-developed countries". Int-Faculty5 points out that referees may be more cautious with papers outside the US and EU when it comes to checking the language. They may require a native English speaker to review the paper before submission, even if a co-author is such a native.

Fourth, the multiyear socialisation and intense training in the physical sciences and various other disciplines (*see section 1.2.2*), as well as their paradigmatic notion of "truth", makes astronomy and its communication infrastructure ***highly exclusive*** (Taubert, 2019). Astronomers use their own jargon and abbreviations – even abstracts are difficult to read for an outsider. This reminds us of Knorr-Cetina's (2003) concept of an "epistemic culture", which are those "amalgams of arrangements and mechanisms – bonded through affinity, necessity, and historical coincidence - which, in a given field, make up how we know what we know" (Heidler, 2017; p.2). They often remain "implicit in the day-to-day research process" and do "not only include the specific epistemic strategies concerning the generation of data, the theoretical approach, and the 'instrumental concepts,' but also the 'collaboration style,' and the way epistemic cultures produce social legitimacy" (Heidler, 2017; p.3).

## 1.5 Summary: Bridge Hypotheses

The study of the material opportunities (*section 1.2*), the prevailing norms (*section 1.3*), the cultural frame (*section 1.4*) and their mutual conditionality are the basis of the *institutional analysis* and therefore "the start and end of any sociological explanation" (Esser, 2000c; p.4, translated from German). It is beyond the scope of this paper to include a full historical derivation of how these external components developed over time. We are interested in the logic of the astronomers' situation in academical research at present today and to explain their actions and resulting aggregated phenomena on its basis. Thereby, the *logic of the situation*, which consists of the three external and the internal component (*section 1.1*) must be translated into *bridge hypotheses* that link the objective situation of an actor to their subjective motives and knowledge. This means that an actors' subjective goal is based on their internal values wishes and oriented around the objective material, institutional and cultural situation. The bridge hypotheses correspond to the variables of the *logic of selection* (*Figure 1*). As motivational factors are the variables the basis that flow into the *action theory* (*Chapter 2*), the general and nomological explanation for the behaviour in the specific situation. In *Chapter 3* we will then turn towards the *transformational rules*, which describe under which circumstances individual behaviour leads to a certain *collective phenomena*.

In this section, we identify the ***bridge hypotheses*** that link the *logic of the situation* of an astronomer in academia to their research behaviour. To do so, we need to translate the *logic of the situation* into variables regarding the knowledge (**k**) & values (**v**) of an astronomer and the expectations (**e**) & utility (**u**) an astronomer assigns to possible outcomes of an action (Esser, 1999). Since this is a qualitative study, we will not quantify those variables, but describe them qualitatively.

### 1.5.1 Variables & Motivational Factors

As we have seen in *section 1.2.2*, successful navigation through an environment requires **knowledge (k)** about the norms and rules prevailing and prescribing the behaviour, whether this is present in tacit or in explicit form. As McCray (2000) points out, "[…] one of the factors determining whether an individual or institution is able to secure access successfully to resources ist the ability to understand and use the unwritten conventions, traditions and rules of astronomy to best advantage." While we pointed out that early career researchers are quite naïve when it comes to the requirements of the evaluation system, we have also seen that PhD students are aware of the importance of publishing and being assessed via quantitative indicators. *Table 5* lists several kinds of **values** (**v**), which define an astronomers' situation and which correspond to the motivational factors which an astronomer bases their decision on. The intrinsic & institutional values both co-define **utility** (**u**) in the sense that those values specify what a certain outcome is worth. The chance to gain more capital also increases **u**. Hence, an outcome of a specific action is worth the most, when both, the primary & secondary code of science are fulfilled, when being compliant with the prevailing norms and when capital is gained at the same time. Capital also increases **u** indirectly by endowing astronomers with the means to perform research. The probability with which a certain outcome can be achieved is called the **expectation** (**e**), which one can better estimate when one has more knowledge (**k**) about institutional norms (i.e. "how the system works"; Int-Faculty1). Due to the ***Matthew effect***, that describes the phenomenon of capital accumulating where capital already exists, capital increases **e** to reach a certain outcome.

*Table 5: Variables & Motivational Factors derived from the logic of an astronomer's situation*
Legend: ↑ = *positive effect*, ↓ = *negative effect*, → = *value (co-)defines the variable*

| Kind of Value (v) | Sub-variable of *v* | General effects on other variables |
|---|---|---|
| Intrinsic Values (*section 1.1*) | Primary Code: "truth" & research out of curiosity | → **u** |
| | Research Quality – Criteria:<br>1. Originality<br>2. Correctness<br>3. Useful for community (and societal impact) | Primary Code ↑ |
| | Secondary Code: recognition | → **u** |
| Capital:<br>(*section 1.2*)<br><br>1. Economic Capital (EC)<br>2. Cultural Capital (CC)<br>3. Social capital (SoC)<br>4. Symbolic capital (SyC) | GENERALLY | Primary Code ↑ (capital makes research possible and is needed to stay in academia);<br>**e**↑ (through Matthew effect)<br>**u**↑ (an outcome is more worth when capital increases as well) |
| | Funding (EC) | Secondary Code ↑ |
| | Telescope (time) (EC) | Secondary Code ↑ |
| | Data (EC) | |
| | Skills & Expert knowledge (CC) | **→ k;**<br>Secondary Code ↑ |
| | Knowledge about norms (CC) | **→ k; → e** |
| | A good team & "working horses": Supervisors, PhDs, Postdocs (SoC) | CC↑; Collaborators↑ |
| | Collaborators (SoC) | EC ↑; CC↑ |
| | Reputation (SyC) | Secondary Code ↑ |
| | Indicator: Publications (SyC) | Secondary Code ↑ |
| | Indicator: Citation rates (SyC) | Secondary Code ↑ |
| | Indicator: Granted telescope time (SyC) | Secondary Code ↑ |
| | Indicator: External grants (SyC) | Secondary Code ↑ |
| | Indicator: Rewards (SyC) | Secondary Code ↑ |
| Institutional values (set by norms & culural frame)<br>(*sections 1.3 & 1.4*) | *via the endowment of meaning function* | **→ u** |
| | *via the orientation function* | Capital ↑ (norms specify how to acquire capital) |
| | Mertonian norms | Primary Code ↑ |
| | Publishability (no data/ code, no negative results), otherwise: Failure | Research Quality ↓ |
| | Evaluation system: indicators (SyC) define what counts as good research performance | Secondary Code ↑<br>Research Quality ↓ |

In summary, the analysis of the three *external components* of the astronomer's situation led to the identification of the necessary resources, rules and cultural goals they can acquire, need and typically commit to in order to perform research. We have seen that norms and cultural goals are related to ***Mertonian norms*** and intrinsic motivations. However, we have also seen that they diverge. The fact that capital is limited and that the publication infrastructure dictates what is publishable and what is not, gives rise to "tensions arising from the co-existence of (potentially) conflicting quality notions" (Langfeldt et al., 2020; p.115); namely those based on the intrinsic values of astronomers and those set by norms in academia. We therefore identify the following potential ***tension relationships***, an astronomer may be subject to.

### 1.5.2 Tension relationship: Competition versus Collaboration

Researchers may "struggle to win their specific means of production" (Bourdieu, 2004; p.57). According to Davoust & Schmadel (1991; p.10), "the increasing number of astronomers" "fierce competition" for limited resources. Also Trimble (1991; p.77) find that "good jobs really are harder to find than they used to be", which many of our established interviewees confirm (*see section 1.3.5*) and the early career astronomers we interviewed are aware of. At the same time, in *sections 1.2 & 1.3* we have learned how important collaboration is. Not only are collaborators a good source of capital, but also often a requirement for big science projects. Hence, like Heuritsch (2018) we find a tension relationship between whether to collaborate or to compete. Int-Faculty8 calls his collaborators not without reason "friendly competitors". Zuiderwijk & Spiers (2019) point out, astronomers value sharing data so that other astronomers can built upon them [also e.g. Int-Postdoc3, Int-Faculty7], while at the same time they may not share them when they find themselves in a race for priority with another team [also Int-Postdoc3]. Int-Faculty7 states that often the ***publication pressure*** doesn't allow for time to archive data properly. Further, by sharing data, astronomers may also make themselves vulnerable to others finding bugs in their reduction code, which would be bad for their reputation (secondary code ↓). Astronomers may reuse data from others, since that increases their performance (Secondary Code ↑; Zuiderwijk & Spiers, 2019). While astronomers may be scared to publish data, because of a potential loss of priority, Taubert (2019) points out that securing priority (Secondary Code ↑) is one of two reasons for astronomers to self-archive pre-prints as early as possible. The other reason is to benefit from feedback from peers to improve their publications (Research Quality ↑).

### 1.5.3 Tension relationship: Demonstrating usefulness versus Risky nature of research

The imperative of accountability in science requires commensurability (cf. Heuritsch, 2019a). Effort or how individual researchers perform cannot be monitored efficiently so assessment cannot be based upon it and therefore, a scientist is rewarded and funded for quantitative achievement instead (Rosenberg & Nelson, 1994). The same holds for usefulness of science. Not only is it difficult to measure usefulness, but there is also a tension between the ever increasing demand for societal relevance (e.g. Bouter, 2008) and the risky nature of basic research. Basic research is the human's endeavour to understand the unknown and as such it is by definition risky (Stephan, 2012). When societal relevance is measured in applicable outputs, and decisions must be made about distribution of researchers amongst research fields, economic pressure to produce such outputs can arise. That leads to a tension between demonstrating its usefulness to society, on the one hand, and not being able to guarantee that due to the risky nature of research, on the other hand. For research fields that perform mainly basic research, such as astronomy, this pressure to justify their societal relevance might be especially high. This is because results of basic research have a delay in leading to ground breaking technological developments or theories. Despite conventional wisdom, however, there are many spin offs[10], such as innovations in dental care and breast cancer detection (cf. Roy & Mountain, 2006) that derive from astronomical discoveries.

There is also the fact that not all research is publishable (see *section 1.3.3*). Data and reduction code as such don't count as a publication and negative results, despite valuable for an astronomer, are mostly not publishable. While risky projects may promise groundbreaking results, it also may not lead to publishable outcomes. As mentioned in *section 1.3.5*, publishability defines the landscape of success and failure. While astronomers define failed research as not having met their ***3 quality criteria***, institutional norms dictate that research is failed when there is no publication. Especially early career researchers may not want to risk the risky projects: "So of course, uh, people that are young now may be still more in the direction of- of the easiest way to- to survive" [Int-Faculty10].

In short, astronomers do strive for groundbreaking results, that have societal impact since their "knowledge forward pushing" effects are naturally bigger than those from incremental steps. However, they also value the latter (*see section 1.1*), despite the fact that reward structures often don't. Therefore, we observe a tension relationship between diverging opinions about what is useful and having to guarantee usefulness on the one hand, while on the other hand there is the need to perform risky projects, that promise groundbreaking research, but have a higher probability to fail in terms of not leading to a publication.

### 1.5.4 Tension relationship: Primary versus Secondary Code of science

As we infer from *Table 5*, both, the primary & secondary code of science (Taubert, 2019) are intrinsic drivers of an astronomer. "Truth" is the primary driver of astronomers to perform science and recognition the secondary code. Recognition is a form of symbolic capital (*section 1.2.4*) and defined through indicators (*section 1.3.5*). We have discussed that while indicators serve as proxies for science (*section 1.2.4*), they diverge from the observed that through defining what is publishable and therefore, the landscape of success and failure, institutional astronomers' notion of quality[11] (*section 1.1*). Therefore, we observe a third potential tension relationship – one of what counts – truth versus indicators.

As elaborated in *section 1.3.1*, institutional norms prescribe their own standards of quality. Similarly, Langfeldt et al. (2020; p.116) who study (co-existing) notions of research quality, "argue that following the advent of research policy, and associated imperatives for increasing levels of accountability and legitimacy, mechanisms for constituting research quality notions that were once reserved for highly professionalised knowledge communities have extended to encompass notions generated within policy and funding domains." The tension relationship at hand therefore is one between co-existing notions of quality; those held by the individual astronomer and those defined by indicators.

### 1.5.5 Publication Pressure & Publish-or-perish

How astronomers respond to the tensions described in this section will be discussed in *section 2.2*. We now turn towards another essential aspect of the logic of an astronomer's situation. The fact that publications are so important for one's career (see *sections 1.2.4 & 1.3.5*) leads to the perception of ***publication pressure***. Our findings agree with Heuritsch (2019a; p.169) that “publication pressure is always ‘at the back’ of an astronomer’s mind.” The author found three factors that may increase the omnipresent pressure due to the ***time-frame*** for the publication process: Publication pressure increases when 1) there is a race for priority, 2) PhDs or Postdoc need to publish during their temporary contract and 3) there is a telescope application deadline for which one needs prove that one can publish. Int-PhD2, for example explains that her field is not a popular one (i.e. not a sexy topic; see *section 1.3.3*), which means less competition and less pressure to be the first one to publish. At the same time she feels an omnipresent pressure to publish:

> “I constantly feel like I have to publish that.”, “So yeah, I really feel the pressure a lot and I hate it but it’s nothing you can do about it, you’re like in this system and you have to-.” [Int-PhD2]

[10] https://www.space.com/731-nasa-spin-offs-bringing-space-earth.html

[11] Cf. Langfeldt et al. (2020; p.131): "That was also when indicators first came into more widespread use and even in early attempts to analyse their emergence and application there was an awareness that they sometimes reflected a drift, or dilution, of research quality (van Raan et al. 1989)."

Publication pressure leads to the ***“publish or perish” imperative***.

> “The number of papers is often used as a criterion for evaluating candidates for a job or promotion and brings about a ‘publish or perish’ mentality among young scientists. It is often necessary to present a communication or a poster (that is eventually published) to obtain funds to attend a scientific meeting. Rapid publication of earlier observations may also be a prerequisite for obtaining more telescope time. These pressures to publish have been increasing simply because of the rising number of competing astronomers. A good bibliography is also helpful in the process of requesting a research grant, and in this case the pressure is not likely to recede, because competitive research in astronomy is increasingly expensive. Finally, the publishing activity of scientists is often used as a quantitative indicator for evaluating basic research.” (Davoust & Schmadel, 1991; p.11)

On the one hand, a good bibliometric record is necessary for astronomers not to “perish”. On the other hand, there is evidence that the “publish or perish” imperative might in turn affect the bibliometric record in random ways. Pluchino et al. (2018; p.2) who study how luck increases a person’s success more than talent, refer to a study from Ruocco et al. (2017) that finds “the distributions of bibliometric indicators collected by a scholar might be the result of chance and noise related to multiplicative phenomena connected to a publish or perish inflationary mechanism”.

# Chapter 2: Action Theory – astronomer's behaviour

The last chapter elaborated on the logic of an astronomers' situation in academic research. We derived the bridge hypotheses that link the objective situation of what material opportunities are present, what norms dictate behaviour and what cultural frame bonds actors, with the internal wishes and goals of an actor. The *logic of the situation* and its *bridge hypotheses* set the basis for what variables flow into an action. By setting the rules of the game and endowing behaviour with meaning, institutions structure expectations, interests and evaluations of actors. They direct behaviour by defining what is good or bad behaviour, what is right or wrong. The action theory we use is the *EU theory* (*Figure 3*), where the action that brings the highest utility (**u**; see *section 1.5*) will be chosen. In order to explain what effects indicator use in research evaluation has on research quality in astronomy we must first study their research behaviour. This chapter will delineate what behavioural contexts the logic of the situation shape, what motivations astronomers de facto follow in performing research, what conflicts they experience and how they respond to them.

We start this chapter by showing how the external components of the *logic of the situation* give an ***extrinsic motivation*** to perform research, in addition to the intrinsic motivation astronomers have due to their intrinsic drive to perform research out of curiosity. As Devoust & Schmadel (1991; p.11) point out, "there may be other motivations for publishing a paper, in addition to that of informing the community of new scientific results or ideas." Taubert (2019) distinguished between two motivational factors for an astronomer to assume the author role: 1) The extrinsic motive of acquiring symbolic capital through securing priority (Merton, 1973 [1957]; in Taubert, 2019). 2) and the intrinsic motive of disseminating new knowledge others can build on (***quality criterion 3***). Taubert (2019) related the extrinsic motive with the secondary code, recognition, and the intrinsic motive with the primary code, truth. Note that, as we pointed out in *section 1.1.3*, praise has been found to be positively correlated with intrinsic motivation as well, as when we are recognized for our work, our work appears more fun (Gagné et al. 2015). However, once recognition is needed as symbolic capital in order to move one's career forward, it becomes an extrinsic motivational factor as well. As we have seen in *section 1.3.5*, this is indeed the case.

> "Because it seems the only thing that is important is number of publications and so you have to do these if you want to continue." [Int-PhD1]

In *section 1.5* we discussed that the need to publish leads to publication pressure and Heuritsch (2019a) points out that this gives an extrinsic motivation to performing research. When the survival of an academic career depends on recognition through publications and citation rates ("publish or perish"), the secondary code recognition may only be defined through those indicators. In the case where indicators constitute what counts as recognition, a tension relationship between primary and secondary code may arise. This can be then translated into a tension between intrinsic and extrinsic motivation. Cronin (2001; p.559) writes that Darwin and "countless other scientists, both then and now, have experienced [the recurrent tension between intrinsic and extrinsic motivation] in the course of their careers". The discussion of how internal and external factors drive astronomers' research behaviour is part of this chapter.

*Figure 7* shows how institutional norms and the personal system shape behaviour: Through the *orientation function* of an institution the actor knows what is right or wrong in a situation and how to behave. Institutions provide *scripts* for certain situations, including what ***mode of action*** to choose. We have already encountered the choice of the mode in *Figure 2*. Most of the times, choosing the mode is not a conscious decision, but prescribed by the norms present in the *logic of the situation*. Only in situations where it is worth it to spend cognitive resources (steps 3-6 in *Figure 3*), for example when there is no script available to the actor, the actor goes into a reflected mode and performs steps 3-6 separately. This may also happen, when the actor disagrees with the norms or the prescribed scripts (as described in *section 1.3.1*). When we look at *Figure 7*, we remember that institutional norms also have a function that endows the situation with meaning; The mere idea of the existence of a legitimate orderliness defines the meaning of behaviour in the specific cultural frame (Esser, 2000c). Because meaning is introduced by institutions, meaning is socially constructed and by legitimating the institution, it functions as the internal anchor of an institution. Those actors who subjectively agree with the institution, its norms and what it represents, support the *internal anchor*. Actors who do not agree with that specific institution do so, because they were socialised differently or decided to follow different, often contradicting, norms. For those actors, an institution needs an *external anchor* in the form of enforcement mechanisms like sanctions, in order to make them follow the rules. However, when the absolute value of the gain from deviating from the norms is higher than that of the loss, actors will choose not to follow the norms. In other words, astronomers will deviate from the mode of normative behaviour, when that deviation has a higher utility.

## 2.1 Anomie

Esser (2000c) refers to the ***theory of anomie*** (Merton, 1949) as a useful concept to explain how actors adapt or deviate from the present structural conditions (i.e. the three components of the logic of the situation). "Anomie" occurs when there is a ***dissociation*** between how to legitimately meet personal or cultural goals and (possibly contradicting) institutional norms. This may be either because it isn't clear what institutionalised means (e.g. capital) are necessary to reach the goal or because the capital is limited in a way that (some) people simply don't have enough capital to meet the requirements set by the institutional norms. In either case, *anomie* weakens the internal anchor of institutional norms and deviant behaviour becomes more attractive. Merton defines five behavioural patterns one may adopt when there is an anomie:

1. **Conformity:** One focusses on those goals, which are attainable with the available capital and under compliance with the institutional norms. Decision making happens in the mode of normative behaviour and doesn't involve much reflection, leading to behaviour according to scripts.
2. **Innovation:** One uses non-institutionalised means in order to attain cultural goals, either because the required capital is lacking, or because institutionalised means are regarded as inefficient. This behaviour may be frowned upon at first, but later it may be acknowledged as innovative. This pattern requires the mode of reflection and deviation from scripts.
3. **Ritualism:** Bone-headed use of institutionalised means, ignoring possible negative consequences and neglecting cultural goals. This pattern displays a truly normative behaviour and doesn't involve any reflection whatsoever, but is a stubborn execution of scripts.
4. **Retreat:** Giving up in trying to meet cultural goals with institutionalised means that an actor (perceives themselves as) withdrawn from (e.g. drug addicts, dropouts from school/ societies, etc). This decision may happen through the mode of reflected, rational acting or affect-driven.

5. **Rebellion:** One rejects cultural goals and institutionalised means in favour of a new, socially not-approved system of goals and means. Like *retreat*, one may display this pattern through reflected and rational, or through affect-driven behaviour.

When looking at those patterns to examine which could be a useful concept to describe the research behaviour of astronomers, we can easily dismiss *ritualism* and *rebellion*. Due to astronomers' intrinsic motivation to perform science, there is little grounds for them to comply with norms, simply for the norms' sake. Early career researchers generally are not in the position to question the institutional norms in the sense of being able to discuss them with older colleagues in a meaningful way [e.g. Int-PhD1, Int-PhD2, Int-PhD3 & Int-Postdoc1] or to not comply with them if they want to survive in academia. Int-Faculty4 also admits that his supervisor was "too important for [his career] to use as a sparring partner, [so he] couldn't really argue with him" whether it was about content or norms. Established research may get away with non-compliance every now and then. To put it in Int-Faculty3's words, he keeps "those norms as broad as possible" and is willing to accept the consequences:

> "So, I always look at data, to me that's the key, on the other hand, frequently I find that my interpretation of data maybe different than other people but I (Coughing) try to give an honest interpretation but I always look for different ways of interpreting it. *Uhm in science just like in culture there are norms. And I'm quite comfortable in keeping those norms as broad as possible.*"

However, in general we do not observe any serious attempts of *rebellion*, since there is not much chance to rebel against performance indicators and surviving in academia at the same time. Which leads us to *retreat.* Retreat is in fact common in science – many astronomers (consider to) leave academia, not least because of their disagreement with institutional norms and/ or cultural goals. However, because we only interviewed those astronomers who remained in science, we could at most capture those who consider to leave academia. That is why our focus won't be on *Retreat*. Now we are left with *conformity* and *innovation*. Let's first have a closer look again, what normative behaviour would mean for an astronomer before we can determine if any of those two patterns fit (better).

As a general approximation, ***normative behaviour*** in astronomy includes sticking to ***Mertonian norms*** and committing to the cultural goals of conducting impactful science. One should write a minimum of one innovative paper a year, which peers thoroughly check for errors and legitimacy. Further, following a "golden child trajectory" (Heuritsch, 2019a; p.166), where one career step follows another smoothly, and having been at prestigious universities (e.g. Pluchino et al., 2018; Heuritsch, 2019a) looks good on paper. When using other people's research that was published in papers (unlike data and software), one has to cite it. Using other's research happens to build on their results in order to push knowledge forward. Collaborators are "friendly competitors" [Int-Faculty8] who support each other in that endeavour by complementing each other's capital.

However, as pointed out at the beginning of this chapter, there may be other motivations to publish a paper than pushing knowledge forward and having societal impact:

> "We are in a society that we have to publish our results from different reasons. Yeah, to be visible as a scientist but also to, uh, share your science." [Int-Postdoc1]

Because quality is difficult to measure, academia has adopted indicators that shall serve as proxies. As we will see in *Chapter 3*, the discrepancy between what astronomers would define as quality (***3 quality criteria***) and what is measured by indicators leads to an ***evaluation gap*** (cf. Heuritsch, 2019a). The need to serve those indicators gives an extrinsic motivation to publishing papers, citing colleagues and acquiring other forms of relevant capital, which may have negative effects on research quality (*Chapter 3*). Hence, astronomers find themselves in an anomie; they want to follow their intrinsic motivation to pursue science in order to push knowledge forward, while at the same time following their extrinsic motivation to comply with institutional norms. That is why Heuritsch (2019a) concluded that astronomers need to perform a *balance act* between both motivational drivers in order to survive in academia, while at the same time not compromising their values. While it was beyond the scope of the author to explicate on the balance act, we dedicate the next sub-section to it. This much can be revealed already: it seems like astronomers use institutionalised means in *innovative* ways in order to keep up the appearance of *Conformity*.

## 2.2 Balance Act

In order to perform the balance act between the intrinsic motivation to push knowledge forward and the extrinsic motivation to serve indicators to have a career in academia, we observe astronomers to deploy various strategies, which is also called "***gaming***". The fact that the need to be accountable can lead to gaming is not new (refer to e.g. Laudel & Gläser, 2014; Rushforth & De Rijcke, 2015), however we frame the "indicator game" (Fochler & De Rijcke, 2017) in terms of the RCT framework. The fact that resources[12] are limited and that control over them is assigned according to the rules of the institution, including what indicators count, leads to a *material bond* (Esser, 2000a) between astronomers. *Figure 8* illustrates such a material bond by depicting a situation where two actors have a certain interest and at the same time exercise as certain control on two resources. The presence of such a material bond implies that actors need to employ strategies in order to maintain or gain control over resources. The scarcer the resources, the more efficient need to be the strategies. Any kind of strategic behaviour can be describes in terms of *game theory* (Esser, 2000a).

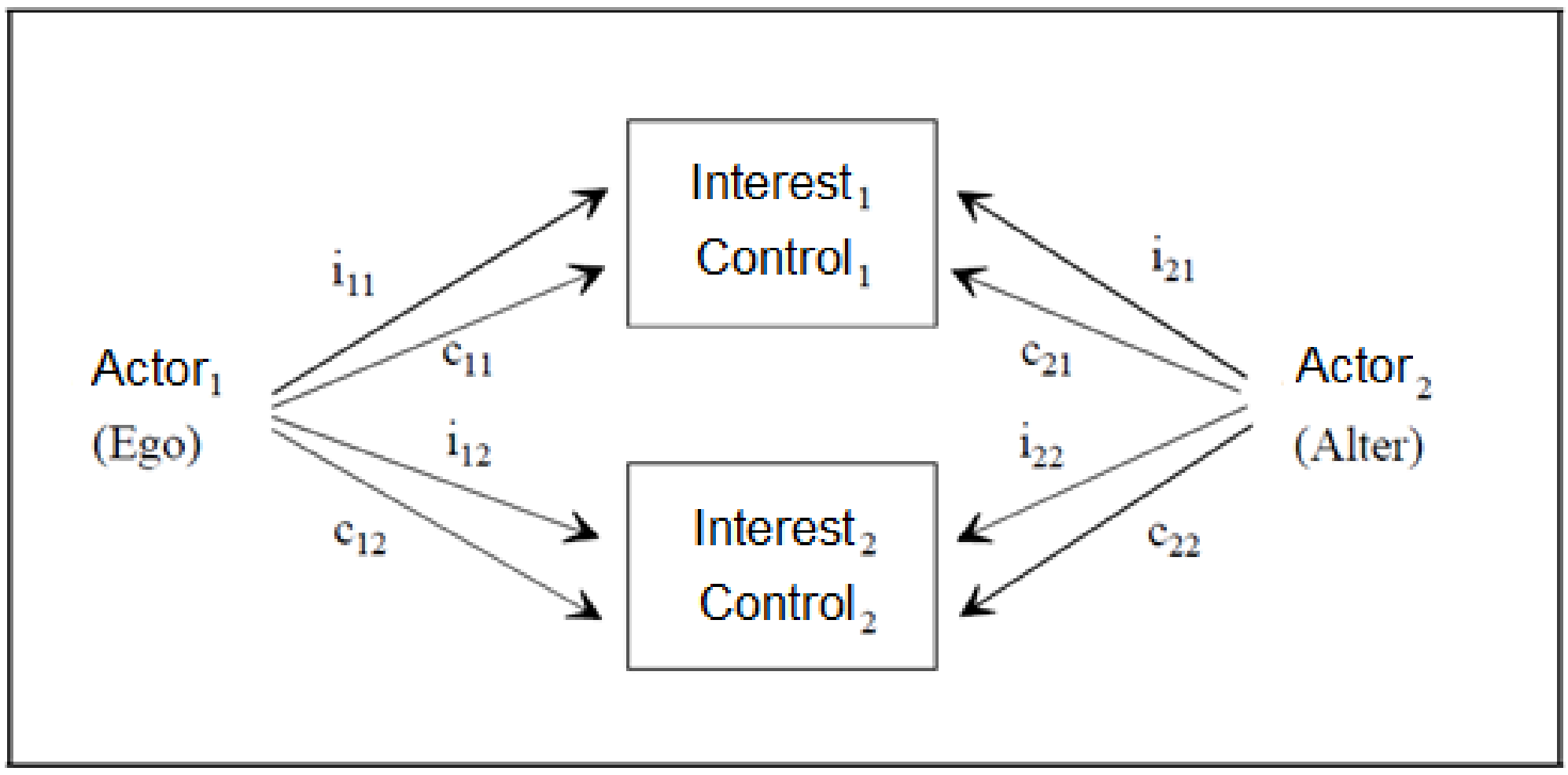

*Figure 8: "The system of a social situation" (modified by the author, based on Esser, 2000a)*

Thereby strategies are rules about actions, available for a player in a certain situation in a specific game. Strategies themselves, according to Esser (2000a) are not actions, but roadmaps that determine actions as re-actions towards the actions of another player. Here, gaming means to serve the indicators for the sake of increase in reputation (Secondary Code ↑) and survival in academia. Thereby, the roadmaps to play this indicator game need to keep up the appearance of *conformity*, but to stay competitive, institutionalised means are used in *innovative ways*. We will now elaborate on such strategies that we observed.

[12] Note here that Bourdieu's concept of capital and Esser's concept of "control of resources" refer essentially the same phenomenon: An actor has or has not access to various forms or capital (resources), such as money, time, social network and has incorporated certain knowledge and values.

Astronomers may not only be in competition for resources with each other, but also with other sciences. They need to convince politicians to spend money on a scientific field, whose aim is not produce any ***economic outputs*** (see *section 1.5.3*). That is why astronomy depends on outreach and popularization of research findings to gain popular support and on adroit lobbying to secure funding. As a consequence, outreach and lobbying have been well-developed in tandem and support each other. As Roy & Mountain (2006; p.20) write, astronomers' "scholarly activities generally prevent them from making fortunes, and they depend on generous donors, or the state, to build the observatories they need." According to the authors, astronomers "have developed unique skill sets" (i.e. strategies) which made them "surprisingly efficient at getting funding from wealthy benefactors and from government agencies". They conclude that "astronomers are salespersons when they push for a given project and deploy strategies to make their project the best in the field" (ibid. p.20). Roy & Mountain (2006) explain that astronomers' draw on four motivational factors to attract and justify funding for their big science projects, such as building a telescope: the quest for knowledge, the quest for achievement, the quest for survival, and the quest for power. The *quest for knowledge* is linked to the astronomers' curiosity and arguments related to this motivational factor draw on "our natural curiosity and a general desire to understand the world around us" (ibid. p.14). The *quest for achievement* is related to the astronomers' intrinsic drive to push knowledge forward: "A second motivation for funding research is to satisfy the need to achieve something big – to do something because it has not been done before, or doing it ten times better than before" (ibid. p.16). As Int-Faculty10 points out in *section 1.1.2.1* pushing knowledge forward helps humanity to survive and to improve the quality of life. When those arguments are not enough, astronomers push their case with arguments concerning the *quest of survival*. They point at examples for spin offs and the "relevance of their work to protect the Earth from the most threatening danger that can face our planet over long time scales" (ibid. p.17).

> "But astronomy really helps the day life.", "We have to invent the day by day devices that are more efficient, that have … , that they use less power." [Int-Faculty10]

Lastly, the *quest for power* is used as a political driver: astronomers "can invoke national pride to show that their communities and respective countries are among the best performers in the world, or are in need of new facilities to maintain their leadership" (ibid. p.18). In the political field, national astronomical communities have developed the ability to formulate consensus priorities and well-organized lobbies make them attractive to politicians and get them funded by the funding agencies. Roy & Mountain (2006; p.18) point out that "in this game, astronomy is at an advantage because […] the public holds a very favourable view of the discipline". Heidler (2017; p.24) writes: "Astronomy draws its support from the high media and public interest, which is based on its ability to produce suggestive pictures and address philosophical issues."

Int-Faculty10 explains that the "politically correct" answer to the question on how to secure funding is to demonstrate the usefulness of the scientific project to the taxpayer in terms of being "transformational". This argument draws on all four quests. However, Int-Faculty10's "politically incorrect" answer is that one needs to "lobby", especially if one needs funding for large instruments. Drawing on the four quests is not enough if resources are scarce. Having friends among politicians helps to convince them why the funding would be a good investment.

> "And usually the benefit is advance in science, uh. You know politicians understand very easy issues, how to build a road to go from Milan to Rome. But if you say I build a telescope to understand, uh, the black hole they do not really understand very well." [Int-Faculty10]

We already showed in *section 1.3.4* that instrument builders are at a disadvantage in receiving funding and citations as compared to those astronomers who merely observe. In order to keep up with the citations of those who merely conduct science and don't build, a scientific target needs to be in mind when the telescope is built. Referring to the target also helps in selling the telescope to funders. Other strategic considerations regarding funding are oriented around how to use it. Int-Faculty8, for example, explains that in New Zealand postdocs are so expensive that one could get five PhD students for one postdoc, which is why the university invests in favour of the former and the latter need to find their own funding.

We also observe gaming strategies when it comes to ***promotions and collaborations***. Int-Faculty2 points out that he "made a mistake" when he didn't search for other job offers in order to be promoted in his current position: "[It] doesn't matter how good you are. You can be very good if you don't have an offer from somewhere else, it's a much higher hurdle [to get promoted]." While the head of his institute said that he isn't "playing dirty enough", he refused to play along:

> "I always refused playing any dirty games in contrast to a lot of other people I know and never pushed anyone else down. I raised everyone who worked with me up."

Int-Faculty2 observed "pushing others down" in collaborations; sometimes leaders of collaborations, who are in a more established position that PhD students, put the blame on students when something goes wrong and at the same time praise themselves with the successes coming out of the collaboration.

Another strategy to get promoted is to work together with prestigious people:

> "Because now what is happening, […] people are trying to game the system. So for example, uh, I know of people who, uh, who work only to *please four or five prominent people in the field*. (I: Mm-hm.) who are their collaborators. So they just work like there's – they are faculty members but they work like postdocs of those guys. […] Uh, because they know that when the, uh, promotion case comes up, uh, they will the-they are going to be asked, uh, to write, uh, letters of recommendation and they're going to say very nice things about you because you've just worked to please them. So I've not done that and I think I have suffered, uh, in my career ( I: Mm-hm. Okay) uh, due to delay in my promotions." [Int-Faculty7]

Int-Faculty12 adds that students also decide for projects, depending on what is "more beneficial" for them, which may include working for more renowned scientists. The same goes for choosing collaboration partners. As we explored in *section 1.2.3*, collaborators are important to get access to resources, enhance one's research portfolio and to publish more papers than one could alone. Int-Faculty2 reports that he doesn't even perceive publication pressure, since publications are "just coming in" due to his collaborators. Those may not be first author papers, but they nevertheless boost one's publication rate. The negotiation of authorship is also strategic; while the lead author is listed at the top and others who contributed listed thereafter, people who didn't write any text, but who made the collaboration, observation or funding possible, are also among the authors [Int-Faculty8].

***Publications*** are one of the biggest grounds an astronomer can play the indicator game on. As Int-Faculty8 points out, the name of the game typically goes as follows: "Publish, publish a lot of papers and get, you know, a huge number of papers under my belt." One strategy to get there is to choose the research topic, such that one can get an "easy publication" [Int-Faculty1] out of it. This includes avoiding risky topics in favour of ***sexy ones***. Int-Postdoc2 reports that flashy topics get attention, especially if published in *Nature* or *Science*. Another strategy involves publishing immature results in order to get papers out more quickly (cf. Heuritsch, 2019a). Int-PhD1, Int-Faculty1 & Int-Faculty2 report that it is normal to publish results even if one is not totally confident about them:

> "Yes, for what I know everyone has to publish, the more, the better position in the short list. (Yes, I see okay.) And that's one aspect that I don't like from researchers. […] because we are maybe it's preferred to publish something that maybe you are not so very sure, or you are not so very confident about what you are saying but it seems that the most important thing is to publish not to do research." [Int-PhD1]

Yet another strategy to get more papers published is salami slicing (e.g. Broad*,* 1981; cf. Heuritsch, 2019a). Int-PhD1, Int-PhD3, Int-Faculty1, Int-Faculty2, Int-Faculty5 & Int-Faculty12 report to have cut up their papers into smaller portions. However, as pointed out by Heuritsch (2019a), astronomers do not only do that in order to get more papers out if their research (which is the definition of salami slicing), but also to increase the communication value (***3rd quality criterion***) of the publications. Int-Faculty2, for example, explains that brief and to-the-point publications are better readable and comprehendible than longer ones:

> "You should put it out promptly, briefly. – What other people call gaming tactics might actually be the way we should go."

Int-Faculty5 & Int-Faculty12 agree with Int-Faculty2's attitude by adding that often it is better to get the research out so that the community can give feedback and can build upon that knowledge. Int-Faculty12 points out however, that often it is hard to draw the line where to cut up the research and where to wait for more data from the telescope to put all the results together in one paper. Because the "barrier to actually going ahead and publishing something is not very high" [Int-Faculty12], Int-Faculty6 complains that there are too many papers containing little scientific value. We may conclude that when cutting up results increases the communication value of a publication, then it is appreciated by the community and it is not, when (immature) results are salami sliced. Int-PhD3 & Int-Faculty1 admit that they committed salami slicing. Int-PhD3 explains that his supervisor wanted him to do that too, so that Int-PhD3 and his Master student could both be first authors and that the latter would be able to apply for a fellowship. Int-Faculty1 explains that he wasn't proud of salami slicing, but that "there is no sainthood in being a scientist." Int-PhD2 observed others doing salami slicing and she finds it "ridiculous", so she wouldn't want to do that herself. We follow Heuritsch's (2019a) conclusion that astronomers approve the publication of immature or cut-up results when this increases the communication value of a publication, in terms of getting valuable feedback and being able to improve the paper or getting the results quicker to the community so that others can build on the published results. Salami slicing, where publication's are cut up solely for the sake of increasing one's publication rate is not approved when that decreases the ***quality*** of the paper.
As pointed out in *section 1.5.5*, Heuritsch (2019a) found that there are ***3 time-frames*** that set the boundary conditions for the publication process: 1) the race for priority, 2) the time-frame of a PhD's or Postdoc's temporary contract and 3) a telescope application deadline. Int-Faculty12 explains that one always needs to find a "sweet spot" between results to be "sufficiently interesting" and the "schedule pressure on the student side". Since results can always be improved, one needs a "judgement call" about such a decision, which is "driven by the job-cycle": "So there's certain times of the year and certain times in a student's career where they need to be more, um, uh, time sensitive about getting the results out". Int-Faculty7, Int-Faculty8 & Int-Faculty10 also emphasise that it's important to make sure that PhDs get enough first author publications during their temporary contract.

An astronomer may not only want to boost their publication rate, but also their citation rate. Int-Faculty6 explains that "the best paper to write is one that has an error in it that's not completely obvious, because then everyone will catch it and will feel obliged to point it out and they have to cite you." Int-Postdoc1 complains that a referee may reject a paper, when the referee works on a similar subject, and doesn't want the author to claim priority.

All of the above are examples for gaming strategies that astronomers employ in order to enhance their career chances. They do so out of the extrinsic motivation to accomplish what bibliometric indicators measure, even though they perceive them as poor measures of quality (see *section 1.3.5*), in order to stay in academia to perform research. Performing research out of curiosity drives their intrinsic motivation. ***The balance act*** then is the art of playing the indicator game, while at the same time compromising ***research quality*** as little as possible.

The balance act comes in many facets. One facet is that the means may justify the ends. Int-Faculty1 & Int-Faculty8, for example, explain that, especially at the beginning of the career, one may had to perform assigned and not freely chosen research, to then find collaborators that do research that is interesting to you. You may also serve the indicators for a while until you are established enough to be freer in your research.

> "Sometimes you have to be opportunistic. Sometimes you have to go for easy publications that you can claim that you are successful, you are good, then you'll get funding. And then you will finally have time to do what you want to do." [Int-Faculty1]

> "So with this comfort buffer [i.e. having many collaborators], the publications come out anyway. They tick all those boxes for the non-important people. So I can focus on what I want, what I think is important. So I spent time on important things. My publication data is over 5,000 total citations and h-index of 36, – which isn't that bad. So you tick the boxes for the stupid people, – who don't understand what the meaning of these numbers is, but they just look good on paper." [Int-Faculty2]

Int-Faculty2 emphasises that he "prepared" for the interview by looking up his bibliometric record. He did that, despite asserting how "stupid" all these metrics are, which shows that he is aware that importance is attributed to such records nevertheless. Int-Faculty3 agrees that a good publication record is relevant, that "certain compromises" are necessary, but that one doesn't need to obey to the evaluation system completely.

> "Uhm […] I wouldn't conclude that you have to in order to succeed, succumb to uhm such things as publish-or-perish. It's relevant but it's just as relevant you know, do

> you like to drive, well you still have to obey the speed limit that maybe a nuisance or you know stopping in a red light even though no one's in the intersection I mean there's certain compromises." [Int-Faculty3]

An important aspect of this study is what these "compromises" (i.e. the balance act) entail and what impact they have on research quality (*Chapter 3*). In *section 1.5*, we have discussed several tension relationships that may result from the *logic of the situation*: 1) Collaboration versus competition; 2) Guaranteeing usefulness versus risky projects; 3) Primary versus secondary code. Part of the balance act of an astronomer is to find ways to deal with those tensions. Often, when humans find themselves stuck between opposing (whether intrinsic or extrinsic) values, expectations or contradicting pieces of information, they tend to harmonise those knowledge- and value systems. In such a case one struggles with the so-called ***cognitive dissonance (***cf. Esser, 1999 & 2000c), a psychological concept that is closely related to the sociological concept of ***anomie*** (see *sub-section above*). While an anomie is a dissociation between the cultural goals and the institutionalised means, a cognitive dissonance may occur when an actor experiences any clash of pieces of information or values – whether normative or not. There are three ways to resolve cognitive dissonance: denial of the incoming information that contradicts available knowledge; the re-valuation of one preference in favour for the opposing one; finding arguments for why pursuing one of the preferences will in the long-term also benefit the pursuit of the other. All of those are aspects of rationalising one's situation, which is to find a rational justification for one's behaviour through highlighting some of the arguments for the behaviour at the cost of the arguments against the behaviour. As Esser points out, what alternative one chooses is an act like any other that happens according to the *EU theory*. In the case of the astronomer, cognitive dissonances arise due to diverging intrinsic values (see *section 1.1*) from what bibliometric indicators measure *(*see *section 1.3.5*), which leads to the above mentioned tension relationships. There is an anomie (see *section 2.1*), due to the ***dissociation*** between how to meet the ***3 quality criteria***, while at the same time serving the bibliometric indicators in a competitive way.

In addition to the three tension relationships, we would like to point out ***two more cognitive dissonances*** we observed. First, it seems that, despite the ***publication pressure*** most interviewees feel quite free in their research. Int-PhD1, Int-PhD2, Int-Postdoc2 & Int-Faculty4 report that during their PhD they feel/felt free to experiment, learn and to pursue their own ideas. At the same time, interviewees mentioned that they were free "as long as publications come out" [Int-Faculty2]. Int-Faculty1 makes a point in not treating his PhDs as "slaves" and "broadcasting a free environment", but putting very much emphasis on "getting into the publishing business" because "this publish-or-perish thing is a real thing":

> "I mean, when they got their PhD, they are among the wild beasts of science. And to be successful they have to have more. I mean, fulfilling the minimum requirement never helps you getting into the real business. You have to be an over-performer, to be honest. So, that's my advice to my students." [Int-Faculty1]

Int-PhD3 reports that he feels like "living from pressurizing moment to pressurizing moment". He feels restricted in his research, because he doesn't feel like he has the time for quality research:

> "It is paper based, so it is quantity based and not quality based.", "Then you have to do it quickly in that way, like everybody does and that is the problem."

Quality of papers is the second cognitive dissonance we would like to point out here. During the interviewees, papers were mentioned as “important” or “impactful” when they received many citations or landed a high JIF. However, when asked what their highest quality paper is, many interviewees mentioned that they would nominate one that didn’t necessarily receive the highest amount of citations or JIF [e.g. Int-Postdoc1, Int-Postdoc3, Int-Faculty5, Int-Faculty8 & Int-Faculty12], but one that instead had impact on the public or when something really had an impact in the sense of pushing knowledge forward.

> “And then so, um-, even though, even though it does-, it doesn't have a particularly high citation, um, at all, uh, it doesn't matter. I reckon-, I look at that and go, that was something which, you know, I came up with myself. And uh, I had that, I had that, that, that fire within to, to pursue it. And that certainly happened about two or three other times, um, in the papers, um, where you've got that real sense of, no, this is something new and exciting and I'm learning stuff. And it's not another run-of-the mill paper, which is good and publishable material, but it's just another incremental advance.” [Int-Faculty8]

> “I mean, I personally know that many researchers do not consider their most cited paper to be their best paper. I myself fall in the same loop. My best paper doesn't have the largest number of citations. And funding agencies don't care about lowly-cited papers. So, there is no clear answer how much, do they support this kind of high-quality research.” [Int-Faculty1]

Int-PhD2 reports that she was on a selection committee for a professorship, where the committee members affirmed that it’s important to look at the quality of papers, but at the end of the day, they always came back to their person's citation and publication rates.

> “So yeah they were always like saying, yeah okay we're not going to focus on publications only because it’s not a qualitative but only quantitative, that’s not the most important thing but in the end it was.” [Int-PhD2]

We conclude that what counts according to performance indicators is so incorporated that in the astronomer’s mind that there are two conflicting definitions of quality present: one where quality is defined through their ***3 quality criteria*** and one where quality is equated with what indicators measure. Furthermore, astronomers may downplay the feeling of publication pressure as something inherent to the system, to be able to feel less restricted in their research process.

Let us come back to the Balance Act. As pointed out above, gaming strategies shall give the appearance of *Compliance*, while institutionalised means how to achieve a good bibliometric record are used in *innovative* ways. **They best fit the third way of how to resolve a cognitive dissonance: by favouring the pursuit of the secondary code (reputation) in the short-term, one hopes to get into a position where one can focus on the primary code in the long-term. The balance act of an astronomer then can be described as the act of choosing how much of the value “primary code” can be compromised in favour of the value “secondary code”.** When facing choices between “high risk, high gain” versus “guaranteed publication”, “sharing data” versus “securing priority”, “increasing the quality of the paper” versus “publishing as quick as possible”, an astronomer will decide according to the *EU-theory*; how much compromising of the primary code is it worth it order to have a gain in the secondary code.

# Chapter 3: Resulting collective phenomenon

The last step of the sociological explanation of how new macro-phenomena come into existence (*see Figure 1*) is to formulate *transformation rules*, which describe how the actions of many individual actors constitute a *collective phenomenon*. In our case – academical astronomy – institutional norms, build the basis for the transformation rules, since they structure research behaviour. It is the very purpose of institutional norms to be followed without questioning in order to guarantee smooth interactions between actors (via ***the 3 functions***) and for the actor to reduce the cognitive effort decision making would otherwise cost (*Figure 3*). Through institutional norms collective phenomena come into being, simply because they prescribe certain social processes that follow from an individual's *situation*. The transformation rule then amounts to assuming regularity of those empirical social processes, which are treated as a logical consequence from institutional norms (Esser, 1999).

Indeed we observed, that astronomers generally comply with institutional norms (see *Chapter 2*). They find it "completely normal" [Int-Faculty6] to be assessed by quantitative indicators (see *section 1.3.5*) and indeed, especially early career researchers are not in the position to discuss evaluation procedures with older colleagues (see *Chapter 2*). Given that astronomers are realists[13], who "like to stress their status as passive observers: they only gather information that can be received from the universe" (Anderl, 2015; p.11), it doesn't come naturally for them to reflect upon the performativity of indicators. Those who do, wouldn't know how to have a better evaluation system (e.g. Int-PhD2, Int-PhD3 & Int-Postdoc3) or give up in pondering about that, since they feel like they can't change anything.

> "And so even if you, fix it [i.e. introducing software citation] by uh-, encouraging citation counting, they haven't fixed the attitudes of the people who are doing the evaluation. The people who are gatekeeping" [Int-Journal]

Rather, we observed that astronomers accept the institutional norms, such as being evaluated by performance indicators, as part of "the system" (Heuritsch, 2019a). Indicators need to be served to climb up the career ladder, in order to do research.

> "Okay. I mean, in the publishing process, um, mostly it's, uh, it's fine. You get used to what, you, you get used to what's expected." [Int-Faculty12]

> "I'm sure that [sacrificing quality for quantity] happens a lot. Which is a bad thing as we all know, but since this is the world we are living in and this is still the system, although they were always saying like no we're not going to look to number of papers, but they always did in the end so. Yeah we're far from alternative ways of choosing people so." [Int-PhD2]

However, because of the discrepancy between what astronomers define as quality research, and what indicators measure, astronomers need to perform a balance act between serving the indicators, while at the same time compromising research quality as little as possible. As we discussed in *Chapter 2*, this includes various gaming strategies and results in short-term favouring the secondary code at the cost the primary code.

[13] "I mean so that's how you do good science you just look at your data. […] What the data are telling you. " [Int-Faculty2]

## 3.1 When the secondary code becomes the primary code

Astronomers affirm that they would do their best ***not to compromise research quality***, despite the ***publication pressure***. Int-Faculty7 reports that he never faced a situation that diverged from his notion of quality. He was always in control and lucky enough that he worked with professors for whom the number of publications was not too important. With his students, he wants them to take full responsibility, but only checks for obvious mistakes. He also doesn't push the students into a certain direction – he communicates "the ethics of the problem [that quality is important], not so much the astronomical skill". Int-Faculty2 reports he only published when he is happy with the results, even when his collaborators push for a publication earlier. Int-PhD3 explains that he didn't publish a paper in his first PhD year, since he likes being thorough and precise and eventually got his paper accepted "as is". Int-Faculty6 asserts that the standard way of doing science is to generate double-checked and reproducible research and that if people in his collaboration were to compromise quality, he would rather take his name off the paper.

> "But I want more than an apparent high quality paper. I want high quality research, high quality data.", "Whereas I know at the back of my mind, for this data point I had to do something inconsistently with something other and okay even though the paper would appear high quality-. That should not be the goal, right? (I: Right.) The goal should be that our research is high quality." [Int-PhD3]
>
> "The best quality I can. It's not high quality but the best as I can. Okay everything is right [i.e. there are no errors]." [Int-PhD1]
>
> "So, I fight hard to have quality over quantity as much as possible, but I often fail because of the pressure." [Int-Faculty1]

Int-PhD2 also reports that she would only want to publish, once that she is happy with the quality of the paper, but that she feels pressured by her supervisor to publish quicker. Hence, despite the good intentions, we do observe a decrease in research quality due to the need to fulfil quantitative goals. As pointed out in *section 1.5*, while methodological standards and the publication infrastructure, with its reviewing system and publication of replicable results, shall support the orientation towards the primary code of truth finding, the evaluation of astronomers through quantitative indicators leads to an orientation towards the secondary code of recognition. As Taubert (2019) points out, advancing one's career opportunities (through secondary code ↑) is a goal that could motivate the publication of research with ***lower quality***. As Davoust & Schmadel (1991) write, the "competition [for resources] is bound to have a strong effect on the publishing activity of astronomers."

Int-PhD2 & Int-Postdoc1 report that truly bad papers wouldn't get accepted and that handing in such a paper would give a bad impression (secondary code ↓). But if it is "okay" [Int-Postdoc1], it usually does get accepted, even if it contains some ***mistakes*** due to the "rush" [Int-PhD1]. Taubert (2019) writes that to ensure priority as early as possible a higher risk for containing errors is accepted for publication in ArXiv. Int-Faculty2 believes that "the publication system completely fails sort of all the time", because the majority of papers shouldn't get accepted – his rejection rate as a reviewer is 2-3 times larger than usual. Next to the potential of containing errors, Int-PhD1 & Int-Postdoc1 explain that there are many articles, that aren't clear in their message (3rd quality criterion ↓).

> "Sometimes it was just ... maybe at the beginning that when I started reading papers that-, there's like nothing in this paper, or why do they publish it? But on the other hand, there are many papers like that. But-, but-, but I think it's-, well, saying that it's fine, or not fine, it's quite complicated because uh, our system of publishing is complicated but that's how it is." [Int-Postdoc1]

The focus on ***sexy topics*** may be another reason for a decrease in quality. As we pointed out in *section 1.3.3*, flashy topics attract more attention, especially if published in journals, like *Nature* or *Science*, which have a higher JIF than the three main journals in astronomy. However, those journals provide less good data retrieval services [Int-Journal]. Moreover, Int-Faculty12 reports that in such high-impact journals, often high quality papers are turned down in favour of papers on sexy topics, which may even contain mistakes.

> "They take these [flashy] results but not the good things. […] So that's the obsession with journal impact factors. We all know how little journal impact factors mean." [Int-Faculty2]

The peer review process is meant to find mistakes and help improve the quality of the paper, but as Int-Faculty9 explains, reviewers are drowned in other tasks, such as publishing, themselves and do not have time to "do the work for the authors". Several interviewees [e.g. Int-PhD1, Int-PhD2, Int-PhD3, Int-Postdoc2, Int-Postdoc3, Int-Faculty2 & Int-Faculty7] report that the decrease in research quality is a result of the ***publication pressure***:

> "There are works that are very important and are published very quickly, and perhaps they are not quality works. There is a lot of publishing pressure. And sometimes the quality goes down because we have to publish like maniacs and because, you know, there is no written standard. But we all know that there must be standards.", "This person has to publish quickly because maybe he or she needs to get more papers out to get a position. So, the quality goes down." [Int-Postdoc2]

As pointed out in *section 1.3.4*, there are no external incentives to publish reduction code or data and due to the publication pressure, the intrinsic motivation to do so is often not sufficient. Zuiderwijk & Spiers (2019; p.233) write that "in the academic system publications are valued more than datasets […]. This may lead to lower quality research". Int-PhD2 confirms that, in her opinion, the publication system doesn't encourage good quality research, because:

> "No journal requires these things like they don't require your codes, they don't require that you actually give them step by step what you are doing. So I think as long as that doesn't change as well so no change."

Int-PhD3 explains that it's difficult to know whether a paper is based on good quality data or an error-free code, when their publication is not encouraged. As a reader you don't know how well check-through all the material that went into the paper is and it is difficult to ***replicate*** the analysis [e.g. Int-PhD1, Int-PhD2, Int-PhD3, Int-Faculty1 & Int-Faculty2]. Errors may propagate (Int-PhD3; cf. Heuritsch, 2019a), and so when one can't check the analysis step-by-step, one can't judge the quality of the analysis. The main reason for low replicability is, again, the ***publication pressure***. Int-Faculty1 guesses that 50% of the papers are not replicable and that replicability a criterion for him as a referee to accept the paper. Int-PhD2 explains that papers generally aren't transparently written and even when one writes an email to the authors for clarification, they often can't answer the questions sufficiently anymore to be able to replicate the study: “So you cannot trust any, even the published papers apparently.” Patat et al. (2017; p.57) refer to the “significant numbers of cases in which negative or inconclusive results do not turn into publications” (see *section 1.3.3*) and regard those as “symptoms of workload pressure in the community”. The authors conclude: “This reflects what may be a growing cultural problem in the community as scientists tend to concentrate on appealing results, especially if they have limited resources, and need to focus predominantly on projects that promise to increase their visibility.”

The ***publish-or-perish imperative***, does not only lead to lower quality papers, but also to an ***information overload*** (Taubert, 2019). Int-Postdoc2 explains that, due to the quantity of papers nowadays, it is difficult to sift through them and decide which ones to use.

> “The only thing I guess the one negative aspect is the, uh, pressure on people to publish. (I: Uh-huh.) And-and I think it's not very healthy because we're getting a lot more, uh, lower quality papers and it becomes hard, uh, harder because you have to read all those papers to find relevant information. (I: Yeah.) Uh, so probably the pressure on publishing is not a good thing.” [Int-Postdoc3]

Herrmann (1986; p.188) writes that the total publication rate showed a particularly steep increase after WW2, but it exhibited a “steady growth tendency” throughout the $20^{th}$ century. Publication pressure is only one of the reasons for increased productivity. Another could be, as pointed out in *section 1.3.3,* that papers are easier to write and publish than in the past, given that there is more available data and that there are more collaborations (Davoust & Schmadel, 1991). Increased availability of data, however, also decreases replicability, since astronomers have less time to replicate each other’s results the more papers and data are available [Int-Faculty8]. Davoust & Schmadel (1991) mention a third reason for the increased total publishing activity of astronomers, which is the increasing absolute number of astronomers. According to the authors, an individual’s relative publication rate may increase with age due to increased competence, more co-authors, and the ***Matthew Effect***.

The fact that the Matthew Effect and luck play a big role in acquiring capital (see *section 1.2.5*) is also a reason why astronomers often ***perceive funding decisions as unfair, biased and not very objective*** [e.g. Int-PhD3, , Int-Postdoc1 & Int-Postdoc2]. For that reason, Int-Postdoc1 explains that one has to learn not to take funding decisions personally. Int-Faculty9 understands why some would find funding decisions unfair, however she points out that the same people who may complain about them, may also serve on a panel and make the exact same decisions. Because allocation of funding also needs to be based on how important the research is for the taxpayer, and whether it fits with strategic considerations with regards to the decadal survey or other national interests, the process may be perceived as biased. Furthermore, as we discussed in *section 1.2.2*, it’s not easy to receive funding when one wants switch subfield, which may lead to ***overspecialisation***.

> “Um, and think, I think it leads to a lot of stove-piping in the field a lot of, a lot of situations where, where all people can do- To stay competitive, all people can do is to continue to dig down deeper in their hole, in their, in their discipline. (I: Okay.) Um, like they have to keep proposing for the thing they’re most expert on, (I: Right.) instead of branching out into other areas. (I: Okay.) Like branching out has to happen on sort of other, uh, you know, side side projects or other funding. It’s very, it can be really difficult to get enough of a name for yourself in that area […] That’s not a good thing. (I: Okay.) That’s a- (I: Yeah.) Yeah, and that’s, that’s a disadvantage of the field like that. Um, I don’t think at this point it’s pretty diversified.” [Int-Faculty12]

In relation to the astronomer's need to be accountable and prove productivity, Fernández (1998; p.62) talks about the "***taylorization*** of scientific work", which is "one of the features of modern science":

> "The scientists becomes a link in a production line (the research group), where his or her level of productivity is measured by certain parameters (number of scientific publications, involvement in research projects, invited talks, citations, etc.)."

Performance indicators define what counts as good research, and what is not measured is not perceived as worth someone's time. For example, fewer astronomers are willing to do something for the common good if it isn't directly measured by some metric [Int-Faculty6]. At the same time, collaborations and networking has become increasingly important during past decades, because they serve as trade markets (cf. Heuritsch, 2019a) for all kinds of capital (*see section 1.2.3*).

Not surprisingly, the pressure to publish and the need to comply with institutional norms, may lead to ***stress***. Many interviewees [e.g. Int-PhD2, Int-Postdoc2, Int-Faculty7 & Int-Faculty8] report that job uncertainty (cf. Waaijer et al., 2017) causes them emotional stress. Int-Faculty8 even suffered from depression, because he felt no permanent position was in sight and Int-Faculty7 describes being rejected for postdoc positions as the "most difficult phase" in his astronomy career. Int-PhD2 claims that this uncertainty is the biggest problem for her in academia – even bigger than publication pressure. Like Int-Postdoc2, she would like to reach tenure, but isn't sure whether she can accept moving to a different country for every postdoc position. Int-Faculty9 reports that the required flexibility to move and working on weekends can be difficult on relationships:

> "But-but it-it does seem like I do have to put in extra time, uh, as a lot of the scientists do. (I: Mm-hm.) To get done what we need to and-and to stay competitive. […] And it could be difficult on relationships because like, uh, I- because I know for my husband he's used to just working a normal workday, and so – (I: Mm-hm.) Whereas for ss- for scientists it's not just a job, it's kind of a lifestyle."

She is aware that she could have advanced her career faster without having kids, but she is happy about the choice she made and accepted that sometimes one has to decide between a happy family life and a good publication-/ citation rate. Int-Faculty8 describes having "heard some horrible stories of how people just make terrible sacrifices" to get tenure:

> "Uh, I mean, I've had several friends of mine, uh, going to the States and, yeah, then go for tenure, but not succeeding. And, um, it's just, doesn't sound like a very pleasant to them, I'll say so."

At the beginning of her PhD, Int-PhD2 feared losing priority, but now she is aware that her topic isn't "sexy" enough for that fear to be justified. If she were working on a sexy topic, she would feel the pressure from the competition, however. Int-Postdoc2 explains that she feels that constant fear of not being good enough, resulting from competition for resources, even though she knows that their allocation also depend on luck. Int-PhD1 reports that he doesn't "feel comfortable in this system". Int-Faculty5 explains out of personal experience that for early career researchers it is difficult not to take negative referee reports personally, because negative feedback often is perceived as failure and that is connected to fear [Int-Faculty11] because every publication counts.

> "So, you really didn't like what you heard and you feel, you, you feel personally attacked, you know. And it's, it's kind of, um, it takes a while to get over that. And it certainly as a, as a young researcher, um, when you see your first referee report of that type, it can be extraordinarily, uh, dispiriting and you need to have a good supervisor who says 'Don't worry, this is quite common. Um, referees will, will, will often, uh, uh, attack, um, weak points, but that's part of the scientific process.' […] Um, and uh, this could have quite a negative effect on people who are trying to start off in the field and who go like, 'Well, he hates me. I'm useless.'" [Int-Faculty8]

While established researchers [e.g. Int-Faculty2] have the opinion that early career researchers shall make their own mistakes to grow and carve out their own niche, this sort of freedom is hardly supported by the publish-or-perish culture. Int-Faculty10 reports that it is "a waste of resources" that the publication pressure cuts the ground under young researcher's free minds. According to Ruocco et al. (2017; in Pluchino et al., 2018; p.2), "innovative ideas are the results of a random walk in our brain network" and publication pressure may supress that kind of creativity. As a result, the pursuit of risky and innovative ideas is not encouraged, even though that is the science needed by society (Thurner et al., 2020).

## 3.2 The Evaluation Gap in astronomy

When looking at the effects that indicator use in evaluation procedures has on research behaviour and knowledge production in astronomy, we conclude that the initial discrepancy between what astronomers define as good research quality, as opposed to what is measured by indicators, which is present in the *logic of the situation*, **produces an *evaluation gap* at the institutional level**. Furthermore, indicators constitute what counts, and individual actions constitute the *collective phenomenon* of an evaluation gap and therefore we can indeed claim that indicator use have ***constitutive effects*** (Dahler-Larsen, 2014) on the knowledge production process in astronomy. At the same time, the evaluation gap is present since the need to serve indicators only alters the priority of the secondary code with respect to the primary code and not the content of the primary code itself (cf. Heuritsch, 2019a). The *transformation rules* that lead to the evaluation gap hence comprise of the need to perform a balance act between primary code versus secondary code (*section 2.2*). As we pointed out in *Chapter 2*, when most actors agree with the norms at play in an institution, transformation rules simply assume a compliant behaviour of the actors. However, as we have shown in *Chapter 2*, in astronomy behaviour is *conform* and *innovative* at the same time. Astronomers use gaming strategies (*section 2.2*); to remain competitive, institutionalised means are used in *innovative ways*, such as salami slicing or going for easy publications, which allows them to prove their performance on paper. In conclusion, we get the following picture (*Figure 9*): The intention of decision makers to use performance indicators is for researchers to be accountable and produce good quality research ("Good quality research & performance" in *Figure 9*). Thereby decision makers assume easy transformation rules, where the institutionalised performance indicators lead to compliant behaviour. **However, indicators quantify quality[14], and thereby transform quality into capital that counts**. Capital can be targeted, by gaming strategies (*section 2.2*). **Hence, gaming targets the measure**. Goodhart's law – "when a measure becomes a target, it ceases to be a good measure" – then is the *first transformation rule* we observe. Further, the discrepancy between the astronomers' intrinsic and extrinsic motivation makes them perform a balance act (*section 2.2*), primary and secondary code are in competition with one another. Astronomers solve this tension relationship (i.e. cognitive dissonance) by rationalising the pursuit of the secondary code at the short-term cost of the primary code, in order to focus on the primary code in the long-term. The art of the balance act is to choose how much of the value "primary code" can be compromised in favour of the value "secondary code". The balance act is the *second transformation rule*. Finally, the *third transformation rule* results from compromising the primary code, which corresponds to a compromise of research quality for the sake of quantity. The resulting *collective phenomenon* is information overload and sacrificed research quality. On a phenomenological level, we hence observe constitutive effects on what counts as good research (see *Figure 9*) and an evaluation gap between the intended ideal and the resulting collective phenomenon.

---

[14] "Yeah so we're back to the thing. Is this the true academic quality? I mean the true academy hasn't change. Quality is quality. Or what people would call quality. Or people, mistake quantity for quality. Any metric measures quantity, it doesn't measure quality." [Int-Faculty2]

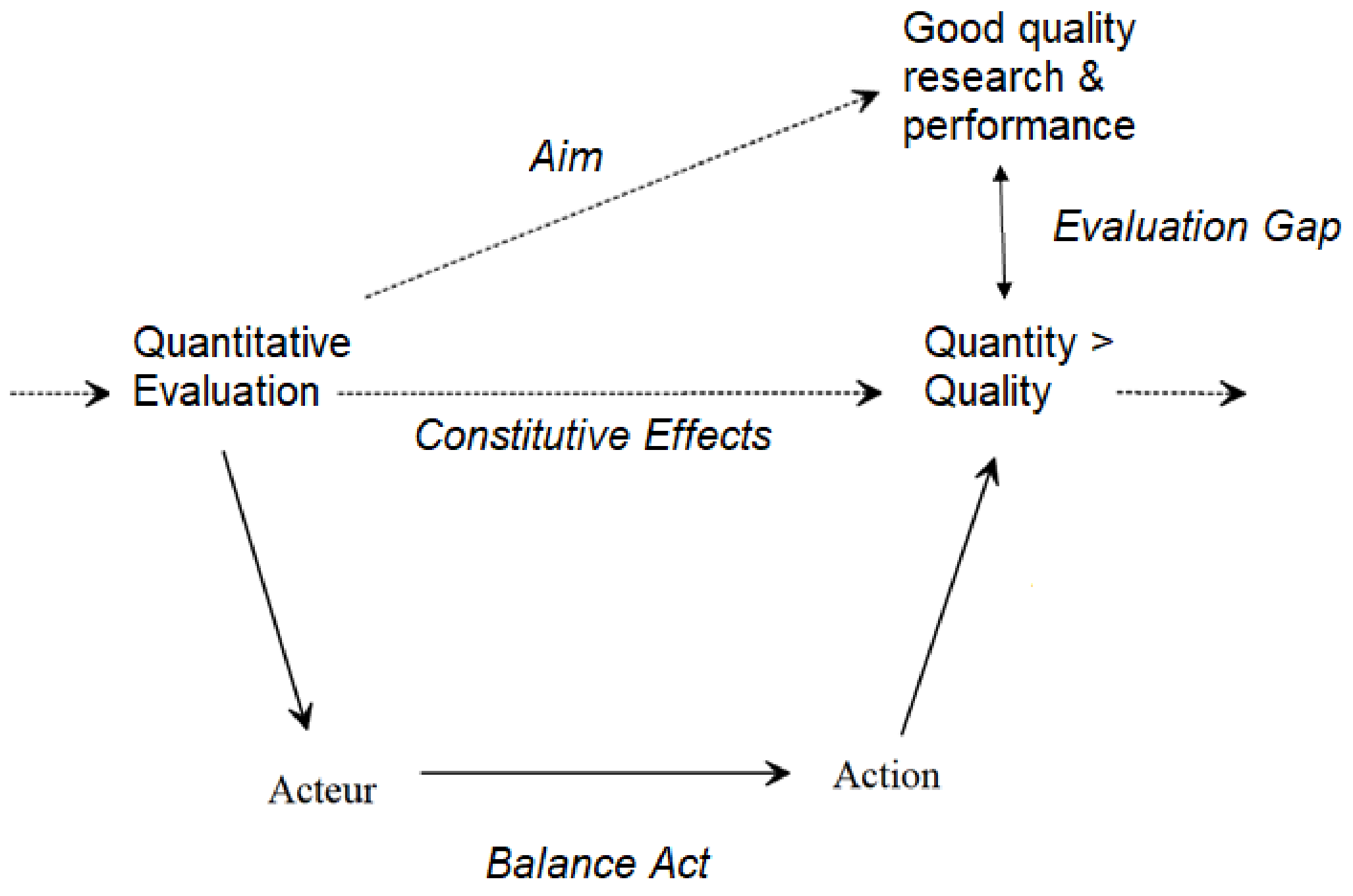

*Figure 9: The Evaluation Gap in astronomy as the resulting collective phenomenon based on a predominant evaluation through performance indicators.*

As a summary, the evaluation gap is characterised through avoiding risky topics in favour of sexy ones; i.e. favouring safe publications over potentially innovative outcomes. That is despite the fact that astronomers point at the fact that in science outcomes cannot be predicted (cf. Heuritsch, 2019b). Further, due to publication pressure and a lack of external incentives to publish data and reduction code in a transparent way, replicability of papers is sacrificed. Immature publications and salami slicing may further add to a decrease in research quality.

> "If I can predict the outcome, it's not science. And several times I've made the right choice to take some risk and just to explore. That's how you find new things." [Int-Faculty2]

> "I basically want to do research where people can base-. There's this saying like, 'I was standing on the shoulders of (I: Giants.) giants.' Basically that's what I want. I hope research would be that you could easily build on previous works. Where you can-. At the moment people just trying to do the same stuff because other people haven't made available their way of methods or codes, and they have to make it up all over again. And then you're only incrementally enhancing the field, and you should really be- […] And to make something even bigger which later on, in ten years becomes yet again, becomes trivial, and you can build something more. I'm basically seeing it as standing on the shoulders and making a pyramid of (I: -knowledge) of knowledge." [Int-PhD3]

"But then there's the frustration of well, you know I'm writing this journal article […] not just to tell a story or to, to improve [knowledge]. But I'm also doing it to advance my career. And so that ends up being like a [bottleneck] on that sharing, right? It ends up being, you know, 'I don't want to share this material because I have to then write another paper on it'. Or uhm-, and which might have some legitimacy or might have more if you said something like, have a student who needs to finish the PhD. And so they need to […] to safely finish their project. […] And so then the journals become uh, uhm feel like-, or at least feel like – I don't know if the journals do – but I feel like uhm that the prioritization is about helping people to publish papers rather than having people to participate in this, this networking of uhm of of resources. And so, when we talk about that inconsistent loop, you know it might be useful if people got credit for the releasing of data, rather than for the writing of papers." [Int-Journal]

"It's not valid to assume that the scientific community wouldn't want to serve the societal benefits. Think about why people are doing it. If researchers in universities were in for the money, – come on, they would better do something else, it's definitely not the best way to become rich. So they want to achieve something, – why do they want to achieve this? Don't they want to make some difference in this world? So isn't that a societal benefit? They want to make the whole process as efficient as possible. The general interests of a lot of researchers are much aligned to the genuine public interests. Why then do we have a system that is completely offset on this? It not only doesn't make sense, it's also very inefficient. You know, we're publishing more and more and more, the quality doesn't get better, and the quality standards drop in my opinion. We have an inflationary system because we attach larger and larger numbers of research excellence to it, it doesn't mean anything. – It's becoming worse and worse year by year." [Int-Faculty2]

# Chapter 4: Discussion & Conclusion

The rational choice framework provided us with the appropriate tool to reconstruct the *structural conditions* present in the *logic of the situation* of an astronomer. In a next step, we translated the *situation* into *bridge hypotheses*, which correspond to the variables and motivational factors flowing into an astronomer's research behaviour. We adopted the *EU-theory* as an *action theory* to explain how, both intrinsic and extrinsic motivational factors, resulting from the four components of the situation, play together to explain astronomer's research behaviour. We have shown that this can be best described as a balance act between the primary and secondary code of science, whereby gaming strategies are employed in order to meet the requirements of the evaluation system. The initial discrepancy between the intrinsic values of an astronomer and what counts as good science as measured by a performance actor then constitutes, by means of the *logic of aggregation*, an evaluation gap on the institutional level.

The astronomers' ***3 quality criteria*** are to push knowledge forward by producing original, methodologically sound & error-free outcomes and communicating those to the community and the public. They are comparable with what Langfeldt et al. (2020; p.120) call the F-type (i.e. field type) notions of quality in science: "originality/novelty, plausibility/ reliability, and value or usefulness". By contrast, S-type (i.e. research-space type) notions of research quality "originate in policy and funding spaces" (ibid. p.119). Those are the ones that are constituted by the use of performance indicators. Scientific excellence and productivity is mostly determined by those indicators; the amount of papers ("the more the merrier, regardless of quality, unfortunately" [Int-Postdoc2]), citation rates, invitations to talks & conferences, being part of review panels for telescope allocations times and journals, supervising students, having a good network of collaborators and luck. As Jansen et al. (2010; p.223) write, "traditional, truth-oriented scientific quality criteria are being replaced by pragmatic, demand-driven criteria of functionality as defined by the stakeholders." To survive in academia, astronomers feel like not having another choice than complying to the institutional norms, such as being evaluated by performance indicators; they accept it as part of "the system" (Heuritsch, 2019a)." The balance act then is the art of playing the indicator game (such that Secondary Code ↑), while at the same time compromising research quality as little as possible (to avoid Research Quality ↓). Gaming strategies shall give the appearance of *compliance*, while institutionalised means how to achieve a good bibliometric record are used in *innovative ways*, such as salami slicing or going for easy publications. They best fit the third way of how to resolve a cognitive dissonance: by favouring the pursuit of the secondary code in the short-term, one hopes to get into a position where one can focus on the primary code in the long-term. This leads to an overall decrease in research quality (Research Quality ↓).

Langfeldt et al. (2020; p.116) proposed to "to shift focus away from questions that essentially address what research quality is, and/or how to measure it, towards the mechanisms through which dominant notions of research quality become established and the social and intellectual tensions arising from co-existence of (potentially) conflicting quality notions". By employing RCT we did exactly that. The tensions we found (*sections 1.5.2-1.5.4, & 2.2*) are the cognitive dissonances between competition versus collaboration, guaranteeing usefulness versus risky & innovative ideas, pressure versus freedom and finally, primary versus secondary value, which amounts to quality versus quantity. By means of the balance act astronomers cope with these tensions.

Before reaching the final conclusions we would like to come back to the concept of a *moral economy*. Atkinson-Grosjean & Fairley (2009) argue that RCT cannot fully explain the moral economy of astronomy. We disagree. The authors draw that conclusion by making a wrong interpretation of their observations; "rational choice theory undermines many of the values and ideals thought constitutive of the practice of science" (ibid. p.162). This is wrong, because, RCT cannot undermine values. Just like "the adjective 'moral' is descriptive not evaluative" (ibid. p.148) in the concept of a moral economy, RCT does not define what action has value and what is not. Assuming that, according to RCT, an act is more valuable when it is rational in our daily use of the word (i.e. high investment of cognitive resources), is a severe misunderstanding of the theory. Instead, RCT advocates for an efficient use of resources. In fact, RCT explains, why in most cases it is actually very rational, not to invest cognitive resources and follow scripts instead. Decreasing the load of cognitive resources we need to activate by re-parametrizing the situation is one of the very functions of institutional norms. Hence, RCT is a tool to explain and not to prescribe how an actor acts, by reconstructing the values the person holds and what external conditions are present. This is why we claim that, by employing RCT, one can explain the moral economy of astronomy. Kohler (1999) points out that studying "scientific practices requires attention to material culture (instruments, research tools, and methods) and moral economy (social rules and customs that regulate the community), and how these two dimensions work together in a particular line of work" (Atkinson-Grosjean & Fairley, 2009; p.151). This is exactly what we have done by applying RCT. McCray (2000; p.685) employs the concept of a moral economy – "the unwritten expectations and traditions that regulate and structure a community" – "as an analytical model to examine how astronomers and science managers allocate resources." This also what we have done. In fact, we have shown how indicator use shapes the moral economy in astronomy. We agree with McCray (2000; p.688), that the "objects of value" of astronomy's moral economy include "an adequate amount of observing time, resources to build and operate new telescope facilities, and funding for one's research". That is how astronomical instruments "shape research opportunities, and they affect careers and institutions" (Baneke, 2019; p.26) and hence they have constitutive effects on how science is done. We further agree with McCray (2000; p.702) that "allocation of resources is not a democratic process" – it is based on political and strategic considerations, as well as performance indicators. We did observe that people from institutes who were involved in building a telescope or who come from prestigious universities have privileged access to telescopes, but did not test whether we find a historical distinction between the "Haves and Have-Nots" like McCray (2000) and Atkinson-Grosjean & Fairley (2009) did. Furthermore, McCray (2000; p.686) writes that "the moral economy of astronomy functions through negotiations and compromises". We found the balance act as an important example for a way to compromise between intrinsic and extrinsic motivation to publish papers.

The aim of this paper was provide an integrated causal theory about what effects indicator use has on research quality in astronomy, by means of employing RCT. We conclude that the current structural conditions in astronomy leads astronomers to focus on quantity rather than quality with regards to their publications. To understand the mechanisms that we have outlined in this paper is important for policy makers and funders if it is their aim to provide conditions that encourage quality research instead. Although employing RCT, this is a solely qualitative study. Future studies may take advantage of RCT's full potential and quantify findings that we have made. Moreover, we suggest to conduct an evaluative inquiry[15], where astronomers are included in development of alternative indicators and incentive structures that make for a more quality encouraging research environment. This poses the challenge that any performance indicator quantifies quality and hence is predestined to fall an easy prey to

[15] https://blogs.lse.ac.uk/impactofsocialsciences/2018/11/29/the-evaluative-inquiry-a-new-approach-to-research-evaluation/

Goodhart's law. However, we are confident that by having reconstructed the structural conditions and mechanisms at play in astronomy today, we have made an important contribution to the development of new ways of resource allocation that encourages quality astronomy over quantity.

## Acknowledgements

First, I would like to extend my gratitude to the 19 interviewed astronomers for their time and openness. Second, Thea Gronemeier and Florian Beng – the students assistants in our junior research group "Reflexive Metrics" – were a huge help all along the way; not only by assisting me with the rather boring and monotonous research tasks, but also being a great mental support. Finally, a big thank you for the people who hosted me during the first COVID-19 lockdown, during which this paper was written. It was an incredibly productive time full of professional and personal growth.